\documentclass[twocolumn]{aastex631}

\usepackage{natbib, amsmath}
\usepackage[]{xcolor}
\usepackage{xspace}

\def\macs{MACS J0018.5+1626 }
\def\macscom{MACS J0018.5+1626, }
\def\macsper{MACS J0018.5+1626. }

\def\E{{\cal E}}
\def\F{{\cal F}}
\def\R{{\cal R}}

\shorttitle{ICM-SHOX I: \macs}
\shortauthors{Silich et al.}

\begin{document}

\title{ICM-SHOX. Paper I: Methodology overview and discovery of a gas--dark matter velocity decoupling in the \macs merger}

\author[0000-0002-1616-5649]{Emily M. Silich}
\affiliation{Cahill Center for Astronomy and Astrophysics, California Institute of Technology, Pasadena, CA 91125, USA}

\author[0000-0001-6411-3686]{Elena Bellomi}
\affiliation{Center for Astrophysics $\vert$ Harvard \& Smithsonian, 60 Garden St., Cambridge, MA 02138, USA}

\author[0000-0002-8213-3784]{Jack Sayers}
\affiliation{Cahill Center for Astronomy and Astrophysics, California Institute of Technology, Pasadena, CA 91125, USA}

\author[0000-0003-3175-2347]{John ZuHone}
\affiliation{Center for Astrophysics $\vert$ Harvard \& Smithsonian, 60 Garden St., Cambridge, MA 02138, USA}

\author{Urmila Chadayammuri}
\affiliation{Center for Astrophysics $\vert$ Harvard \& Smithsonian, 60 Garden St., Cambridge, MA 02138, USA}

\author[0000-0002-1098-7174]{Sunil Golwala} 
\affiliation{Cahill Center for Astronomy and Astrophysics, California Institute of Technology, Pasadena, CA 91125, USA}

\author{David Hughes}  
\affiliation{Instituto Nacional de Astrofísica, Óptica, y Electrónica (INAOE), Aptdo. Postal 51 y 216, 7200, Puebla, Mexico}

\author[0000-0003-4229-381X]{Alfredo Montaña}
\affiliation{Instituto Nacional de Astrofísica, Óptica, y Electrónica (INAOE), Aptdo. Postal 51 y 216, 7200, Puebla, Mexico}

\author[0000-0003-3816-5372]{Tony Mroczkowski}
\affiliation{European Southern Observatory, Karl-Schwarzschild-Str. 2, D-85748, Garching, Germany}

\author[0000-0002-6766-5942]{Daisuke Nagai} 
\affiliation{Physics Department, Yale University, New Haven, CT 06520, USA}

\author{David Sánchez} 
\affiliation{Consejo Nacional de Ciencia y Tecnología-Instituto Nacional de Astrofísica, Óptica, y Electrónica (CONACyT-INAOE), Luis Enrique Erro 1, 72840 Puebla, Mexico}

\author[0000-0003-0122-0841]{S. A. Stanford} 
\affiliation{Department of Physics and Astronomy, University of California, Davis, CA 95616, USA}

\author{Grant Wilson}
\affiliation{University of Massachusetts, Amherst, MA 01003, USA}

\author[0000-0001-8253-1451]{Michael Zemcov}
\affiliation{Rochester Institute of Technology, Rochester, NY 14623, USA}

\author[0000-0002-0350-4488]{Adi Zitrin}
\affiliation{Ben-Gurion University of the Negev, P.O. Box 653, Be’er-Sheva 8410501, Israel}

\begin{abstract}
Galaxy cluster mergers are rich sources of information to test cluster astrophysics and cosmology. However, cluster mergers produce complex projected signals that are difficult to interpret physically from individual observational probes. Multi-probe constraints on the gas and dark matter cluster components are necessary to infer merger parameters that are otherwise degenerate. We present ICM-SHOX (Improved Constraints on Mergers with SZ, Hydrodynamical simulations, Optical, and X-ray), a systematic framework to jointly infer multiple merger parameters quantitatively via a pipeline that directly compares a novel combination of multi-probe observables to mock observables derived from hydrodynamical simulations. We report a first application of the ICM-SHOX pipeline to \macscom wherein we systematically examine simulated snapshots characterized by a wide range of initial parameters to constrain the \macs merger geometry. We constrain the epoch of \macs to the range $0$--$60$ Myr post-pericenter passage, and the viewing angle is inclined $\approx 27$--$40$ degrees from the merger axis. We obtain constraints for the impact parameter ($\lesssim 250$~kpc), mass ratio ($\approx 1.5$--$3.0$), and initial relative velocity when the clusters are separated by 3~Mpc ($\approx 1700$--3000~km~s$^{-1}$). The primary and secondary clusters initially (at 3 Mpc) have gas distributions that are moderately and strongly disturbed, respectively. We discover a velocity space decoupling of the dark matter and gas distributions in \macscom traced by cluster-member galaxy velocities and the kinematic Sunyaev-Zel’dovich effect, respectively. Our simulations indicate this decoupling is dependent on the different collisional properties of the two distributions for particular merger epochs, geometries, and viewing angles. 
\end{abstract} 


\section{Introduction} \label{sec:intro}

The standard $\Lambda$ cold dark matter ($\Lambda$CDM) cosmological model predicts that structures in the universe form hierarchically \citep[e.g.,][]{millenium}. In the early universe, weak positive fluctuations in the cosmic density field formed small overdensities, which overcame cosmic expansion and collapsed via gravity. From these density perturbations, larger structures form throughout cosmic time by merging and smooth accretion. The current ($z \sim 0$) stage of cosmological structure formation in the universe is the formation and growth of galaxy clusters\footnote{As the expansion of the Universe continues to accelerate under the standard $\Lambda$CDM model, the largest expected virialized structures are only a few times larger than the most massive clusters currently known \citep[e.g.,][]{largestobjs}.}. With total masses of $\sim10^{14}$--$10^{15}$ M$_{\odot}$, galaxy clusters are comprised of two main mass components: dark matter (DM; $\sim80$--$90$\% of the total mass) and baryonic matter ($\sim10$--$20$\%), which is dominated by hot, diffuse plasma in the intracluster medium \citep[ICM; see][for reviews]{voitrev,kbreview}.

While some growth of galaxy clusters can be attributed to smooth accretion of streams of material from the cosmic web and lower mass structures (e.g., galaxies), the primary hierarchical growth mechanism of clusters is major mergers between similar mass systems \citep{muldrew15, Molnar2016}. Such events can drive bulk motions and large-scale turbulence in the ICM with velocities of $\sim 10^3$ km~s$^{-1}$ and heat the gas up to temperatures of $\sim10^7$--$10^8$ K \citep{Markevitch2007}. Studies of galaxy clusters can, therefore, play two key scientific roles. First, because of the enormous scales of distance \citep[a few Mpc;][]{voitrev} and energy \citep[$\lesssim 10^{64}$ erg;][]{Molnar2016} that clusters probe, generic physical processes like turbulence, shocks, and accretion play out to an unprecedented degree in mergers. Consequently, complementary observational and simulation-based studies of major mergers are a powerful means to further our understanding of these ubiquitous physical phenomena and the role they play in cosmological structure formation. Second, population statistics of major mergers between massive clusters of galaxies are a sensitive diagnostic of the underlying cosmological model  \citep[e.g.,][]{lacey1993, Fakhouri2010, Thompson2012}. For example, major mergers trace the velocity distribution of halos at the extreme end of the mass function, so observed velocities from well-characterized samples can be compared to velocity distribution predictions derived from cosmological simulations in order to test various cosmological parameterizations \citep{Molnar2016}. Exceptional mergers \citep[e.g., the ``Bullet Cluster'' 1E 0657-558;][]{Markevitch2002} have the potential to provide powerful tests of the $\Lambda$CDM paradigm, and much simulation work has been undertaken to understand the cosmological implications of these mergers \citep[see][]{thompson2015}.  

Observational diagnostics of galaxy cluster mergers have included X-ray imaging and spectroscopy, gravitational lensing (GL) mass reconstructions, radio relics, thermal Sunyaev-Zel’dovich (tSZ) effect imaging, and positional offsets between these observables \citep{clowe2006, russell22, vanWeeren2010, dimascolo2021, Molnar2012, Molnar2016}, which can generally be reproduced by both cosmological and idealized simulations \citep[see][]{ZuHone2018}. However, observational comparisons to simulations have primarily been limited in two ways. Most analyses involve only a single object because of the need for deep, multi-probe data in order to characterize a merger adequately. Furthermore, these analyses focus on mergers occurring in or near the plane-of-sky (POS), so that the morphological signatures associated with the merger can be clearly identified, thus making them more straightforward to interpret. To overcome these limitations, one can also make use of line-of-sight (LOS) velocity information, which expands the set of mergers that can be analyzed and may also be critical to obtaining conclusive results from comparisons to simulations 
\citep[e.g.,][]{Chadayammuri22}. 

LOS velocity information in mergers has typically been obtained from redshifts of cluster-member galaxies, which are believed to reliably trace the DM distribution \citep{Ma2009, Owers2011, Boschin2013}. However, as mergers evolve, the DM velocities are expected to decouple from the ICM velocities, since the collisional ICM is affected by hydrodynamical processes, while the DM is collisionless, interacting only via gravity \citep{Poole2006}. While the existence of a velocity space difference between the ICM and DM can be inferred from positional offsets of the two components \citep[e.g.,][]{merten2011}, a direct measurement of this velocity decoupling has not yet been achieved observationally, since this requires spatially resolved measurements of both the ICM and DM LOS velocities. In the X-ray band, such observations can be carried out with microcalorimeter instruments, as shown by the observations of bulk velocities in the Perseus cluster with \textit{Hitomi} \citep{hitomi2016, hitomi2018}. While this technique will be made available by upcoming instruments on \textit{XRISM} \citep[][]{xrism}, \textit{Athena} \citep{athena}, and \textit{LEM} \citep{lem_white_paper}, it is not currently available. Recently, ICM LOS velocities of individual clusters have been measured using the kinematic Sunyaev-Zel’dovich (kSZ) effect \citep{Sayers2013, Adam2017, sayers2019}, which is a Doppler shift of the cosmic microwave background (CMB) signal due to the bulk motion of galaxy clusters (\citealt{Sunyaev1980}; see \citealt{Mroczkowski2019} for a review). By incorporating LOS ICM velocities derived from current kSZ measurements, it is now possible to directly probe both the POS morphology and LOS velocity features of both mass components (DM and ICM) in mergers.

In this paper, we introduce the ICM-SHOX (Improved Constraints on Mergers with SZ, Hydrodynamical simulations, Optical, and X-ray) sample, which comprises a set of deep, multi-probe data for eight massive, intermediate-redshift galaxy clusters. From these data we are able to characterize the POS morphology and LOS velocity structure of the constituent ICM and DM components for the first time in such a merger analysis. ICM-SHOX contains four galaxy cluster mergers likely occurring primarily along the LOS, three mergers likely occurring in the POS, and one relaxed object as a control (see Table \ref{tab:mergers_table}). The quality of these multi-probe data is relatively uniform across each object in the sample. The ICM-SHOX sample consists primarily of exceptional (e.g., high mass, actively merging) objects, which allows us to test for deviations from $\Lambda$CDM in the most extreme regime of structure growth in the universe. 

In order to determine the geometry of these mergers and obtain robust population statistics of cluster component masses, gas profiles, initial impact parameters, initial relative velocities, merger epochs, and viewing angles, we compare observables derived from the multi-probe data to analogous mock observables generated with tailored idealized hydrodynamical binary galaxy cluster merger simulations. We have developed an automated pipeline which enables us to reduce the observational data for each cluster, generate analogous mock observables from simulations characterized by a vast array of initial parameters, and directly compare the datasets. Our framework then jointly infers the above merger parameters for each cluster quantitatively via a frequentist statistical analysis. In this study, we demonstrate the capabilities of this pipeline applied to our novel combination of multi-probe data for one member of ICM-SHOX, \object{MACS J0018.5+1626}. 

In section \ref{sec:obsdata}, we describe the observational data for the full ICM-SHOX sample and detail the data reduction and analysis procedures for \macsper Section \ref{sec:sims} contains an overview of the hydrodynamical simulations we employ and describes our process of generating mock observables for direct comparison to the observational data. In sections \ref{sec:comparisons} and \ref{sec:results}, we describe methods for constraining the \macs merger parameters with the ICM-SHOX pipeline and the resulting constraints, respectively. We give conclusions and applications to the full ICM-SHOX sample in section \ref{sec:conclusions}. We assume a spatially flat $\Lambda$CDM cosmology with $H_0 = 70$ km s$^{-1}$ Mpc$^{-1}$ and $\Omega_{m, 0} = 0.3$, unless otherwise specified. 

\section{Observational data} \label{sec:obsdata}
\begin{deluxetable*}{lccccc}[t]  \label{tab:mergers_table}
\caption{Details of the galaxy clusters in ICM-SHOX$^{*}$}
\tablehead{\colhead{Cluster name} & \colhead{RA (HMS)} & \colhead{Dec (DMS)} & \colhead{$z$} & \colhead{M$_{500}$ ($10^{14}$ M$_{\odot}$)} & \colhead{Dynamical state}}
\startdata
    Abell 0697  & 08:42:57.6  & +36:21:57  & 0.28  & 17.1 &  likely LOS merger \\
    Abell 1835  & 14:01:01.9  & +02:52:40  & 0.25  & 12.3   & relaxed \\
    \macs  & 00:18:33.4  & +16:26:13  & 0.55  & 16.5 &  likely LOS merger \\
    MACS J0025.4$-$1222  & 00:25:29.9  & $-$12:22:45  & 0.58  & $\phantom{1}$7.6  & likely POS merger \\
    MACS J0454.1$-$0300  & 04:54:11.4  & $-$03:00:51  & 0.54  & 11.5 &  likely POS merger \\
    MACS J0717.5+3745  & 07:17:32.1  & +37:45:21  & 0.55  & 24.9 &  likely LOS merger \\
    MACS J2129.4$-$0741  & 21:29:25.7  & $-$07:41:31  & 0.59  & 10.6 &  likely LOS merger \\
    RX J1347.5$-$1145  & 13:47:30.8  & $-$11:45:09  & 0.45  & 21.7 &  likely POS merger 
\enddata 
\tablenotetext{}{$^{*}$Summarized from Table 1 of \cite{sayers2019} and references therein.}
\end{deluxetable*}

\subsection{The ICM-SHOX sample}
For each object in the sample, we characterize the POS DM morphology using projected total mass ($\Sigma$) maps from strong GL models fit to imaging data from the \textit{Hubble Space Telescope} (\textit{HST}; e.g., \citealt{zitrin2011,zitrin2015}). The POS ICM morphology is described by two observables. First, we construct X-ray surface brightness (XSB) and temperature (kT) maps using observations taken with the \textit{Chandra X-ray Observatory}. Then, we obtain projected ICM density maps using the SZ effect \citep{sayers2019} measured via a combined analysis of data from \textit{Bolocam}, \textit{AzTEC}, and \textit{Planck}. The LOS DM velocity is traced using spectroscopic redshifts ($z_{\text{spec}}$) of cluster-member galaxies obtained primarily with the \textit{DEIMOS} instrument \citep{faber2003} at the \textit{Keck Observatory}. We characterize the LOS ICM velocity (ICM v$_{\text{pec}}$) using projected ICM velocity maps from the SZ effect, again based on an analysis of \textit{Bolocam}, \textit{AzTEC}, and \textit{Planck} observations \citep{sayers2019}. 

\begin{figure*}[t] 
    \centering 
        \includegraphics[width=1.0\textwidth]{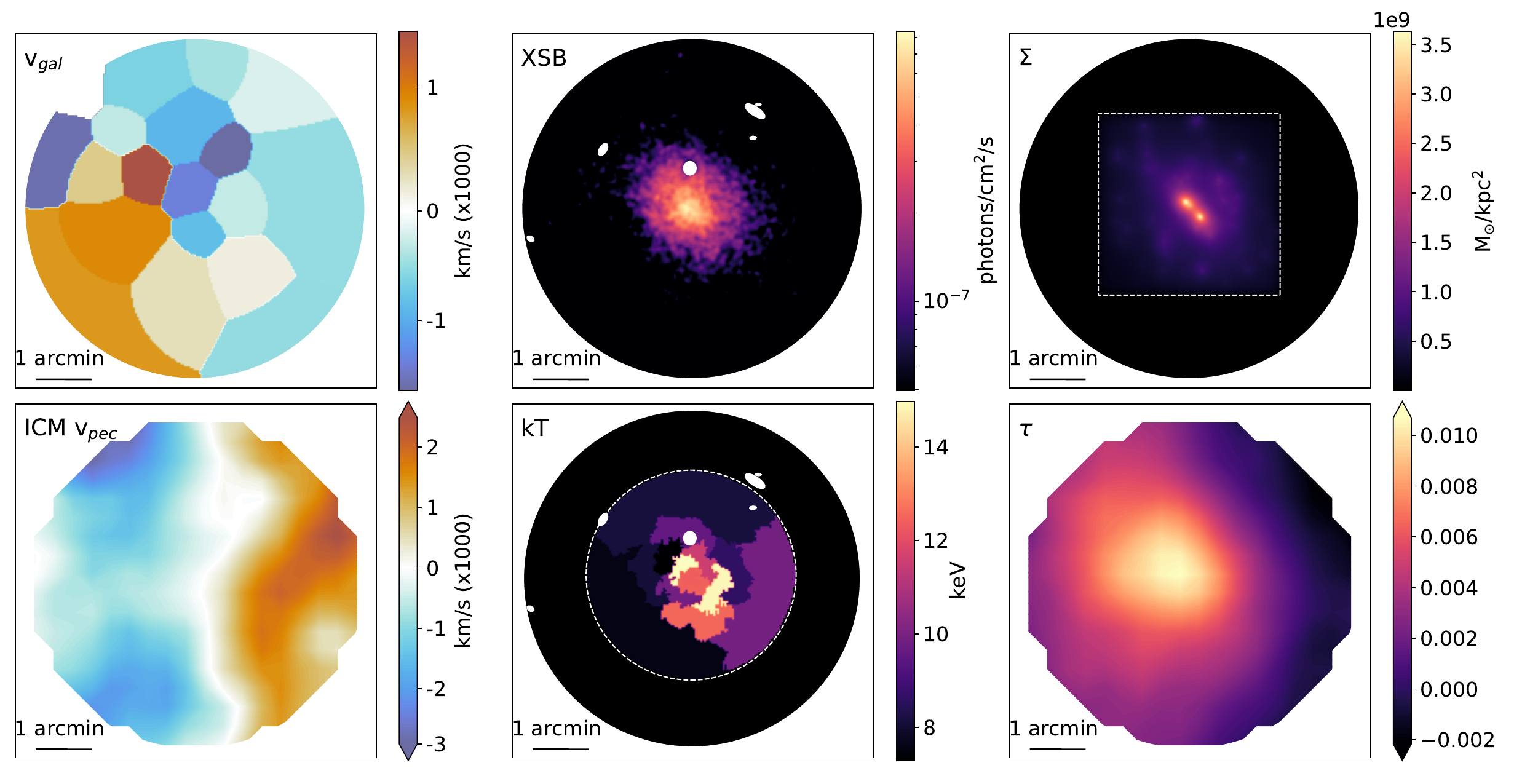}
        \caption{\macs multi-probe observations: \textit{top row, from left:} LOS galaxy velocities (v$_\text{gal}$), X-ray surface brightness (XSB), GL projected total mass ($\Sigma$); \textit{bottom row, from left:} LOS ICM velocities (ICM v$_{\text{pec}}$), X-ray derived temperatures (kT), LOS ICM electron optical depth ($\tau$).} \label{fig:obsdata}
\end{figure*}

Each dataset probes $> 3'$ from the center of the cluster, with the exception of the GL maps (which are limited by the \textit{HST} field of view) and the kT maps (which are only constructed within a central region where the cluster counts are comparable to background counts; see section \ref{subsec:xray}). At the redshift of the ICM-SHOX sample, this angular extent corresponds to $\sim1$ Mpc in the POS, which is typically large enough to capture primary features of the merger (e.g., separate cluster mass peaks and merger-driven shocks). In this analysis, we focus on one member of the ICM-SHOX sample, \macscom which is a massive, extensively studied cluster merger \citep[see, e.g.,][]{Solovyeva2007} at $z = 0.546$ \citep{redshifts} that is likely elongated along the LOS \citep{Piffaretti2003, sayers2019}. We describe each observational dataset (see Figure \ref{fig:obsdata}) and corresponding data processing for \macs below. 

\subsection{X-ray}  \label{subsec:xray}
The X-ray data reduction was performed with \texttt{CIAO} version 4.14 \citep{ciao}. \macs observations correspond to \dataset [Chandra ObsId 520]{https://doi.org/10.25574/00520}. The raw ACIS-I data for ObsID 520 were calibrated using the CalDB version 4.9.4 with \texttt{chandra$\_$repro}. The data were exposure-corrected with \texttt{fluximage}. Point sources were identified using the \texttt{CIAO} implementation of a wavelet source detection method \citep[\texttt{wavdetect};][]{Freeman2002} and excluded from the exposure-corrected images. The light curve was flare-filtered with \texttt{deflare} to identify good-time-intervals (GTI), which were applied to the data. The total GTI after data filtering is $63.8$ ks. We used the \texttt{flux\_obs} tool to generate a filtered, exposure-corrected $0.5$--$7$ keV XSB map free of point sources, which, having $1.7 \times 10^4$ source counts, is sufficiently deep to resolve morphological features. 

The \texttt{blanksky} routine was used to construct blank-sky background event files scaled and reprojected to match the \macs data. Since they are derived from deep, point source-free observations averaged across large regions of the sky, the background files nominally account for contributions from the particle-induced instrumental background \citep{acis_bkg}, as well as astrophysical foreground components (e.g., the Local Hot Bubble and the hot Galactic halo) and background components (e.g., the contributions from unresolved extragalactic point sources to the cosmic X-ray background). 

In order to generate robust kT maps, we first identified a circular region centered on \macscom within which the cluster source counts ($N_{\text{cluster}}$) are comparable to or larger than the background counts within each pixel ($N_{\text{BKG}}$). We estimated $N_{\text{cluster}} + N_{\text{BKG}}$ for each pixel in the $0.5$--$7$ keV filtered events image (binned to a pixel size of 4$''$ to reduce statistical variation between pixels), and we derived $N_{\text{BKG}}$ with the analogously binned blank-sky background image. The resulting circular region within which $N_{\text{source}} \sim N_{\text{BKG}}$ for $\gtrsim 75$\% of the pixels has $r = 1.8$\arcmin. 

Then, the $0.5$--$7$ keV filtered events image was contour binned within the 1.8\arcmin$\;$ circular region using the \texttt{contbin} algorithm \citep{Sanders2006}. The $0.5$--$7$ keV blank-sky background image was specified for the signal-to-noise (S/N) calculations used to determine the bin edges. We extracted source spectra from the filtered event file and background spectra from the blank-sky event file for each of the regions defined by the contour bins with \texttt{specextract}. We tailored the contour binning algorithm input parameters so each cluster bin has $\sim1000$ counts after the blank-sky background subtraction to maximize the kT map spatial resolution while maintaining sufficient counts to obtain a high-quality spectral fit. Each extracted spectrum was binned to have a minimum of 25 counts per spectral bin to allow the use of $\chi^2$ statistics in the spectral fitting. 

All spectral fitting was performed with \textit{Sherpa} over an energy range of $0.5$--$7$ keV. The blank-sky background-subtracted cluster bins were fit with a model for a single collisionally ionized plasma modified by interstellar absorption \citep[\texttt{tbabs $\times$ apec};][]{tbabs, Smith2001} with fixed $n_{\text{H}} = 4 \times 10^{20}$ cm$^{-2}$ \citep{absmap}, $z = 0.546$, and $Z = 0.3\; Z_{\odot}$ using abundances from \citet{angrabunds}. The plasma temperature and normalization were fit as free parameters. For \macscom the contour binning produces 14 spatial bins within $1.8'$ of the cluster center, within which temperature enhancements due to shocks produced in the merger are easily identifiable.

We then inspected the \macs XSB and kT maps for discontinuities associated with merger-driven shocks, which, if present, could be used to help constrain the merger geometry \citep[e.g.,][]{Markevitch2002} and evolutionary state \citep[e.g.,][]{Russell12}. We first applied a Gaussian gradient magnitude filter \citep{sanders2016, ggm2} to the \macs XSB map and identified enhancements which would be indicative of merger-driven shocks. For each of these enhancements, we generated radial surface brightness profiles and inspected them for evidence of surface brightness jumps. In a study of the Abell 2256 merger, \citet{Breuer2020} note that for highly inclined mergers (occurring near the LOS), temperature jumps can be more reliable indicators of the presence of a shock front. Therefore, in any regions which indicated the possible presence of a radial surface brightness jump, we fit downstream and upstream plasma temperatures to photons extracted from regions below and above the jump position, respectively. We applied the Rankine-Hugoniot jump conditions as a function of Mach number for the ratio of the downstream and upstream plasma temperatures across a plane-parallel shock \citep{LandL1959} to each enhancement-identified region of interest. Within uncertainties, all Mach numbers were consistent with one (i.e., no shock). Given our relatively low S/N, we do not use the presence (or lack) of clear shocks in the X-ray observables as a diagnostic of the merger geometry. 

\subsection{Gravitational lensing}  \label{subsec:lensing}
The projected total mass distribution of \macs was modeled using the Light Traces Mass (LTM) GL approach of \cite{zitrin2015} applied to \textit{HST} imaging. A full description of the method is given in \cite{zitrin2015}; in short, the LTM model assumes that cluster-member galaxy masses scale like their luminosities, and the total DM distribution follows a similar shape, so that it can be represented by a smoothed map of the cluster-member galaxies. The model is constrained by minimizing the distances of predicted multiple images from their actual observed locations with a Markov chain Monte Carlo (MCMC) minimization utilizing a $\chi^2$ function. 

A preliminary strong lensing model of \macs based on older \textit{HST} data was detailed in \citet{zitrin2011}. Here, we use a revised model which incorporates publicly available VLT/MUSE data (Program ID 0103.A-0777(B), PI: Edge) as well as dedicated Gemini/GNIRS data (Program ID: GN-2021B-Q-903, PI: Zitrin) for the cluster. These data reveal UV and optical emission lines as well as a (double-peaked) Ly$\alpha$ line at $z=3.21$ for the main lensed system \citep{furtak2022}. \textit{HST} imaging data from the RELICS survey \citep{Coe2019, relics_ref} has yielded two new multiple image systems, which are also incorporated in the revised version. In  total, the current model, which is similar to the one presented in \citet{furtak2022}, has ten free parameters and uses six sets of multiple image constraints from five background galaxies up to redshift $z\sim5$ as constraints. The final image reproduction RMS of the model is $0.86''$. In the resulting projected total mass distribution map, two primary mass peaks are clearly identifiable. 

\subsection{SZ effect}  \label{subsec:SZ}
\cite{sayers2019} produced projected LOS ICM density and velocity maps for each merger in our sample, including \macscom using SZ effect observations from \textit{Bolocam}, \textit{AzTEC}, and \textit{Planck}. The data were used to generate SZ effect maps at 140 and 270 GHz with a common point spread function with a full-width at half-maximum (PSF$_{\text{FWHM}}$) of 70\arcsec. The data from \textit{Planck} were used solely to set the absolute additive normalization of the SZ effect maps, since, while the ground-based data provide much better angular resolution than \textit{Planck}, the data processing procedures required to subtract atmospheric fluctuations also remove the normalization information. The 140 and 270 GHz data were combined with an X-ray-derived kT map to produce the LOS ICM density and velocity maps. We note that the kT map used in the SZ analysis is derived from the same data as the \macs kT map generated in this work (see Sec. \ref{subsec:xray}) but was calculated with an older X-ray analysis procedure. Differences between the X-ray reduction used by \cite{sayers2019} and that employed in the current work produce negligible changes in the SZ-derived quantities relative to the measurement noise. The S/N for the ICM density map is $\sim$5--10, and the uncertainty within each $70''$ resolution element in the ICM v$_{\text{pec}}$ map is $\sim1000$ km s$^{-1}$. 

\subsection{Cluster-member optical spectroscopy}  \label{subsec:specs}

Previous studies have shown that averaging redshifts of cluster-member galaxies is a useful measure of the bulk LOS DM velocity structure in a galaxy cluster \citep[e.g.,][]{Ma2009, Boschin2006, Dehghan2017}. Furthermore, cosmological simulations have indicated that the bias between the velocity dispersions of cluster-member galaxies and DM is relatively low \citep{Anbajagane2022}. Therefore, throughout this work, we assume that the LOS cluster-member galaxy velocities trace the LOS DM velocity structure.

We have combined both existing literature catalogs \citep{obsz_dressler, obsz_ellingson, obsz_crawford} with new \textit{DEIMOS} observations obtained by our group to collect 156 cluster-member $z_{\text{spec}}$ across the face of \macs within a $6.7' \times 6.7'$ region around the cluster center. The \textit{DEIMOS} data which were observed prior to August 2021 were reduced with the \texttt{spec2d} \textit{DEIMOS} data reduction pipeline \citep{spec2d_cooper, spec2d_newman}, and those observed from August 2021 onwards were reduced with a modern Python implementation of the \textit{DEIMOS} data reduction pipeline \citep[\texttt{PypeIt};][]{pypeit}. We used \texttt{SpecPro} \citep{specpro} to extract redshifts from the 1D reduced cluster-member galaxy spectra by identifying either the Ca II H$+$K absorption lines or the [O II] 3727$\textup{~\AA}$ doublet, with some objects additionally showing the H$\delta$ and G-band absorption lines. We provide tables of the \macs literature cluster-member redshifts, as well as (non-)cluster-member redshifts that were observed in our \textit{Keck} program in Appendix \ref{sec:appB}. 

Next, we converted each redshift relative to the median redshift of the cluster (0.546) to a velocity relative to the average SZ-derived bulk ICM LOS velocity of \macs ($v_{\text{bulk}} \simeq -100$ km s$^{-1}$; \citealt{sayers2019}). In other words, we define a cluster-member galaxy at $z = 0.546$ to be at rest relative to the bulk ICM LOS velocity. We thus enforce that the overall average observed bulk velocity of the ICM and the DM are equal. Then, we applied the Weighted Voronoi Tessellation (WVT) algorithm of \cite{Diehl2006} to bin the cluster-member velocities within the $6.7' \times 6.7'$ region into a high-fidelity spatial map. We use a publicly-available Python implementation of the WVT algorithm \citep[XtraAstronomy/Pumpkin;][]{rhea2020} for the spatial binning, and we apply estimators from \cite{Beers1990} to calculate the velocity ($v_\text{gal}$) within each spatial bin. The \macs LOS galaxy velocity map has 19 spatial bins across the face of the cluster.  

\section{Hydrodynamical binary merger simulations}  \label{sec:sims}

\begin{figure}[b] 
    \centering 
        \includegraphics[width=0.47\textwidth]{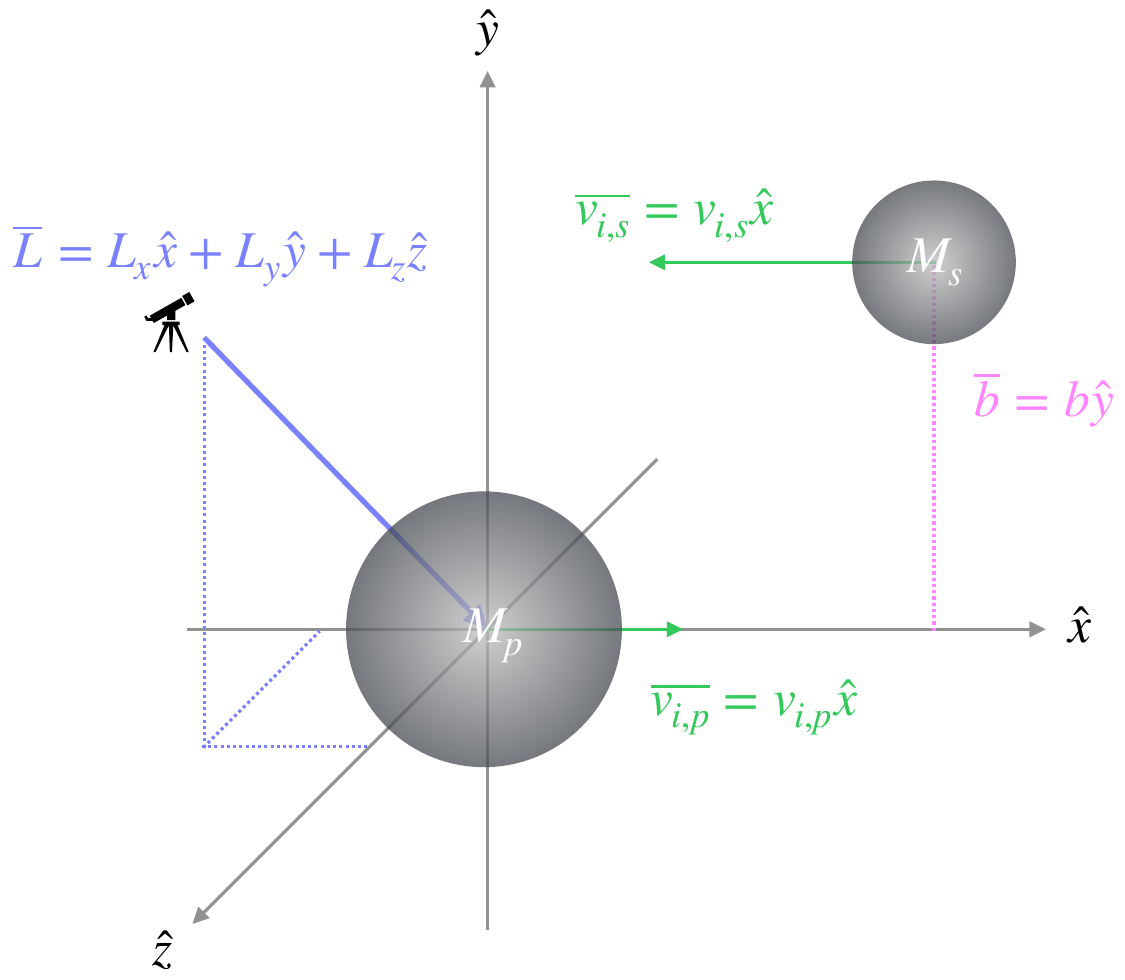}
        \caption{Schematic of the merger geometry initial configuration \citep[based on Figure 1 of][]{Chadayammuri22}. $M_P$ and $M_S$ are the primary and secondary cluster masses, respectively. $b$ is the impact parameter (defined along the $\hat y$ axis). $v_{i, p}$ and $v_{i, s}$ are the primary and secondary cluster initial velocities (defined in the $\hat x$ axis), where $v_i = |v_{i, p} - v_{i, s}|$. ${\bf L}$ is the viewing angle vector for simulation projections, defined with components in ($\hat x$, $\hat y$, $\hat z$).} \label{fig:diagram} 
\end{figure} 

\subsection{Simulated datasets}
We generate simulated observables analogous to the multi-probe \macs data by tailoring a suite of idealized hydrodynamical galaxy cluster merger simulations, after the manner of previous works \citep[e.g.,][]{Ricker2001,Poole2006,ZuHone2011,Chadayammuri22}. We use the GPU-accelerated code GAMER-2 \citep{Schive2018} to run the simulations, which solves the equations of hydrodynamics, N-body interactions, and gravity on an adaptive mesh refinement (AMR) grid. Each simulation is initialized as a binary cluster merger for a given choice of total mass, mass ratio $R = M_P/M_S$ of the primary to the secondary cluster (where both masses are defined as the enclosed mass within the cluster's $r_{200c}$), impact parameter $b$, and relative velocity of the infalling galaxy clusters $v_i$. 

We define the total \macs mass derived from the GL mass reconstruction map ($M_{500} = 1.1 \times 10^{15}$ M$_{\odot}$) in all simulations and use this to scale the masses of the two clusters pre-merger. We note that this GL mass estimate is lower than that derived from an X-ray scaling relation (see Table \ref{tab:mergers_table}). However, we select the GL mass estimate for our simulations since using the X-ray derived mass estimate results in inconsistencies when comparing the total surface density maps. 

The clusters are initially separated by a distance of $3$ Mpc, and $v_i$ is defined along the merger axis ($\hat x$) while $b$ is defined along the orthogonal $\hat y$ axis. A schematic of the merger initialization is shown in Figure \ref{fig:diagram}. Whenever projections of a simulation are taken, the viewing angle vector (${\bf L}$) represents the position of the `observer' (i.e., the vector whose perpendicular plane is the projection plane). Each cluster is comprised of DM and star particles and gas defined on the grid distributed by choices of total mass, gas, and stellar profiles; all assumed to be spherically symmetric and in hydrostatic and virial equilibrium. Any given deviation from these assumptions (e.g., triaxiality) would expand the parameter space beyond what is feasible for study, and furthermore, we expect deviations from spherical symmetry or HSE to produce effects which are far sub-dominant relative to the primary merger-driven processes. We describe the overall setup in more detail below. %

\subsubsection{Total mass profiles} \label{sec:totalmass}
Each cluster component begins in hydrostatic equilibrium (HSE) with the total mass following a truncated Navarro–Frenk–White (NFW) profile \citep{truncatedNFW}: 

\begin{equation} \label{eq:totalmass}
    \rho_{\text{tot}} (r) \propto \frac{1}{(r/r_s)(1+r/r_s)^2} \left( \frac{1}{1+(r/r_t)^2} \right)
\end{equation}
where $r_s = r_{200c}/c_{200c}$ is the scale radius and $r_t = 2r_{200c}$ is the truncation radius. For each cluster, the concentration parameter $c_{200c}$ is computed with the \texttt{colossus} package \citep{diemer2018} assuming a concentration$-$mass relation model from \citet{diemer2019} for the cosmological parameters calculated in \citet{planck18}. 

\subsubsection{Gas profiles} \label{sec:gasprofs}
The gas density profiles follow a modified $\beta-$profile as given in \citet{gasprof}:
\begin{equation}
    \rho_g (r) \propto \frac{(r/r_c)^{-\alpha/2}}{\left[ 1 + (r/r_c)^2 \right]^{3\beta/2 - \alpha/4}} \cdot \frac{1}{\left[ 1 + (r/r_s)^{\gamma} \right]^{\epsilon/(2\gamma)}}
\end{equation} where $r_c$ is the core radius, $\alpha$ is the inner slope parameter, $\beta$ is the $\beta$-profile parameter, $r_s$ is the scale radius ($> r_c$), $\gamma$ controls the width of the outer transition, and $\epsilon$ is the outer logarithmic slope parameter. Here, we neglect the additional additive term introduced by \citet{gasprof}, which controls the inner core shape. The gas mass is normalized to the total mass within $r_{500}$ through the assumed gas fraction from \citet[][]{sayers2019} based on the analysis of \citet[][$f_{\text{gas}} = 0.115$]{Mantz2010}. Assuming HSE and the total mass profile, this gas density profile uniquely determines the temperature, pressure, and entropy profiles. 

In this work, we utilize three gas profiles which vary the degree of disturbedness of the gas in each galaxy cluster: strongly disturbed (``non-cool-core''; NCC), moderately disturbed (``intermediate-cool-core''; Int), and undisturbed (``cool-core''; CC). The parameters for these profiles are given in Table \ref{tab:gas_params}. In the case of a simulation initialized with an undisturbed primary cluster gas profile and a secondary cluster gas profile which has been strongly disturbed, the total gas profile combination is notated CC$+$NCC, and the same logic applies for other combinations of gas profiles. 

\begin{deluxetable}{c|ccc}[h]  \label{tab:gas_params}
\caption{Merger simulation gas profile parameters}
\tablehead{\colhead{Parameter} & \colhead{NCC value} & \colhead{Int value} & \colhead{CC value}}
\startdata
    $r_c$ [r$_{2500}$] & 0.5 & 0.3 & 0.2 \\
    $\alpha$ & 0.1 & 0.3 & 1.0 \\
    $\beta$  & 0.67 & 0.67 & 0.67 \\
    $r_s$ [r$_{200}$] & 1.1 & 1.1 & 1.1 \\
    $\gamma$ & 3.0 & 3.0 & 3.0 \\
    $\epsilon$ & 3.0 & 3.0 & 3.0 
\enddata 
\end{deluxetable} \vspace{-2em}

\begin{figure}[t] 
    \centering 
        \includegraphics[width=0.48\textwidth]{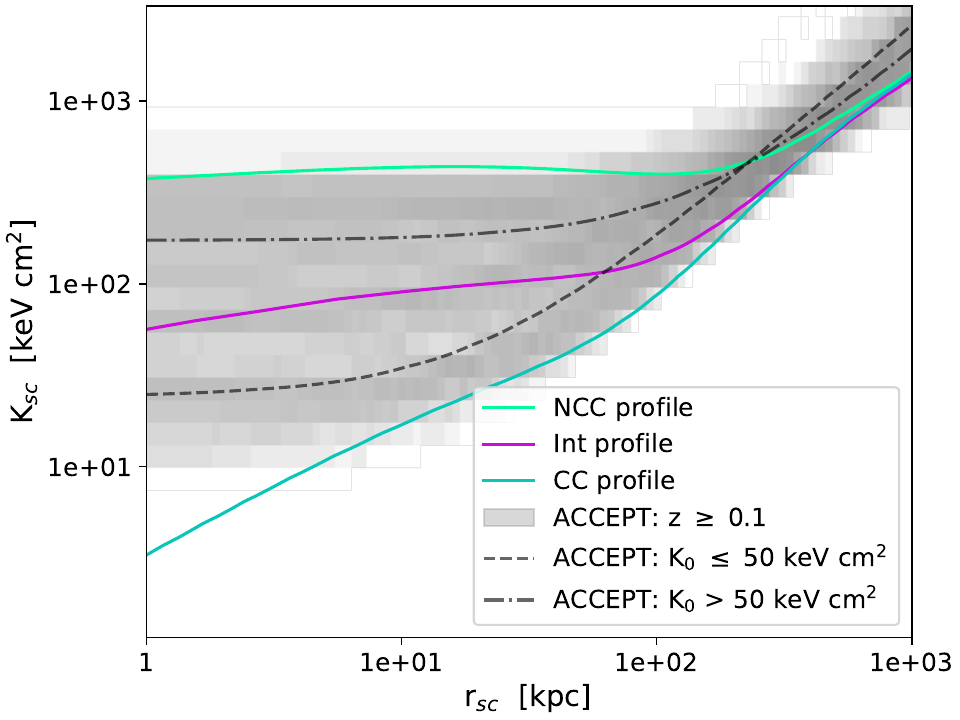}
        \caption{Scaled entropy distribution for ACCEPT galaxy clusters at $z \geq 0.1$. Simulated entropy profiles corresponding to our selection of initial gas profiles (NCC, Int, and CC) are overplotted. The simulated entropy profiles span the range of the observed distribution. } \label{fig:entropyfig} 
\end{figure}

Analogs to the entropy profiles of these three gas distributions can be found in the observed profiles from galaxy clusters in the ``Archive of Chandra Cluster Entropy Profile Tables'' \citep[ACCEPT;][]{entropy_accept} database. To illustrate this, in Figure \ref{fig:entropyfig}, we plot the distribution of scaled best-fitting entropy profiles for 131 ACCEPT galaxy clusters at $z \geq 0.1$.

In order to minimize the effects of mass differences between the ACCEPT clusters and \macscom we utilize scalings between $K_{500} - T_{500}$ and $r_{500} - T_{500}$ to normalize the $K$ and $r$ axes of each calculated ACCEPT cluster entropy profile. We apply scalings derived from the self-similar model of \citet{kaiser1986} in the form outlined by \citet{nagai2007}, namely that 
\begin{equation}
    K_{500} \propto M_{500}^{2/3} \cdot E(z)^{-2/3},
\end{equation} 
\begin{equation}
    T_{500} \propto M_{500}^{2/3} \cdot E(z)^{2/3}, 
\end{equation} and
\begin{equation}
    M_{500} \propto r_{500}^{3}, 
\end{equation} where $E(z)$ is the redshift scaling of the Hubble parameter. 

We take the average cluster temperature reported by \citet{entropy_accept} as an approximation of $T_{500}$ for each ACCEPT cluster, and scale the $K$ and $r$ axes of each profile to $K_{sc}$ and $r_{sc}$ according to the above scalings as follows: 
\begin{equation}
    r_{sc} = \frac{r}{(T_{\text{acc}} / T_{\text{cl}}) \cdot (E(z_{\text{cl}}) / E(z_{\text{acc}}))^{4/3}} 
\end{equation} and 
\begin{equation}
    K_{sc} = \frac{K(r)}{(T_{\text{acc}} / T_{\text{cl}})^{1/2} \cdot (E(z_{\text{cl}}) / E(z_{\text{acc}}))^{1/3}}
\end{equation} 
where $T_{\text{acc}}$ is the average cluster temperature of the ACCEPT cluster being scaled, $T_{\text{cl}}$ is the average cluster temperature of \macs as reported in the ACCEPT database, and $E(z_{\text{cl}})$ and $E(z_{\text{acc}})$ are the relative expansion rates at the redshifts of \macs and the relevant ACCEPT cluster, respectively. We further indicate the two populations identified by \citet{entropy_accept} by plotting the mean best-fit entropy profiles for clusters with core entropy $K_0$ above and below $50$ keV cm$^2$ scaled by the average $T_{\text{acc}}$ and $E(z_{\text{acc}})$ of each population. Figure \ref{fig:entropyfig} indicates that the NCC and CC entropy profiles are near the edges of the observed population, with Int approximately in the middle of the observed population.

With the gas profiles thus determined, the gas cells on the grid are assigned densities and pressures from these profiles, with their initial velocities set to zero in the rest frame of the cluster.

\subsubsection{DM and star particle initialization}
Once we have both the gas mass and total mass profiles, the difference between these two yields the mass profile of DM and stars. In our simulations, these are represented by particles that interact between themselves and the gas only via gravity. DM and star particles are identical in our simulations, the latter being tagged as ``star'' type only to serve as a tracer of the center of each cluster as the merger evolves. This is done by setting the stellar mass profile as a truncated NFW profile (Equation \ref{eq:totalmass}) with $\rho_{\rm s, star} = 0.9\rho_s$, $r_{\rm s, star} = 0.5r_s$, and $r_{\rm t, star} = r_t$, which represents the large concentration of stars at the cluster potential minimum from the brightest cluster galaxy. To ensure that there is not a significant stellar contribution to the mass at large radii, we apply an exponential cutoff to the stellar profile at $r = 20$~kpc. 

The particle positions and velocities are set up as discussed in previous works, most recently by \citet{Chadayammuri22}. For each of the particle positions, a random deviate $u = M(<r)/M_{\rm total}$ is uniformly sampled in the range [0, 1]. The mass profile $M(<r)$ for that particular mass type is inverted to give the radius of the particle from the center of the halo. 

For the DM and star particles, their initial velocities
are determined using the procedure outlined in \citet{Kazantzidis2004}, where the energy distribution function is calculated via the Eddington formula \citep{eddington1916}:
\begin{equation}
\F(\E) = \frac{1}{\sqrt{8}\pi^2}\left[\int^\E_0{d^2\rho \over d\Psi^2}{d\Psi \over \sqrt{\E - \Psi}} + \frac{1}{\sqrt{\E}}\left({d\rho \over d\Psi}\right)_{\Psi=0} \right] 
\end{equation} 
where $\Psi = -\Phi$ is the relative potential and $\E = \Psi - \frac{1}{2}v^2$ is the relative energy of the particle. We tabulate the function $\F$ in intervals of $\E$ to solve numerically for the distribution function at a given energy. Given the radius of the particle, particle speeds can then be chosen from this distribution function using the acceptance-rejection method. Once particle radii and speeds are determined, positions and velocities are determined by choosing isotropically distributed random unit vectors in $\Re^3$.

\subsubsection{Epoch readout} \label{sec:epochs}
Once the mass profiles have been determined and the particles initialized, each galaxy cluster component is assigned an initial position and velocity, and data products are generated for a set of epochs as the merger evolves. We read out the simulations in increments of $0.02$ Gyr, but only consider snapshots beginning 0.2 Gyr after the simulation initialization (defined as $t = 0$ Gyr) through $t = 2.1$ Gyr (well past the epoch of pericenter passage of a given merger). 

For each merger simulation, the epoch at which pericenter passage occurs ($t_{\text{peri}}$) is dependent on the initial conditions, most notably $v_i$. We therefore calculate $t_{\text{peri}}$ as the epoch at which the distance between the positions of the primary and secondary halos is minimized, where the position of each halo is calculated simply as the median position of the star particles associated with the halo. For any simulation epoch $t$, $(t - t_{\text{peri}})$ thus serves as an epoch proxy which can be compared from one simulation to another in a way that references time from the standard pericenter passage event, rather than the initial conditions.

\subsection{Mock observables} \label{subsec:mocks}
The best method of constraining galaxy cluster merger geometries from simulations is by comparing observational data to simulated observables \citep[e.g.,][]{Springel2007, Lage2014, Chadayammuri22}. We derive mock observables from the merger simulations using the \texttt{yt} software \citep{Turk2011} to project the XSB, total mass density, ICM density, kSZ effect amplitude, and galaxy velocities (using a subset of DM particles as proxies for the galaxies) into 2D images. For each mock observable, we use the \macs redshift to determine the transformation from physical units to angular units in the sky. We then process these mock observables to make them nominally equivalent to the quality of the observational data. This consists of convolving the SZ-derived ICM projected density and LOS velocity maps with a $70''$ FWHM Gaussian kernel, smoothing the XSB map to an effective resolution of $1.96''$, convolving the projected total density map with a $1.8''$ FWHM Gaussian kernel, and using the WVT and \cite{Beers1990} estimators described in section \ref{subsec:specs} to bin the cluster-member galaxy velocities into a spatial map. 

We use the \texttt{pyXSIM} package \citep{zuhone2016} to derive X-ray event files from the simulation data products in order to construct kT maps across the face of the cluster. We tailor the X-ray event files to include galaxy cluster emission, astrophysical foreground emission, and the \textit{Chandra}/ACIS-I particle background by implementing the \texttt{SOXS} \texttt{instrument$\_$simulator} tool \citep{soxs}. \texttt{pyXSIM} generates the simulated cluster emission using the cluster density, temperature, and velocity from the 3D simulations. In our case, we assume that the X-ray emission originates from an absorbed collisionally ionized plasma (\texttt{tbabs $\times$ apec}) with $n_{\text{H}} = 4 \times 10^{20}$ cm$^{-2}$, at $z = 0.546$ (as in the \macs observational data; see section \ref{subsec:xray}). As the simulation does not include metallicity, we assume $Z = 0.3\; Z_{\odot}$ throughout the cluster with \cite{angrabunds} abundances. The 3D events are projected along a specified LOS and written to a 2D event file. In \texttt{SOXS}, we create a blank sky background file by additionally writing the background/foreground events to a separate event file so that it can be subtracted in the same way as the observational blank sky background files in the X-ray temperature fitting (see section \ref{subsec:xray}). In this way, we construct simulated kT maps in an analogous way to the \macs observables, following the observational spectral fitting methods within the same $r = 1.8'$ circular region. 

\begin{deluxetable*}{cccl}[t]  \label{tab:features_table}
\caption{Key features in each \macs observational probe}
\tablehead{\colhead{Observable} & \colhead{Wavelength} & \colhead{Cluster component} & \colhead{Features}}
\startdata
    ICM v$_{\text{pec}}$ & SZ $+$ X-ray & LOS ICM velocity & -- $\sim$E--W dipole with magnitude scale $\sim4000$--$5000$ km s$^{-1}$ peak- \\ 
    & & & $\;\;\;$to-peak ($\pm \sim 1000$ km s$^{-1}$)  \\
    \rule{0pt}{4ex} v$_\text{gal}$ & Optical ($z_{\text{spec}}$) & LOS DM velocity & -- $\sim$NW--SE dipole with magnitude scale $\sim3000$ km s$^{-1}$ peak-to-peak \\
    \rule{0pt}{4ex} XSB & X-ray & POS ICM morphology & -- Extended/diffuse emission elongated along NE--SW axis \\ 
    & & & -- Lacking two well-separated emission peaks \\ 
    \rule{0pt}{4ex} kT & X-ray & POS ICM morphology & -- Hot, centrally-peaked temperature distribution elongated along \\ 
    & & & $\;\;\;$NE--SW axis with maximum cell at $\sim15$ keV \\ 
    & & & -- Lacking two well-separated cool cores \\ 
    \rule{0pt}{4ex} $\tau$ & SZ $+$ X-ray & POS ICM morphology & -- Centrally-peaked distribution with tail extending NE \\ 
    \rule{0pt}{4ex} $\Sigma$ & Optical (GL) & POS DM morphology & -- Two centrally-located mass peaks separated by $\sim22''$ along \\
    & & & $\;\;\;$NE--SW axis
\enddata 
\end{deluxetable*}

\section{Matching simulations to observations} \label{sec:comparisons}
In order to determine both the merger simulation initial conditions and the relevant observational parameters (e.g., the viewing angle of our LOS relative to the merger axis) which produce reasonable matches to the \macs observational data, we apply a set of matching criteria to mock observables generated with a wide range of simulation parameters (described in Section \ref{sec:matchingapp}). The matching criteria we apply to the resulting mocks are designed to select key features from the \macs observables, so that physical information (e.g., the merger epoch or viewing angles) can be determined. We manually calibrate these matching criteria on each \macs cluster observable, with the exception of the ICM v$_{\text{pec}}$ map. In this case, the low S/N of the ICM v$_{\text{pec}}$ map prevented us from identifying a categorical matching criterion, and we, therefore, developed a more formal statistical comparison for this observable (see Section \ref{subsec:ICMmap} below). The key features of each observational probe are summarized in Table \ref{tab:features_table}, and the associated matching criteria are described in the following sections. 

\subsection{ICM v$_{\text{pec}}$ map} \label{subsec:ICMmap}
The \macs kSZ-derived ICM v$_{\text{pec}}$ map morphologically resembles a $\sim$E--W dipole with a velocity difference of $\sim4000$--$5000$ km s$^{-1}$ peak-to-peak (though the uncertainties in this map are significant at $\sim 1000$ km s$^{-1}$). In order to maximally utilize the \macs ICM v$_{\text{pec}}$ information to place constraints on the simulated ICM v$_{\text{pec}}$ maps while taking these substantial observational uncertainties into account, we apply the following method. 

We begin by generating 1000 noise realizations ($N_{\text{X}}$) of the \macs observational ICM v$_{\text{pec}}$ map ($M_{\text{obs}}$). These realizations fully capture all of the noise fluctuations, including correlations, due to both the instrument and residual astrophysical contamination. In brief, the single-band noise realizations described in \citet{sayers2019} are created for both the 140 and 270~GHz SZ images. These include jackknife images to estimate noise due to the instrument and atmosphere, along with random realizations of primary CMB fluctuations and the relevant galaxy populations (dusty star-forming and radio AGN). We then obtain a v$_{\text{pec}}$ map for each of these realizations (i.e., the $N_{\text{X}}$) using the same procedure applied to the observed data. Finally, we compute the RMS of the 1000 realizations within each pixel to obtain $M_\text{RMS}$, which is a map that provides the positionally-dependent uncertainty per pixel. While $M_\text{RMS}$ is not a complete description of the noise due to pixel-pixel correlations, it can be used as a proxy to estimate the goodness of fit, as detailed below.

For each simulated ICM v$_{\text{pec}}$ map ($M_{\text{sim}}$), we begin by binning $M_{\text{sim}}$ to an identical pixel scale as $M_{\text{obs}}$ and trim both maps to a circular region of $r = 3'$ centered on \macs for consistency between the maps. For each $N_X$, we generate a noisy simulated map $M_{\text{noisy}} = (M_{\text{sim}} + N_{\text{X}}) - w$, where $w$ represents the uncertainty weighted mean of $(M_{\text{sim}} + N_{\text{X}})$. The value of $w$ is similarly calculated and subtracted from the observed map $M_{\text{obs}}$ in order to remove the overall bulk velocity of the system. This procedure gives us 1000 noisy simulated maps, from which we empirically determine a simulated $\chi^2$ distribution defined by the statistic:
\begin{equation}
\begin{split}
    \chi^2_{\text{sim}} &= \sum_i \frac{(M_{\text{noisy}, i} - M_{\text{sim}, i})^2}{(M_{\text{RMS}, i})^2} 
\end{split}
\end{equation} for $i$ pixels in each map.

\begin{figure*}
    \centering 
    \includegraphics[width=0.99\textwidth]{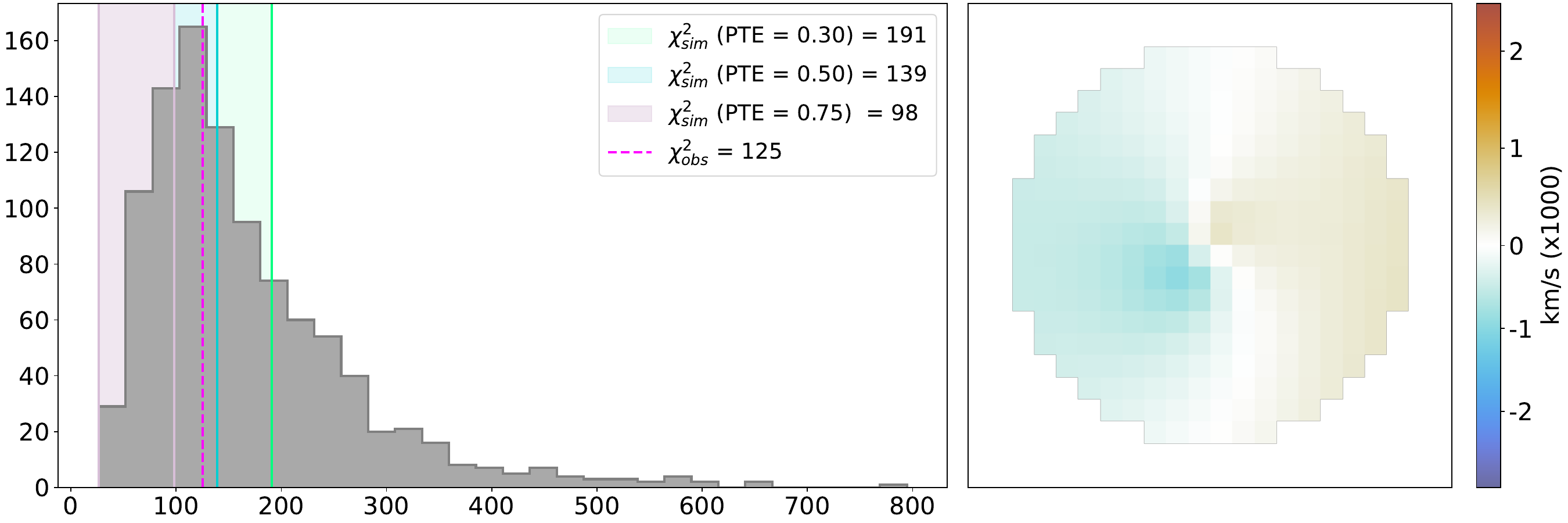}
    \includegraphics[width=0.99\textwidth]{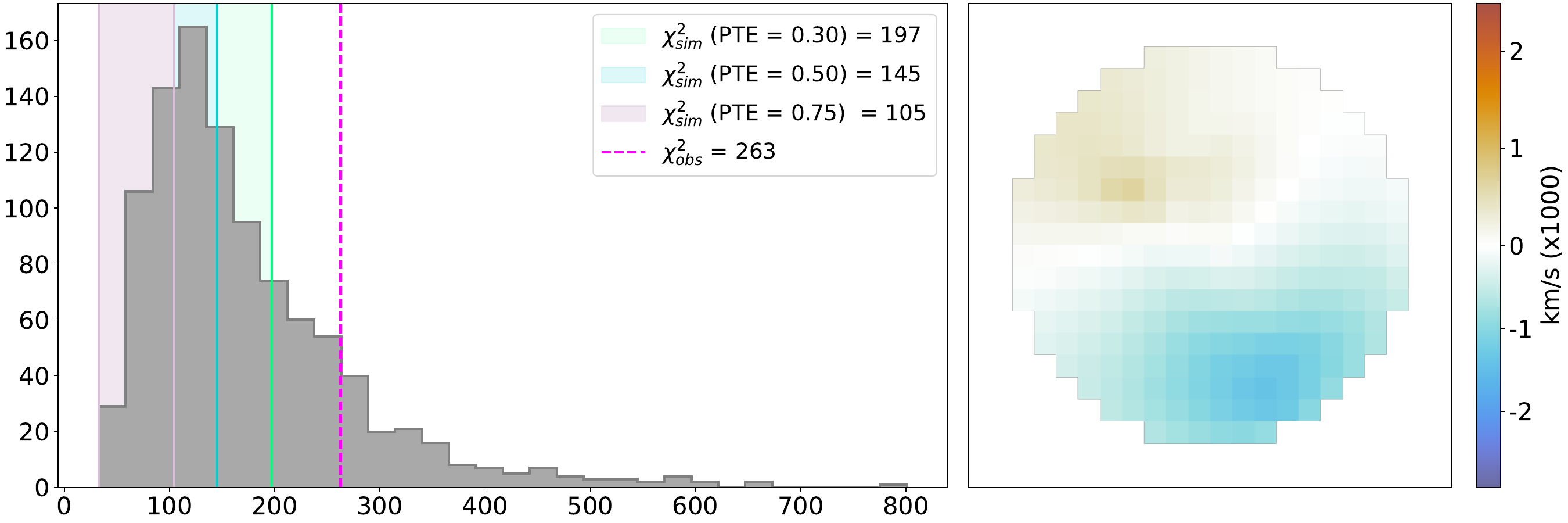}
    \caption{\textit{Top:} simulated ICM v$_{\text{pec}}$ map and empirically determined distribution of $\chi^2_{\text{sim}}$ values with $\chi^2_{\text{obs}}$ indicated for a simulation with with $R = 1.5$, $b = 250$ kpc, $v_i = 3000$ km s$^{-1}$, gas profile: Int$+$NCC, and $|{\bf L} \cdot \; \hat x| = 0.88$ at an epoch 0.06 Gyr after pericenter passage. This mock ICM v$_{\text{pec}}$ passes the matching criterion. \textit{Bottom:} same, at an epoch 1 Gyr after pericenter passage. This mock ICM v$_{\text{pec}}$ map fails the matching criterion. In each figure, the shaded regions associated with each PTE threshold represent the range of $\chi^2_{\text{obs}}$ values that would be allowed by the threshold.} \label{fig:ICM_comparison}  \vspace{1em}
\end{figure*}

We then define for each $M_{\text{sim}}$ an observational $\chi^2_\text{obs}$ assuming $M_{\text{sim}}$ is the true underlying velocity distribution: 
\begin{equation}
    \chi^2_{\text{obs}} = \sum_i \frac{(M_{\text{obs}, i} - M_{\text{sim}, i})^2}{(M_{\text{RMS}, i})^2} 
\end{equation} for $i$ pixels in each map. Next, we use the empirically-determined distribution of $\chi^2_{\text{sim}}$ values to identify the benchmark where the probability-to-exceed (PTE; often referred to as a p-value) of the $\chi^2_{\text{sim}}$ distribution is $0.30$. Our matching criterion therefore requires that PTE $> 0.30$ for a given $\chi^2_{\text{obs}}$. In Figure \ref{fig:ICM_comparison}, we plot examples of simulated ICM v$_{\text{pec}}$ maps, one of which passes the matching criterion (i.e., PTE $> 0.30$ for the corresponding value of $\chi^2_{\text{obs}}$), and one of which fails (PTE $< 0.30$ for $\chi^2_{\text{obs}}$). 

\subsection{GL mass reconstruction map}
The \macs GL mass reconstruction map reveals two centrally-located mass peaks separated by $\sim22''$ ($\simeq 140$ kpc at $z = 0.546$) along the NE--SW axis. We associate these peaks with the centers of the mass distributions of the primary and secondary cluster components in the merger. In order to reproduce this morphology, we require simulation snapshots to exhibit primary and secondary cluster mass peaks separated by $\pm 50$\% of this observational angular separation ($11'' \leq \theta_{\text{GL}} \leq 33''$).

\subsection{XSB map}
The \macs XSB map indicates the presence of extended X-ray emission elongated along the NE--SW axis. Unlike the GL mass map, this emission lacks two clearly identifiable peaks, so from our viewing angle, the gas appears as a single-peaked distribution. This implies that the merger has progressed to a point where significant interactions between the gas in the original systems have already occurred. To select for this quality, we exclude simulation snapshots wherein we identify two distinct intensity peaks (with \texttt{peak\_local\_max} from \texttt{scikit-image}; \citealt{scikit-image}) separated by $> 3''$ ($\simeq 20$ kpc at $z = 0.546$).

\subsection{kT map}
The \macs X-ray derived kT map indicates a hot (kT$_{\text{max}}\sim15$ keV), centrally-peaked temperature distribution elongated along the NE--SW axis. The kT map lacks any evidence for an intact cool core, indicating that either the original gas distributions have already been significantly impacted by the merger, or that neither cluster initially possessed a cool core. We therefore employ two matching criteria to the simulated kT maps. First, in order to exclude snapshots where the temperature distribution is globally much higher than that of \macscom we require simulation snapshots to have a median temperature across the kT map of $< 15$ keV. Second, to identify simulation snapshots that exhibit a centrally-peaked temperature distribution, we require that simulated kT maps have kT $> 9.5$ keV plasma in every cell within a central map region defined by a circle with $r = 20''$. These two criteria jointly exclude simulated snapshots which exhibit clear cool cores and/or global temperatures inconsistent with the \macs kT map. 

\subsection{v$_\text{gal}$ map} \label{sec:vgal_subsec}
The \macs $z_{\text{spec}}$-derived v$_\text{gal}$ map morphologically resembles a $\sim$NW--SE dipole with magnitude scale $\Delta v_{\text{gal, obs}} \approx 3100$ km s$^{-1}$ peak-to-peak. In order to identify simulation snapshots where the galaxy velocities replicate this dipole magnitude, we require that the absolute magnitude difference across the simulated v$_\text{gal}$ maps ($\Delta v_{\text{gal, sim}}$) be within the $1\sigma$ statistical uncertainty ($\pm 940$ km s$^{-1}$) of the \macs v$_\text{gal}$ absolute magnitude difference. This enforces the criterion $\Delta v_{\text{gal, sim}} = 3100 \pm 940$ km s$^{-1}$. 

In addition, the \macs observables reveal a rotational offset in projection between the LOS ICM and DM velocity dipoles (by way of the ICM v$_{\text{pec}}$ and v$_\text{gal}$ maps), which we attribute to the decoupling of the DM and gas velocities during the merger. In order to identify simulation snapshots which similarly exhibit a projected rotation between the ICM v$_{\text{pec}}$ and v$_\text{gal}$ maps, we first fit a 2D $1^{\text{st}}$ degree polynomial to each of the observationally-derived \macs ICM v$_{\text{pec}}$ and v$_\text{gal}$ maps with \texttt{lmfit}. Using these models, we identify for each observable the axis on which the velocity extrema lie (i.e., the vector that goes through the median positions of the 10\% highest and lowest velocity values in the best-fit model map). Then, we calculate the absolute value of the angular offset between the axes of velocity extrema in the ICM v$_{\text{pec}}$ and v$_\text{gal}$ maps. The resulting \macs $\Delta \theta_{v, \text{obs}}$ is $\approx135$ deg. We then apply the same formalism to the simulated ICM v$_{\text{pec}}$ and v$_\text{gal}$ maps for each simulated snapshot and require that $\Delta \theta_{v, \text{sim}}$ is within 45 degrees of $\Delta \theta_{v, \text{obs}}$. This excludes simulated snapshots where the ICM and DM velocities are either fully coupled or are decoupled in a manner which is inconsistent with the \macs observables. 

\subsection{Applications of the matching criteria} \label{sec:matchingapp}
In order to constrain the primary and secondary cluster gas profiles, we first generated a set of simulations with reasonable initial guesses for $R$, $b$, and $v_i$ while individually varying the primary and secondary cluster gas profiles (see Table \ref{tab:sim_params1}). We sample the simulations at the epochs defined in section \ref{sec:epochs} over a grid of ${\bf L}$ vectors. Here, ${\bf L}$ is defined by a set of vectors spaced evenly in $|{\bf L}$ $\cdot \; \hat x|$ space with randomly populated ($\hat y$, $\hat z$) components. 

The simulations are initialized with combinations of gas profiles assigned to the primary and secondary cluster components (see section \ref{sec:gasprofs}). We sample simulations where the primary cluster gas profile has been either moderately or strongly disturbed (by, e.g., recent merger activity) and the secondary cluster gas profile is any one of undisturbed, moderately, or strongly disturbed. We do not include simulations where the primary cluster gas profile is undisturbed in this suite due to the lack of evidence in the \macs observables for a strong cool core in the X-ray derived observables. 

\begin{deluxetable}{l|l}[t]  \label{tab:sim_params1}
\caption{Merger simulation parameter space sampled in the first application of matching criteria to determine the primary and secondary cluster gas profiles.}
\tablehead{\colhead{Parameter} & \colhead{Value(s)}}
\startdata
    \multicolumn{2}{c}{Fixed} \\ \hline
    $R$ & 1.5 \\
    $b$ & 250 kpc \\
    $v_i$ & 3000 km s$^{-1}$ \\ 
    Epoch & $\{0.20, 0.22, 0.24, 0.26, ..., 2.10\}$ Gyr \\ 
    $|{\bf L} \cdot \; \hat x|$ & $\{0.65, 0.69, 0.73, ..., 1\}$ \\ \hline 
    \multicolumn{2}{c}{ }\\[-2ex]
    \multicolumn{2}{c}{Varied} \\ \hline
    Gas profile & NCC$+$CC, NCC$+$Int, NCC$+$NCC, \\
                 & Int$+$CC, Int$+$Int, Int$+$NCC 
\enddata 
\end{deluxetable} \vspace{-2em}
\begin{deluxetable}{l|l}[t]  \label{tab:sim_params2}
\caption{Merger simulation parameter space sampled in the second application of matching criteria to determine the epoch, viewing angle, $R$, $b$, and $v_i$ based on gas profile results from the first application of matching criteria.}
\tablehead{\colhead{Parameter} & \colhead{Value(s)}}
\startdata
    \multicolumn{2}{c}{Fixed} \\ \hline
    Gas profile & Int$+$NCC \\
    Epoch & $\{0.20, 0.22, 0.24, 0.26, ..., 2.10\}$ Gyr \\ 
    $|{\bf L} \cdot \; \hat x|$ & $\{0.65, 0.69, 0.73, ..., 1\}$ \\ \hline 
    \multicolumn{2}{c}{ }\\[-2ex]
    \multicolumn{2}{c}{Varied} \\ \hline
    $R$ & $\{1.2, 1.5, 2.0, 3.0, 5.0\}$ \\
    $b$ & $\{0, 50, 100, 250, 500\}$ kpc \\
    $v_i$ & $\{1000, 1700, 2400, 3000, 3800, 5000\}$ km s$^{-1}$ 
\enddata 
\end{deluxetable} 

\begin{figure*}[t] 
    \centering 
        \includegraphics[width=1\textwidth]{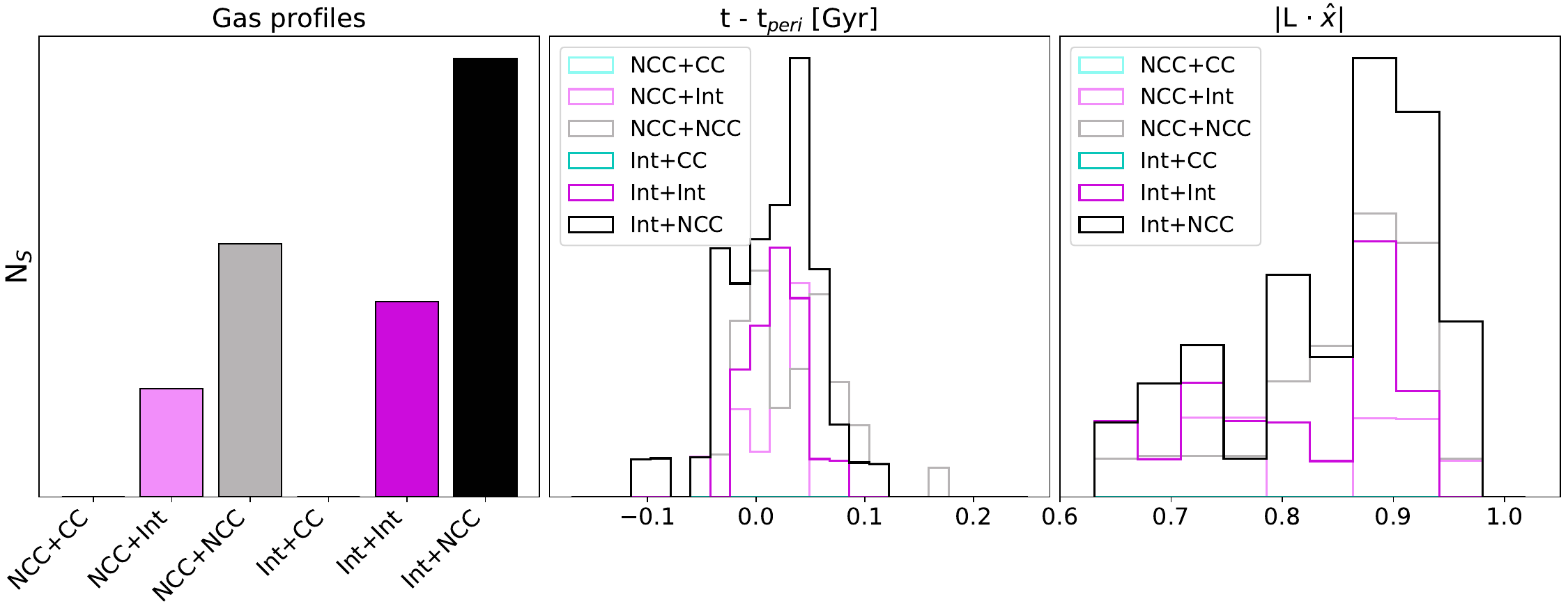}
        \caption{Weighted distribution of gas profiles, epochs, and viewing angles for simulated snapshots with $R = 1.5$, $b = 250$ kpc, $v_i = 3000$ km s$^{-1}$ that pass the matching criteria. The Int$+$NCC gas profile combination is preferred by the matching algorithm. The epoch distribution peaks slightly after the pericenter passage, and viewing angles offset from the merger axis by $\approx 20$--$30$ degrees are most strongly preferred.}   \label{fig:filtfigure} \vspace{1em}
\end{figure*} 

When consolidated into a single matching algorithm and applied to the simulations, this first application of the matching criteria reduces the simulated parameter space to $107$ out of $4.4 \times 10^{4}$ snapshots ($0.2$\%). In Figure \ref{fig:filtfigure}, we plot weighted distributions of gas profiles, epochs, and viewing angles associated with the simulation snapshots that pass the application of the matching criteria. In order to account for biases on mock observable evolution rates that different initial conditions can produce (e.g., simulations initialized with higher $v_i$ will be sampled less finely in space relative to those initialized with lower $v_i$ due to the fixed epoch readout increment), each snapshot's contributions to the distributions are weighted by the relative velocity of the cluster component cores at that snapshot. The cluster relative velocities are calculated using the median velocity of the star particles associated with each halo. From this weighted gas profile distribution, we select the Int$+$NCC combination as the profile that best fits the observational data given our fixed initial conditions. We reserve a full discussion of the gas profile variation matching results for section \ref{sec:gasresults}, and choose to fix the Int$+$NCC gas profiles for further simulation runs. 

Next, in order to constrain $R$, $b$, and $v_i$ for the \macs system, we generated a set of simulations with Int$+$NCC gas profiles, while individually varying $R$, $b$, and $v_i$ (see Table \ref{tab:sim_params2}). The epoch and viewing angle sampling is identical to the first set of simulations described above. We then performed a second application of the matching criteria to this set of simulations, which reduces the simulated parameter space to $432$ out of $1 \times 10^{5}$ snapshots ($0.4$\%). We illustrate the resulting weighted distributions of $R$, $b$, $v_i$, epochs, and viewing angles that pass the matching criteria in Section \ref{sec:results} (Figure \ref{fig:analysis_figure}). 

\section{Parameter variation results} \label{sec:results}
In the following sections, we describe the results of the first and second applications of the matching criteria (see section \ref{sec:matchingapp}) to our simulation suite for each of the initial conditions that were varied in the simulation initializations (gas profiles, mass ratio, impact parameter, and initial relative velocity), in addition to key observable parameters (epoch and viewing angle). As described in Section \ref{sec:matchingapp}, we individually vary one initial parameter for each simulation rather than marginalizing over all correlations between merger parameters (with e.g., an MCMC) due to computational limitations. We, therefore, do not expect the resulting weighted parameter distributions to be exactly equivalent to rigorously-determined posterior probability distributions, which would allow us to compute formal confidence limits on each parameter. Instead, when reporting ranges of parameters consistent with the \macs observables, we identify for each initial parameter any value which produces a set of snapshots greater than $50$\% as likely as the most likely sampled value determined by the weighted distributions resulting from the matching criteria applications.

\subsection{Gas profiles} \label{sec:gasresults}
In the first application of the matching criteria to the simulations with varying gas profiles, simulations with either a moderately or strongly disturbed secondary gas profile are strongly preferred to those with an undisturbed secondary gas profile. No simulation with an undisturbed secondary gas profile produces any snapshots which pass the matching criteria. This is sensible, since there are no clear indicators of a prominent cool core in any of the \macs observables with which the matching criteria are empirically calibrated. We note that the cool core can be destroyed as a result of the merger, although the initial conditions and epochs associated with this scenario are excluded by our matching criteria.

In addition, simulations with a moderately disturbed primary gas profile are preferred to those with a strongly disturbed gas profile. We find that simulations which utilize a strongly disturbed primary gas profile generally produce mock observables which are more extreme than the \macs observables. For example, the resulting simulated kT maps are much hotter (kT$_{\text{max}}\gtrsim20$ keV) at epochs near pericenter passage than the \macs kT map. There exists a moderate preference for a strongly disturbed secondary gas profile that is evident both when the primary gas profile is moderately disturbed and when it is strongly distrurbed. The only simulation initialized with gas profiles other than Int$+$NCC that produces snapshots greater than $50$\% as likely as the most likely sampled gas profile combination (Int$+$NCC) is that with an NCC$+$NCC gas profile (only $\simeq 58$\% as likely). Based on these preferences from the first matching criteria application, we select the Int$+$NCC gas profile combination for use in the simulations wherein $R$, $b$, and $v_i$ are varied.

\subsection{Merger epoch}
\begin{figure}[t] 
    \centering 
    \includegraphics[width=0.47\textwidth]{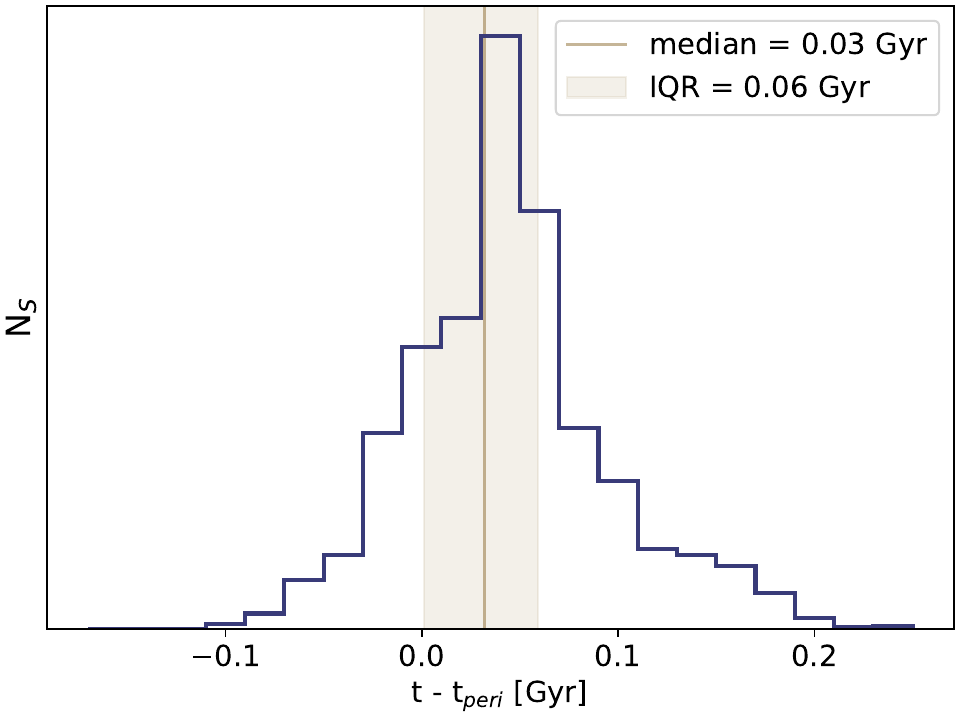}
        \caption{Weighted distribution of epochs for simulation snapshots that pass the matching criteria. The outputs of both applications of the matching criteria are used to generate the full weighted epoch distribution here. The median of selected epochs is $0.03$ Gyr after the pericenter passage with an IQR of $0.06$ Gyr.} \label{fig:epochfig} 
\end{figure}

In both the first application of the matching criteria to the simulations with varying gas profiles and the second application to the simulations with varying $R$, $b$, and $v_i$, there is a strong selection for simulation snapshots within $-0.1 \; (0.2)$ Gyr before (after) the calculated epoch of pericenter passage in each merger simulation. We plot the weighted distribution of selected epochs for simulation snapshots that pass the matching criteria for all the simulation variations described above in Figure \ref{fig:epochfig}. The distribution rises with a roughly Gaussian shape beginning at $\approx 0.1$ Gyr before the pericenter passage, peaks at $t - t_{\text{peri}} \approx 0.03$ Gyr, then falls off similarly in a roughly Gaussian-shape within $\approx 0.1$ Gyr of the pericenter passage, before a later-time wing takes over and falls to zero at $t - t_{\text{peri}} \approx 0.2$ Gyr. Quantitatively, the median of selected epochs is $0.03$ Gyr after the pericenter passage with an inter-quartile range (IQR) of $0.06$ Gyr. The \macs merger epoch $t - t_{\text{peri}}$ is, therefore, likely in the range $0$--$60$ Myr after the pericenter passage. 

\begin{figure*}[!htb]
    \centering 
        \includegraphics[width=1\textwidth]{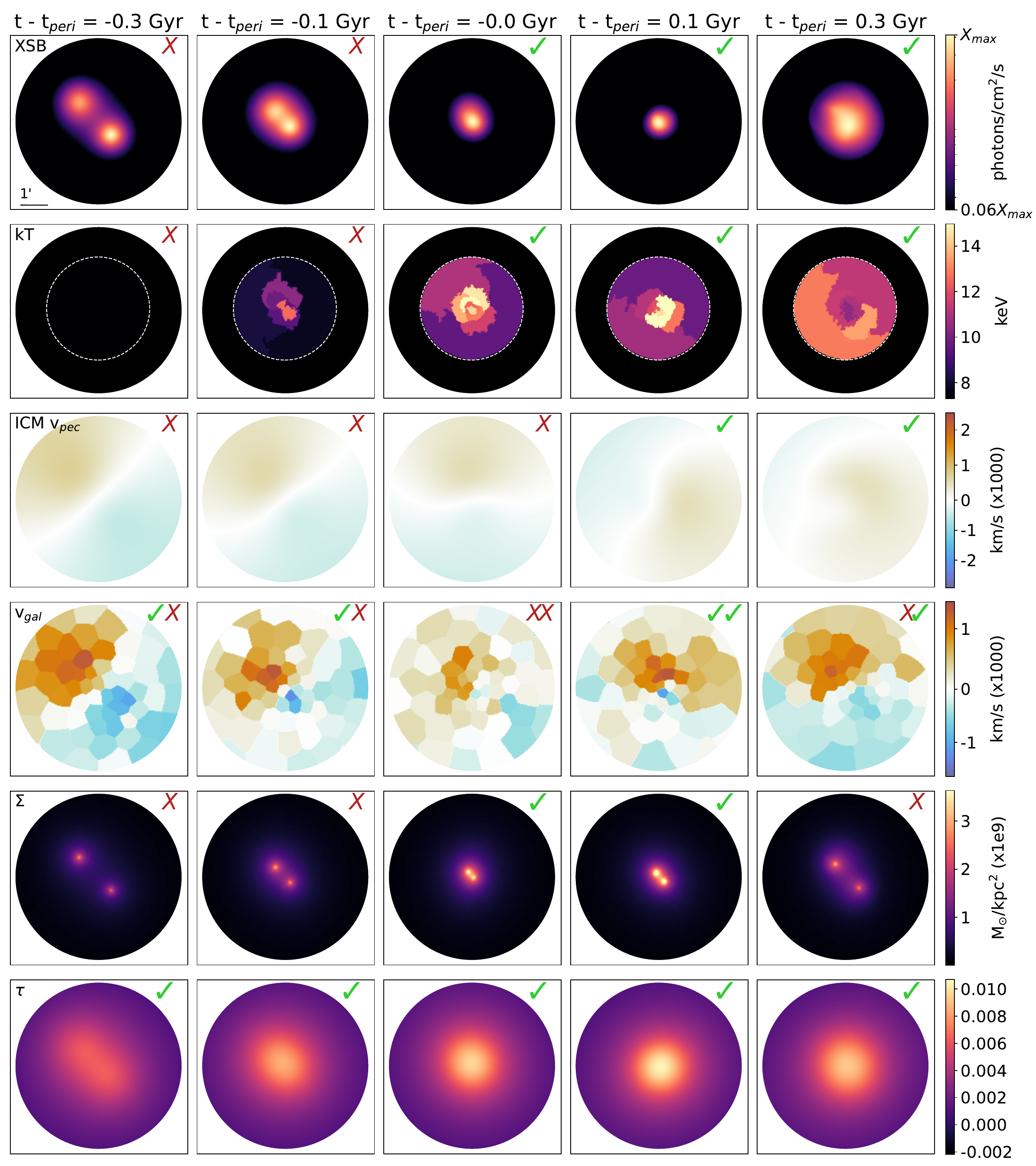}
        \caption{Evolution of mock observables as a function of epoch with indicators for the pass/fail status of each mock observable to the corresponding matching criterion. \textit{Row 1:} XSB evolution. The 99$^{\text{th}}$ percentile in XSB ($X_{\text{max}}$) sets the upper bound of the colorbar, while the lower bound is fixed in order to span the same relative flux scale as the \macs XSB map. \textit{Row 2:} kT evolution. \textit{Row 3:} ICM v$_{\text{pec}}$ evolution. \textit{Row 4:} v$_{\text{gal}}$ evolution. The leftmost indicator in each panel represents the pass/fail status of the $\Delta v_{\text{gal, sim}}$ magnitude matching criterion. The rightmost indicator represents the pass/fail status of the $\Delta \theta_{v, \text{sim}}$ rotational offset with the ICM $v_{\text{pec}}$ map. \textit{Row 5:} $\Sigma$ evolution. \textit{Row 6:} $\tau$ evolution. Note that we do not apply an individual matching criterion to mock $\tau$ maps due to the minimal constraining power of the \macs $\tau$ map. The only simulation snapshot shown here which passes all of the matching criteria is at $t - t_{\text{peri}} \approx 0.1$ Gyr (i.e., \textit{column 4}).} \label{fig:xsbprogr} 
\end{figure*} 
\begin{figure*}[!htb] 
    \centering 
        \includegraphics[width=1\textwidth]{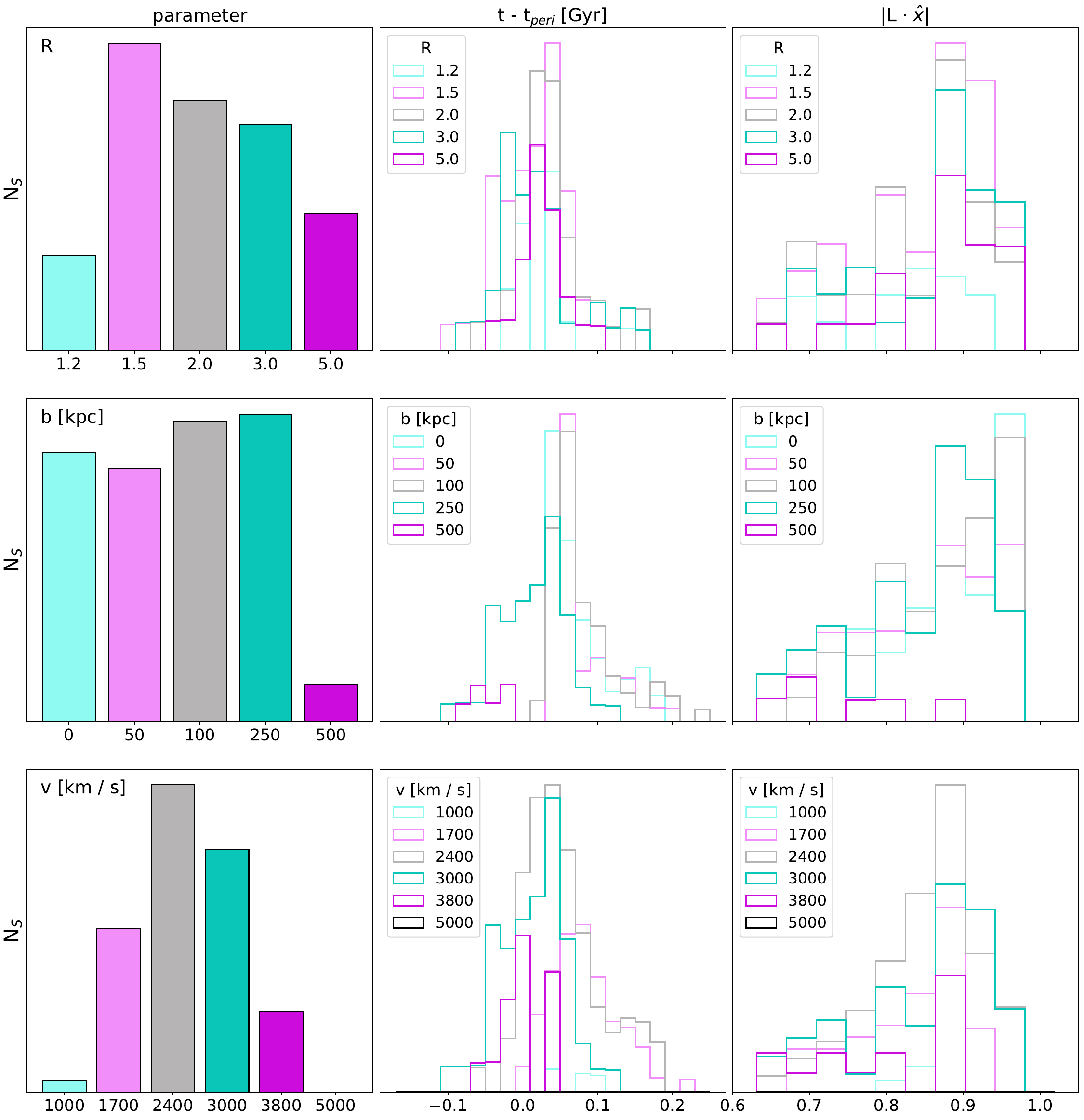} 
        \caption{Weighted distributions of varied parameters (\textit{top:} mass ratio, \textit{middle:} impact parameter, \textit{bottom:} initial relative velocity), epochs, and viewing angles from simulated snapshots that pass the matching criteria. The default parameters which are held fixed unless explicitly varied are $R = 1.5$, $b = 250$ kpc, $v_i = 3000$ km s$^{-1}$, and gas profile: Int$+$NCC. Impact parameters of $b \lesssim 250$ kpc, mass ratios of $R \approx 1.5$--$3.0$, and initial relative velocities of $v_i \approx 1700$--$3000$ km s$^{-1}$, are preferred by the matching criteria. The weighted epoch distributions peak around or shortly after the pericenter passage, and viewing angles offset from the merger axis by $\approx 15$--$35$ degrees are preferred by the matching criteria.}  \label{fig:analysis_figure} \vspace{1em}
\end{figure*} 

To further illustrate this result, we plot the evolution of mock maps at $t - t_{\text{peri}} = -0.3$, $-0.1$, $0$, $0.1$, and $0.3$ Gyr for a simulation with initial conditions and observational parameters that are reasonably favored by the matching algorithm in Figure \ref{fig:xsbprogr}. In each panel, we indicate whether the observable passes its corresponding matching criterion with a green check mark (pass) or a red $X$ (fail). In the XSB observable progression, the clearly identifiable emission peaks associated with each cluster are disrupted, and the maps become more spatially extended as the merger evolves beyond the pericenter passage. Only XSB maps after the pericenter passage epoch pass the XSB matching criterion. The kT progression indicates that the global temperature rapidly increases around the pericenter passage of the clusters when central regions of gas are heated by merger-driven shocks, and cools globally shortly thereafter. Only snapshots from the pericenter passage to $t - t_{\text{peri}} = 0.3$ Gyr pass the temperature matching criteria for this simulation. In the ICM v$_{\text{pec}}$ observable progression, the ICM velocity dipole structure is highly disrupted around the pericenter passage, after which it rotates in the POS relative to the v$_{\text{gal}}$ dipole. Only snapshots after the pericenter passage epoch, when the dipole orientation matches that of \macscom are selected by the ICM v$_{\text{pec}}$ matching criterion. The v$_{\text{gal}}$ progression shows a high galaxy velocity magnitude difference at early epochs that is reduced at pericenter passage as the DM halos pass the point of their closest approach (projected along the LOS). All snapshots prior to the pericenter passage epoch are therefore selected by the v$_{\text{gal}}$ matching criterion, in addition to one snapshot immediately after the pericenter passage at $t - t_{\text{peri}} = 0.1$ Gyr. The rotational offset between the ICM v$_{\text{pec}}$ and v$_{\text{gal}}$ dipoles is only of sufficient magnitude to pass the $\Delta \theta_{v, \text{sim}}$ matching criterion at epochs after pericenter passage, meaning that the only snapshot shown which passes all three velocity criteria is at $t - t_{\text{peri}} = 0.1$ Gyr. In the $\Sigma$ observable progression, snapshots from the pericenter passage epoch until $t - t_{\text{peri}} = 0.1$ Gyr produce total projected mass peaks which are separated within the bounds set by the matching criteria. Finally, the $\tau$ progression is shown for reference. We do not apply any matching criteria to the $\tau$ map, since any reasonable choice of input parameters yields $\tau$ maps that are consistent with the \macs $\tau$ map, given the large observational uncertainties.

These evolutionary trends validate the result of the matching algorithm applications, i.e., that the mock observables favor a merger occurring near or shortly after pericenter passage. For a simulation with $R = 1.5$, $b = 250$ kpc, $v_i = 1700$ km s$^{-1}$, gas profile: Int$+$NCC, and $| {\bf L} \cdot \; \hat x| = 0.84$, the snapshots which pass the matching criteria are at $t - t_{\text{peri}} = 0.04$, $0.06$, and $0.08$ Gyr. In Figure \ref{fig:movie}, we show an animation of the evolution of mock observables as a function of epoch for a simulation with $R = 1.5$, $b = 250$ kpc, $v_i = 1700$ km s$^{-1}$, gas profile: Int$+$NCC, and $| {\bf L} \cdot \; \hat x| = 0.92$.

\begin{figure}[t] 
    \centering 
        \includegraphics[width=0.47\textwidth]{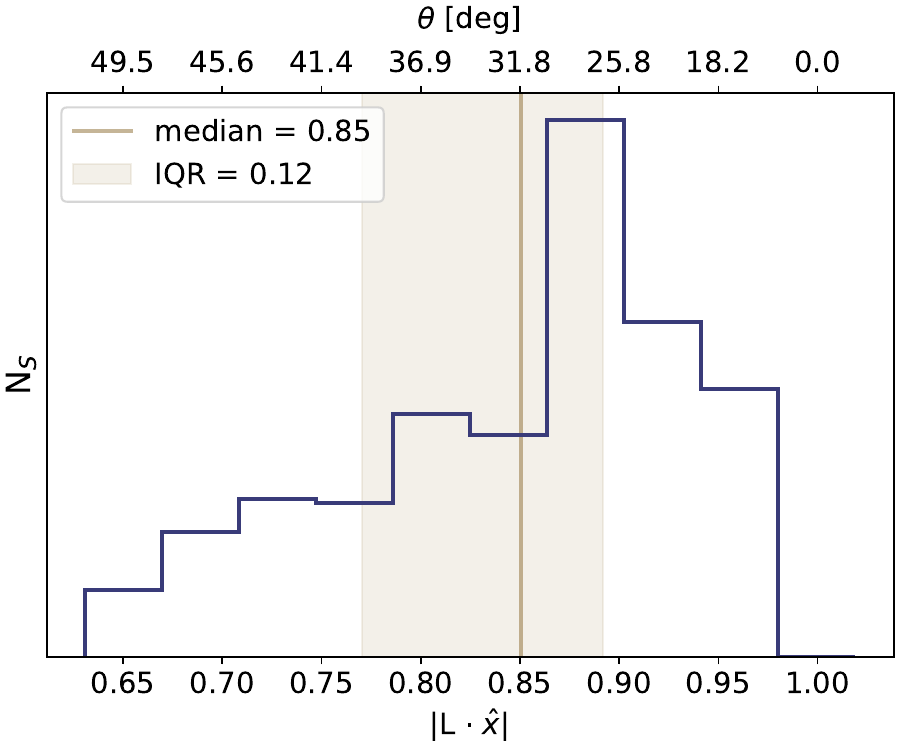}
        \caption{Weighted distribution of viewing angles for simulation snapshots that pass the matching criteria. As in Figure \ref{fig:epochfig}, the outputs of both applications of the matching criteria are used to generate the full weighted viewing angle distribution here. The median of the selected viewing angles is $|{\bf L} \cdot \; \hat x| = 0.85$ ($\approx 32$ degrees) with an IQR of 0.12 ($\approx 13$ degrees).} \label{fig:Lfig} 
\end{figure} 

\subsection{Viewing angle}
In both the first and second applications of the matching criteria to the simulations, there is a strong selection of simulation snapshots observed at viewing angles marginally offset from the merger axis. We plot the weighted distribution of selected viewing angles for simulation snapshots that pass the matching criteria for all the simulation variations in Figure \ref{fig:Lfig}. In particular, the matching algorithm prefers values of $|{\bf L} \cdot \; \hat x|$ that peak at  $\approx 0.88$ (i.e., inclined $\approx 28$ degrees from the merger axis). Quantitatively, the median of the selected viewing angles is $|{\bf L} \cdot \; \hat x| = 0.85$ ($\approx 32$ degrees) with an IQR of 0.12 ($\approx 13$ degrees). The \macs viewing angle is therefore likely between $0.77 \lesssim | {\bf L} \cdot \; \hat x| \lesssim 0.89$ (inclined $\approx 27$--$40$ degrees). In no case does a merger oriented along the line of sight ($|{\bf L} \cdot \; \hat x| = 1$) pass the matching criteria. 

Furthermore, the degree of decoupling in the POS between the gas and dark matter velocity dipole structures, as traced by the ICM v$_{\text{pec}}$ and $v_{\text{gal}}$ maps, varies strongly as a function of the viewing angle. We describe these dependencies in detail in Section \ref{sec:icmdm}. 

\subsection{Mass ratio}
The mass ratio is similarly constrained in the second application of the matching criteria, assuming a fixed gas profile, $b$, and $v_i$. The resulting weighted distribution of $R$ values selected by the matching algorithm is shown in Figure \ref{fig:analysis_figure}. While the distribution peaks at $R = 1.5$, simulations initialized with mass ratios of $R = 2.0$ and $3.0$ are greater than $50$\% as likely as those with $R = 1.5$. The likelihood of reproducing the data falls off rapidly (less than $50$\% as likely as the most likely sampled mass ratio) for larger and smaller mass ratios ($R = 1.2$ and $5.0$), which indicates that the mass ratio of \macs is likely between $R \approx 1.5$--$3.0$. 

\subsection{Impact parameter}
The \macs impact parameter is constrained in the second application of the matching criteria, assuming a fixed gas profile, $R$, and $v_i$. The resulting weighted distribution of $b$ values selected by the matching algorithm is shown in Figure \ref{fig:analysis_figure}. The distribution formally peaks at $b \approx 250$ kpc, and the likelihood of reproducing the data falls off sharply for $b > 250$ kpc. However, all sampled impact parameters less than this value, while less likely, are still plausible (greater than $50$\% as likely as $b \approx 250$). We therefore place an upper limit on the \macs impact parameter of $b \lesssim 250$ kpc. At the epoch of pericenter passage, the separation between the cluster cores (i.e., the pericenter distance) is $\simeq 90$ kpc for a simulation initialized with $R = 1.5$, $b = 250$ kpc, $v_i = 1700$ km s$^{-1}$, and gas profile: Int$+$NCC.

However, we note that this impact parameter constraint is highly dependent on the degree of observed rotational offset between the ICM v$_{\text{pec}}$ and v$_{\text{gal}}$ maps. In Section \ref{sec:icmdm}, we illustrate the dependencies of the constrained impact parameter on the velocity decoupling. 

\subsection{Initial relative velocity}
The initial relative velocity of the \macs cluster components is similarly constrained in the second application of the matching criteria, assuming a fixed gas profile, $b$, and $R$. The resulting weighted distribution of $v_i$ values selected by the matching algorithm is shown in Figure \ref{fig:analysis_figure}. The distribution peaks at $v_i \approx 2400$--$3000$ km s$^{-1}$. The most likely initial relative velocity based on the weighted distribution is $v_i = 2400$ km s$^{-1}$, but simulations initialized with initial relative velocities of $v_i = 3000$ and $1700$ km s$^{-1}$ are greater than $50$\% as likely as those with $v_i = 2400$ km s$^{-1}$. Initial relative velocities of $v_i = 1000$, $3800$, and $5000$ km s$^{-1}$ produce snapshots less than $50$\% as likely as the most likely sampled initial relative velocity. We, therefore, conclude that \macs likely has an initial relative velocity between $v_i \approx 1700$--$3000$ km s$^{-1}$. 

\begin{figure}[t] 
    \centering 
        \includegraphics[width=0.45\textwidth]{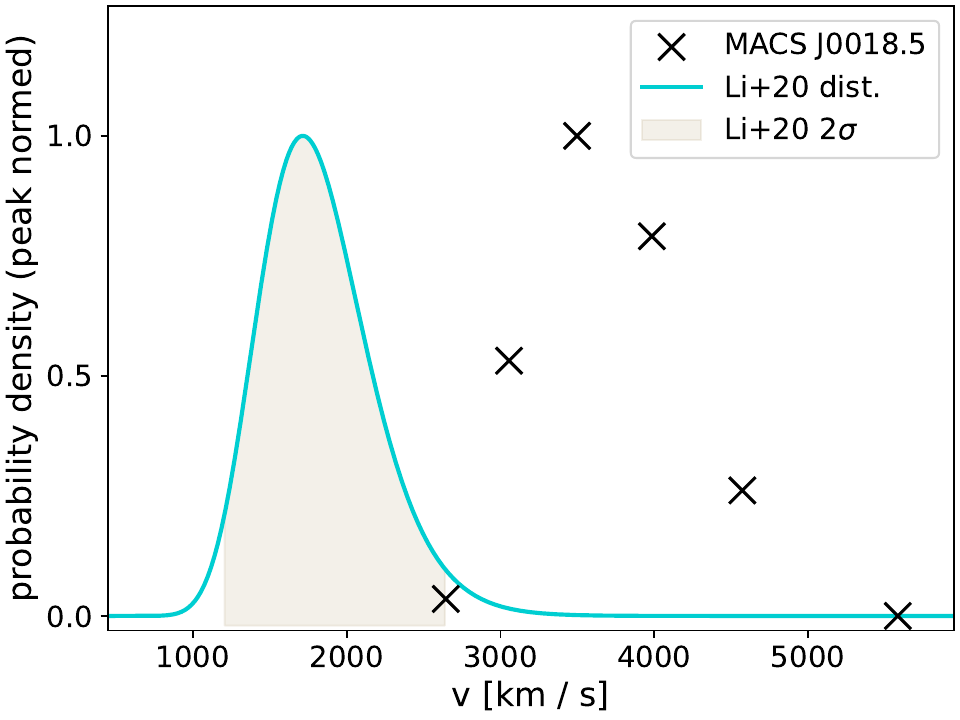}
        \caption{Comparison of the calculated \macs infall velocity weighted distribution (Figure \ref{fig:analysis_figure}) to the subhalo infall velocity distribution from \citet{Li2020}. All values have been peak-normalized for straightforward comparison. The \macs velocities indicate that \macs is a relatively exceptional object compared to those studied by \citet{Li2020}.} \label{fig:velcomps} 
\end{figure}

Using a suite of N-body simulations in a $\Lambda$CDM cosmology, \citet{Li2020} constructed a distribution of the velocity ($v$) of infalling subhalos as a function of mass and redshift. They found a nearly universal log--normal distribution which peaks near the virial velocity ($V_h$) of the primary subhalo ($v/V_h \approx 1.15$). For a primary subhalo at the mass and redshift of \macs assuming a mass ratio of $R \approx 1.5$, the \citet{Li2020} subhalo infall velocity distribution has a median velocity of $\approx 1800$ km s$^{-1}$ with a 2$\sigma$ range from $\approx 1200$--$2600$ km s$^{-1}$ (see Figure \ref{fig:velcomps}). It is important to note that the underlying cause of the spread of initial velocities in our analysis is measurement uncertainty, while for the \citet{Li2020} analysis it is due to cosmic variance. Thus, there is no expectation that the spreads will be comparable to each other. 

The range of \macs cluster core initial relative velocity values calculated in this work is $v_i \approx 1700$--$3000$ km s$^{-1}$ when the primary and secondary cluster are separated by 3 Mpc. The \citet{Li2020} infall velocities are measured when the subhalo centers are separated by the virial radius of the primary subhalo. We therefore calculate for each sampled $v_i$ the corresponding velocity when the clusters are separated by the virial radius of the primary cluster ($v_{r_{200, p}}$ where $r_{200, p} \simeq 1.1$ Mpc). After applying this correction, we find that the analogous range of \macs cluster infall velocity vaues is $v_{r_{200, p}} \approx 3100$--$4000$ km s$^{-1}$. While this range is on average higher than the 2$\sigma$ range of \citet{Li2020}, there is nonzero overlap between the two distributions. Since our galaxy cluster sample was constructed to maximize the number of high-mass, actively merging systems, we would expect more extreme initial relative velocities on average than in \citet{Li2020}. Furthermore, we emphasize that the \macs cluster core initial relative velocity values are not expected to be exactly equivalent to a rigorously-determined posterior probability distribution, and likely yield an under-estimation of the true uncertainties since we have not marginalized over all correlations between merger parameters in this analysis due to computational limitations.  This comparison indicates that \macs is a reasonably exceptional object relative to those studied by \citet{Li2020}, which is in agreement with our selection. 

\begin{deluxetable}{l|l}[t]  \label{tab:sim_results}
\caption{\macs merger parameters constrained in this analysis}
\tablehead{\colhead{Parameter} & \colhead{Value(s)}}
\startdata
    Merger epoch & $0 \lesssim t - t_{\text{peri}} \lesssim 60$ Myr \\ 
    Mass ratio & $R \approx 1.5$--$3.0$ \\
    Impact parameter & $b \lesssim 250$ kpc \\
    Initial$^{a}$ relative velocity &  $v_i \approx 1700$--$3000$ km s$^{-1}$ \\ 
    Viewing angle & $0.77 \lesssim | {\bf L} \cdot \; \hat x| \lesssim 0.89$$^{b}$  \\ 
    Initial$^{a}$ cluster gas profiles & Int$+$NCC$^{c}$ 
\enddata 
\tablenotetext{}{$^{a}$when the cluster components are separated by 3~Mpc}
\tablenotetext{}{$^{b}$(inclined $\approx 27$--$40$ degrees from the merger axis)}
\tablenotetext{}{$^{c}$(moderately $+$ strongly disturbed)} \vspace{-2em}
\end{deluxetable}

Early hydrodynamical simulation studies designed to reproduce the merger geometry of the Bullet Cluster \citep[e.g.,][]{Milosavljevic2007, Springel2007, Mastropietro2008} constrained initial merger parameters by qualitatively matching a handful of characteristic morphological features in mock XSB and total mass distributions to observational datasets.  However, none of these works attempted to place uncertainty estimates on the inferred merger parameters. Several works have since improved upon these methods \citep[e.g.,][]{Molnar2012,Lage2014,Chadayammuri22}. For example, \citet{Lage2014} made more quantitative estimates of the Bullet Cluster merger parameters by performing a $\chi^2$ fit between mock and observed datasets. Based on their optimization, \citet{Lage2014} characterizes an initial relative cluster velocity of the Bullet Cluster, in terms of a percent increment relative to the velocity acquired by the clusters while falling in from infinity, with moderate precision ($-10.9 \pm 15$\%). 

In this work, we have systematically compared mock observables from a suite of hydrodynamical simulations to the \macs system with a set of quantitative matching criteria. As a result, we have presented distributions of merger parameters that are consistent with the observed data. In the case of the initial relative velocity, we report this distribution in terms of a physical velocity when the \macs cluster components are separated by $3$ Mpc to be $\approx 1700-3000$~km s$^{-1}$ (i.e., corresponding to an approximate fractional uncertainty of 30\%). We do not expect these distributions to be exactly equivalent to rigorously-determined posterior probability distributions, but we provide these empirically-calibrated distributions of likely merger parameters for the first time in such a merger analysis. A summary of all constrained merger parameters for \macs is shown in Table \ref{tab:sim_results}.

\begin{figure*}[t] 
    \centering 
        \includegraphics[width=0.77\textwidth]{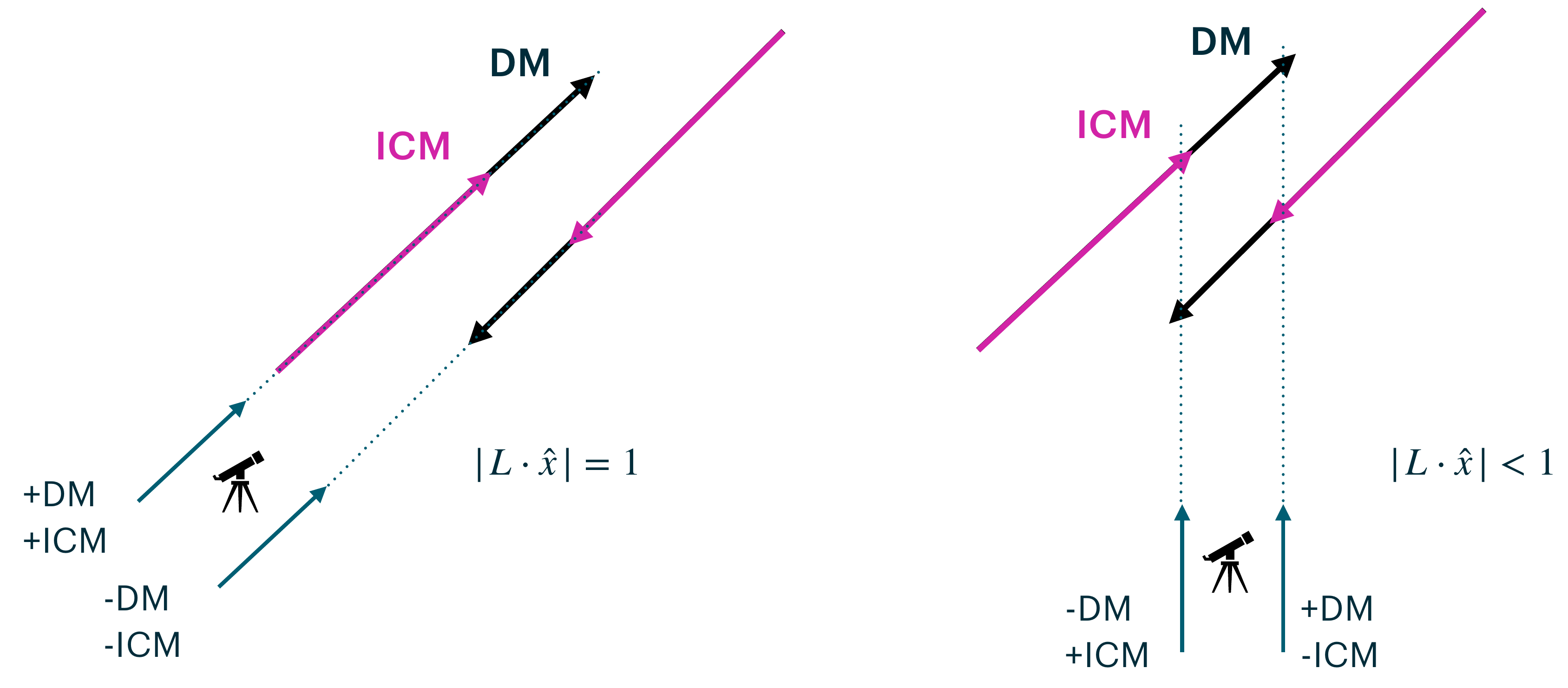} 
        \caption{Simple 2D illustration of an induced rotational offset between the DM and ICM velocity dipoles in projection when observed along the merger axis (\textit{left}) and offset from the merger axis in the plane of the merger (\textit{right}). In this simple schematic, which assumes only that the ICM distribution lags the DM distribution as the merger evolves near the pericenter passage, the DM/ICM dipole alignment flips from 0 degrees to 180 degrees.} \vspace{1em} \label{fig:twistillust}
\end{figure*} 

\begin{figure*}[t]
    \centering 
        \includegraphics[width=1\textwidth]{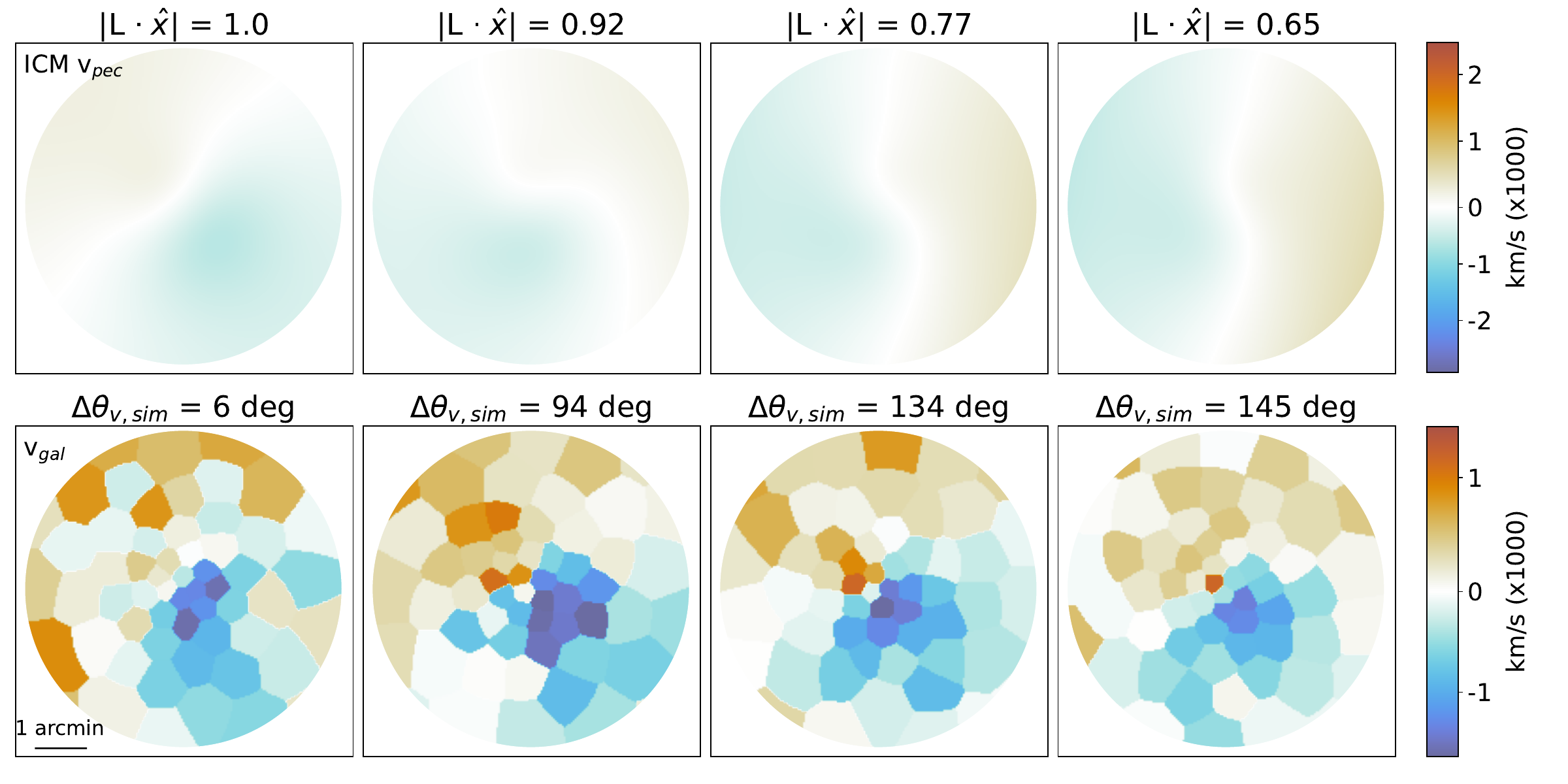}
        \caption{Demonstration of the ICM/DM velocity dipole rotational offset as a function of $| {\bf L} \cdot \; \hat x|$ for a simulation with $R = 1.5$, $b = 250$ kpc, $v_i = 3000$ km s$^{-1}$, gas profile: Int$+$NCC at $t - t_{\text{peri}} = 0.02$ Gyr (after pericenter passage). The dipoles are roughly aligned when the viewing angle is aligned with the merger axis, and magnitude of the rotational offset increases as the viewing angle moves farther from the merger axis.}   \label{fig:twistprog} \vspace{1em}
\end{figure*} 

\section{Gas--dark matter velocity decoupling} \label{sec:icmdm}

\begin{figure*}[t] 
    \centering 
        \includegraphics[width=1\textwidth]{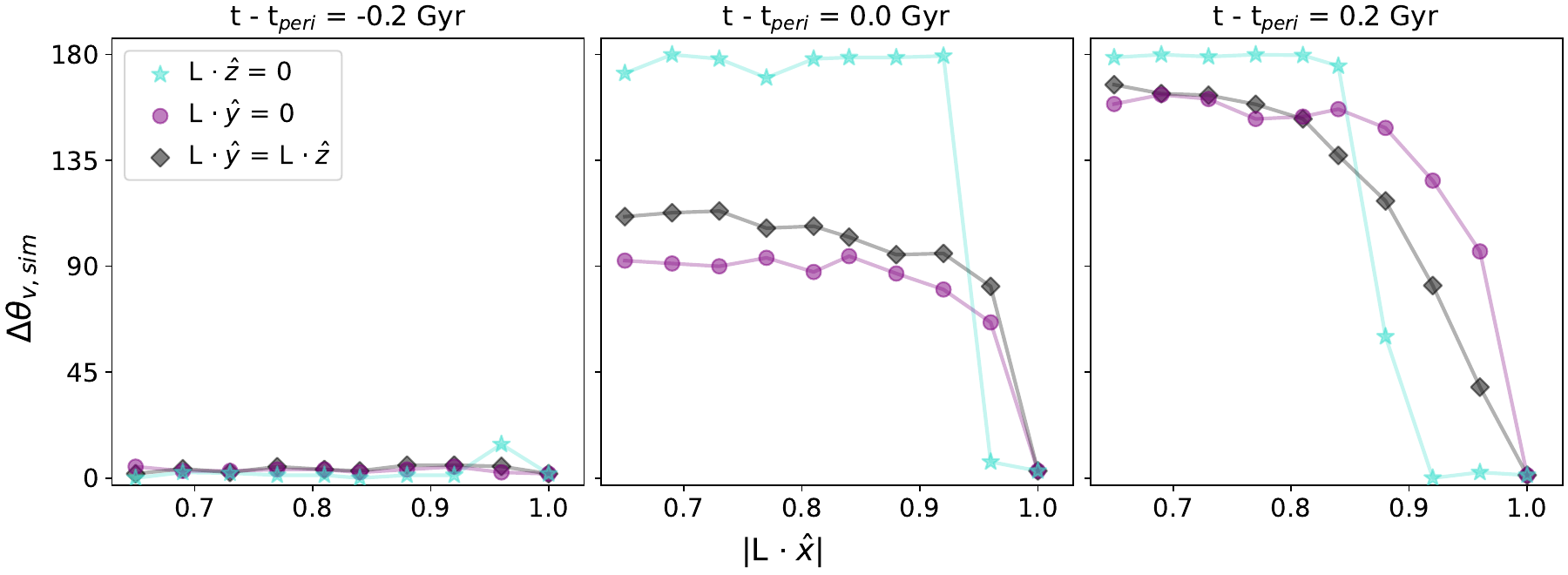} 
        \caption{$\Delta \theta_{v, \text{sim}}$ values for simulation snapshots at $t - t_{\text{peri}} = -0.2$ Gyr (\textit{left}), pericenter passage (\textit{center}), and $t - t_{\text{peri}} = 0.2$ Gyr (\textit{right}) as a function of $| L \cdot \; \hat x|$. The simulation was initialized with $R = 1.5$, $b = 250$ kpc, $v_i = 3000$ km s$^{-1}$, gas profile: Int$+$NCC. Here, $\Delta \theta_{v, \text{sim}}$ is plotted as an absolute-valued angle from 0 to 180 degrees. Prior to pericenter passage, the DM and ICM LOS velocity dipoles are aligned for all viewing angles. At the pericenter passage epoch, there is an induced rotation of the velocity dipoles, and $\Delta \theta_{v, \text{sim}}$ moves from 0 degrees for all $| {\bf L} \cdot \; \hat x|$, excepting where $| {\bf L} \cdot \; \hat x| = 1$. After pericenter passage, the DM and ICM LOS velocity dipoles approach anti-alignment for viewing angles moving farther from the merger axis, while $\Delta \theta_{v, \text{sim}}$ remains $\approx 0$ for $| {\bf L} \cdot \; \hat x| = 1$. Simulations viewed with larger components of ${\bf L}$ in $\hat y$ exhibit higher degrees of rotation, particularly near pericenter passage.}  \vspace{1em}\label{fig:twisting}
\end{figure*} \vspace{1em}

\macs exhibits a decoupling in projection between the ICM v$_{\text{pec}}$ and the v$_{\text{gal}}$ maps, which trace the LOS gas and DM velocity structure, respectively. This decoupling manifests as a rotational offset in the POS between the two velocity dipoles of $\Delta \theta_{v, \text{obs}} \approx135$ deg. As described in section \ref{sec:vgal_subsec}, we applied a matching criterion enforcing that the mock observables exhibit a similar degree of rotational decoupling in order to determine the \macs merger geometry. In this section, we present a more detailed exploration of this velocity decoupling behavior in the simulations and highlight its dependencies on the merger epoch, viewing angle, and impact parameter. 

\subsection{Epoch \& viewing angle}
In Figure \ref{fig:twistillust}, we provide a simple qualitative 2D schematic of a galaxy cluster merger evolving around the pericenter passage. Since the ICM is affected by collisional processes as the merger progresses, the ICM distribution will begin to lag the DM distribution spatially. When the merger is observed along the merger axis, the ICM and DM velocity dipoles will appear to be well aligned, i.e., $\Delta \theta_{v, \text{obs}} \approx 0$ degrees. However, when the viewing angle of the system is offset from the merger axis in the plane of the merger, the dipoles become misaligned ($\Delta \theta_{v, \text{obs}} \approx 180$ degrees). As the viewing angle moves out of the 2D plane of the merger, we would expect to observe velocity dipole rotational offsets at intermediate angles between 0 and 180 degrees, which we characterize below. 

Figure \ref{fig:twistprog} illustrates the dependence of the rotational offset between the ICM and DM LOS velocity dipoles as a function of viewing angle at a fixed epoch for a given set of initial merger parameters. In the full simulation analysis, the components of ${\bf L}$ are evenly sampled in $\hat x$ and randomly sampled in ($\hat y, \hat z$). To more clearly depict the rotational offset dependence on viewing angle and minimize effects from merger asymmetries when randomly sampled in ($\hat y, \hat z$), the viewing angle vector \textbf{L} in this progression is defined to have equal component magnitudes in ($\hat y, \hat z$) as the component along $\hat x$ is varied. At $| {\bf L} \cdot \; \hat x| = 1$, the velocity dipoles are aligned, and $\Delta \theta_{v, \text{sim}}$ increases for viewing angles farther from the merger axis (until the boundary of the sampled parameter space, $| {\bf L} \cdot \; \hat x| = 0.65$). 

The observed rotational offset between the ICM and DM LOS velocity dipoles is further dependent on the merger epoch. In Figure \ref{fig:twisting}, we plot tracks of $\Delta \theta_{v, \text{sim}}$ as a function of $| {\bf L} \cdot \; \hat x|$ at three epochs for the same simulation initialized above. We here identify three tracks in ($\hat y, \hat z$): where $ {\bf L} \cdot \; \hat y = 0$, $ {\bf L} \cdot \; \hat z = 0$, and where $ {\bf L} \cdot \; \hat y = {\bf L} \cdot \; \hat z$. This removes any fluctuations between sets of mock observables that could be attributed to a random sampling of merger asymmetries in ($\hat y, \hat z$). These three tracks indicate that prior to pericenter passage ($t - t_{\text{peri}} = -0.2$ Gyr), the DM and ICM LOS velocity dipoles are aligned for all viewing angles. At the pericenter passage epoch, there is an induced rotation of the velocity dipoles, and $\Delta \theta_{v, \text{sim}}$ moves from 0 degrees for all values of $| {\bf L} \cdot \; \hat x|$ excepting where $| {\bf L} \cdot \; \hat x| = 1$ (i.e., looking down the merger axis). There is a further dependence on the specific ($\hat y, \hat z$) track being followed near the pericenter passage, where simulations viewed with larger components of ${\bf L}$  in $\hat y$ exhibit higher degrees of rotation. After the pericenter passage ($t - t_{\text{peri}} = 0.2$ Gyr), there is a trend towards anti-alignment of the DM and ICM LOS velocity dipoles for viewing angles moving farther from the merger axis (decreasing $| {\bf L} \cdot \; \hat x|$), while $\Delta \theta_{v, \text{sim}}$ remains $\approx 0$ for $| {\bf L} \cdot \; \hat x| = 1$. 

\begin{figure*}[t] 
    \centering 
        \includegraphics[width=0.75\textwidth]{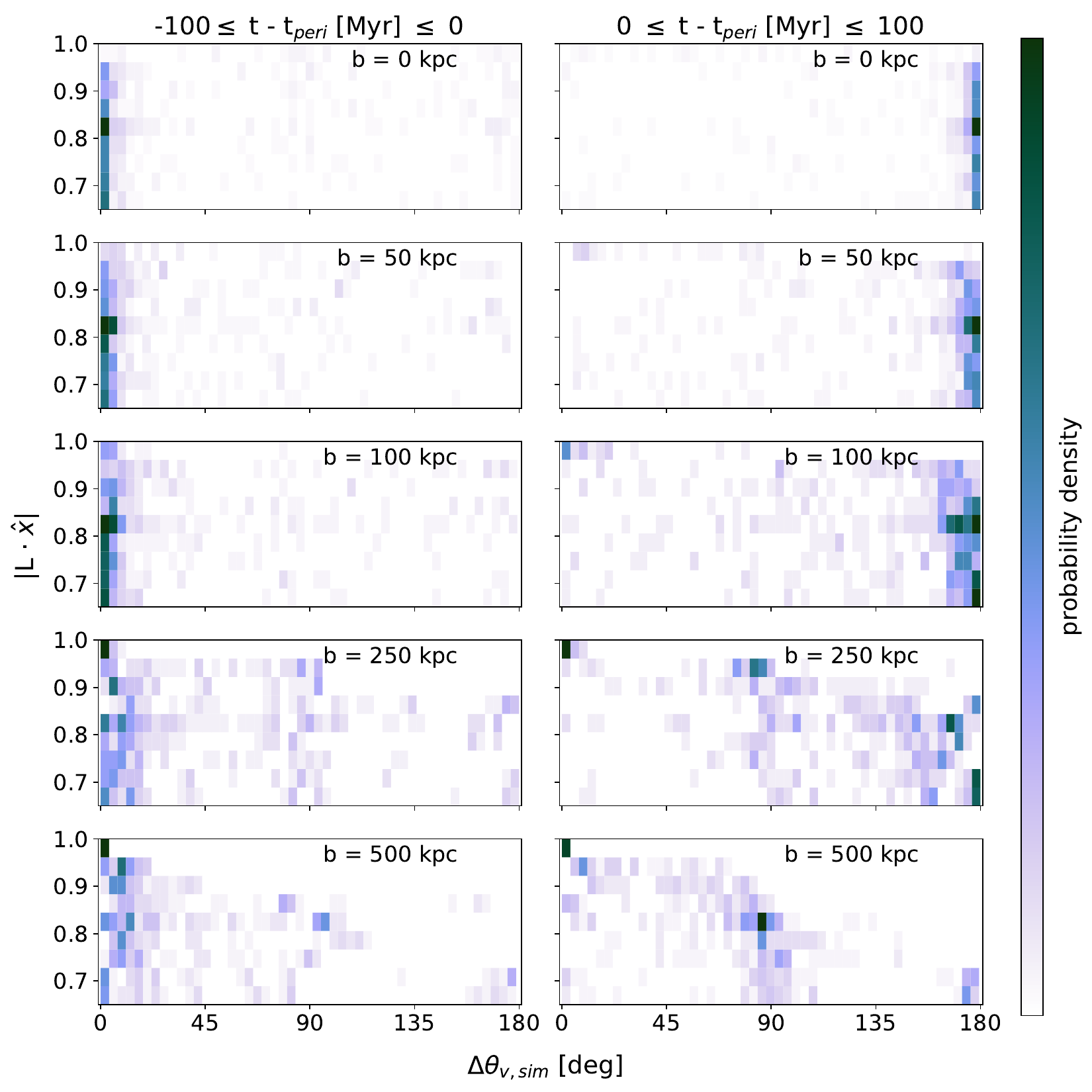} 
        \caption{Distributions of $\Delta \theta_{v, \text{sim}}$ as a function of $| {\bf L} \cdot \; \hat x|$ for simulations initialized with each of the sampled impact parameters ($b = 0$, $50$, $100$, $250$, and $500$ kpc) with a set of initial conditions reasonably favored by the matching criteria ($R = 1.5$, $v_i = 3000$ km s$^{-1}$, and gas profile: Int$+$NCC) for epochs prior to and post-pericenter passage ($|t - t_{\text{peri}}| \leq 100$ Myr). Simulations initialized with small impact parameters undergo a rapid transition from ICM/DM LOS velocity dipole alignment to anti-alignment with few intermediate rotational offset values. Larger impact parameters exhibit more continuous distributions of $\Delta \theta_{v, \text{sim}}$ from alignment to anti-alignment, and can additionally remain at semi-stable intermediate $\Delta \theta_{v, \text{sim}}$ values that are dependent on the merger viewing angle and impact parameter (e.g., at $ \Delta \theta_{v, \text{sim}} \sim 90$ degrees for $| {\bf L} \cdot \; \hat x| = 0.84$ and $b = 500$ kpc).} \vspace{1em} \label{fig:2dhist}
\end{figure*} 

\subsection{Epoch \& impact parameter}
The impact parameter further influences the degree of  rotational offset between the ICM and DM LOS velocity dipoles. In Figure \ref{fig:2dhist}, we plot distributions of the angular value of the velocity dipole rotational offset as a function of the viewing angle component along the merger axis ($| {\bf L} \cdot \; \hat x|$) for simulations initialized with each of the sampled impact parameters (with a set of initial conditions reasonably favored by the matching criteria: $R = 1.5$, $v_i = 3000$ km s$^{-1}$, and gas profile: Int$+$NCC) at a short range of epochs prior to and post-pericenter passage ($|t - t_{\text{peri}}| \leq 100$ Myr). There is rarely any case for which $\Delta \theta_{v, \text{sim}} \neq 0$ for simulations initialized with $b = 250$ kpc and viewed at $|{\bf L} \cdot \; \hat x| = 1$, in agreement with Figures \ref{fig:twistprog} and \ref{fig:twisting}. 

Prior to the epoch of pericenter passage ($-100 \leq t - t_{\text{peri}} \leq 0$ Myr), simulations initialized with low impact parameters ($b \lesssim 100$ kpc) show a large degree of spatial correlation between the ICM and DM velocity dipoles, i.e., $\Delta \theta_{v, \text{sim}} \approx 0$. The rotational offset is more continuously initialized as the merger approaches pericenter passage for simulations initialized with larger impact parameters. The distributions of $\Delta \theta_{v, \text{sim}}$ for larger sampled impact parameters ($250 \lesssim b \lesssim 500$ kpc) exhibit a broader spread of $\Delta \theta_{v, \text{sim}}$ values in the same short range of epochs prior to pericenter passage ($-100 \leq t - t_{\text{peri}} \leq 0$ Myr).

\begin{figure*}[t] 
    \centering 
        \includegraphics[width=1\textwidth]{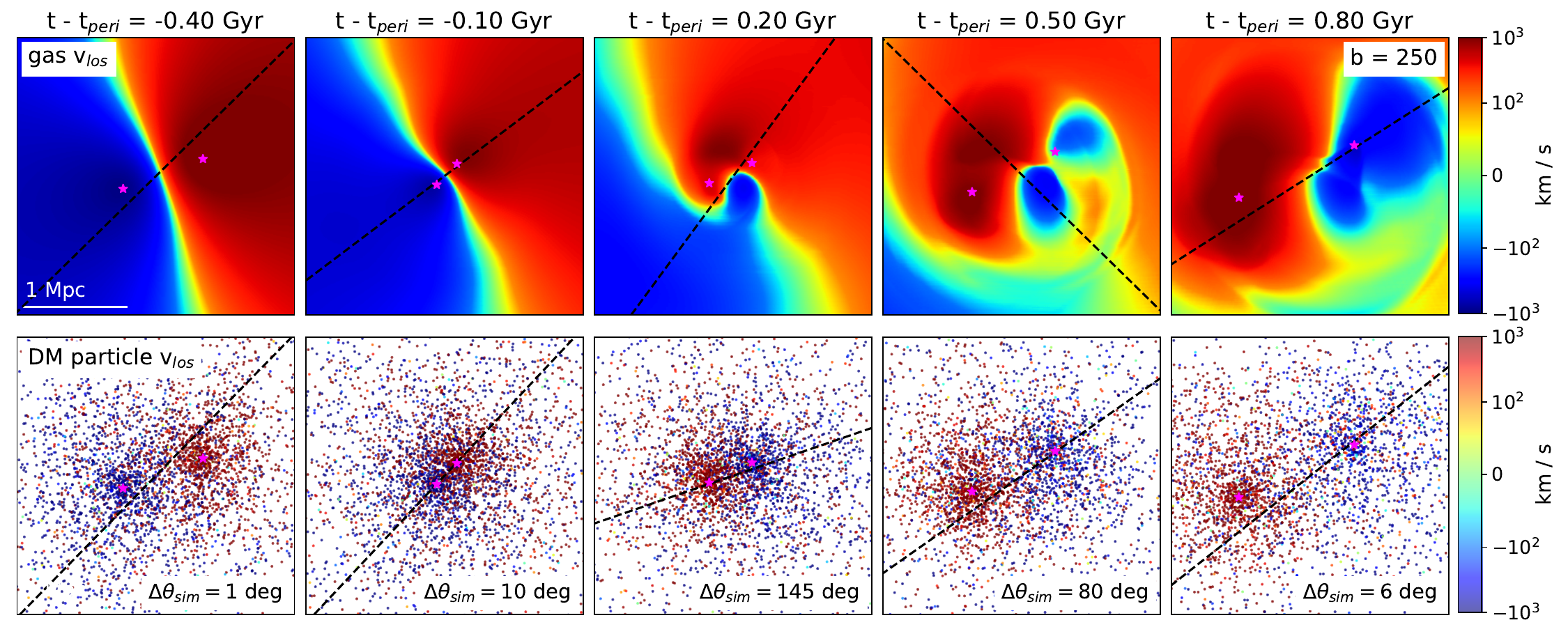} 
        
        \vspace{2em}
        
        \includegraphics[width=1\textwidth]{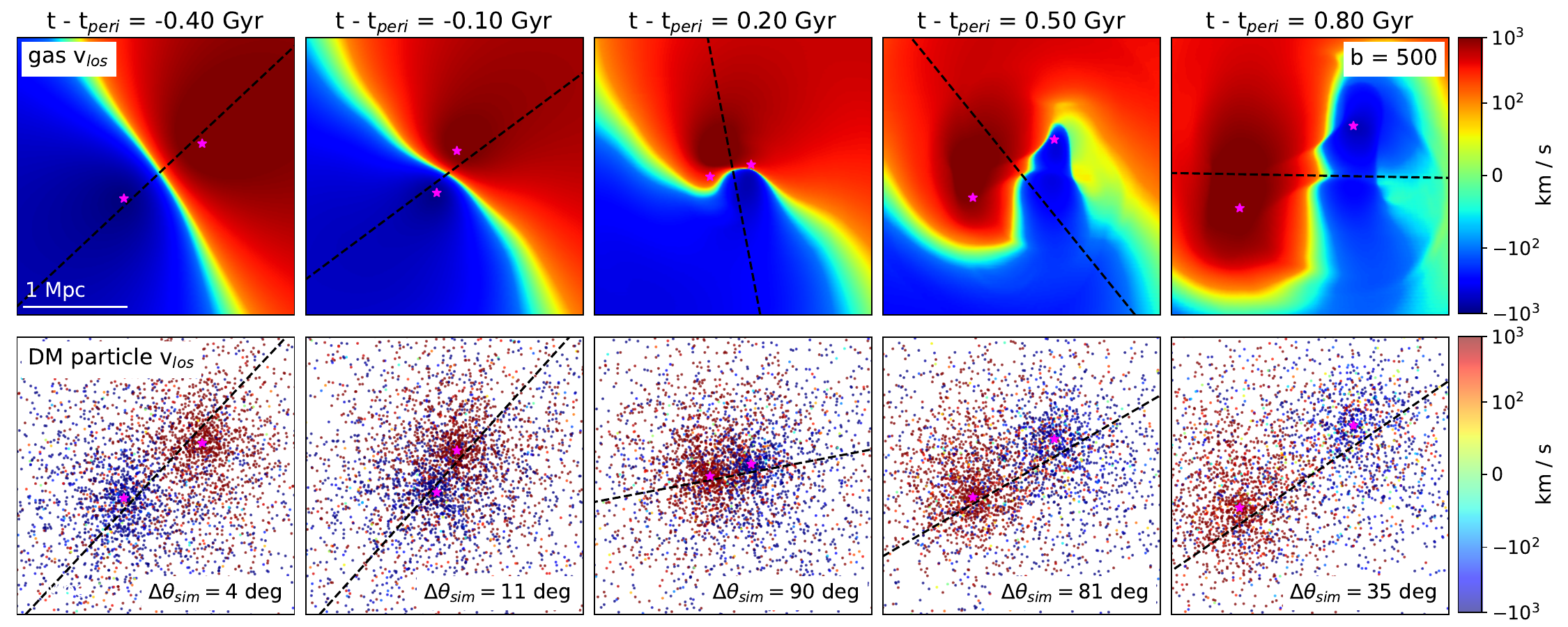}
        \caption{Evolution of the ICM/DM LOS velocity dipole decoupling (measured as $\Delta \theta_{v, \text{sim}}$) as a function of epoch for simulations initialized with impact parameters of $b = 250$ (\textit{top}) and $500$ kpc (\textit{bottom}), with $R = 1.5$, $v_i = 3000$ km s$^{-1}$, and gas profile: Int$+$NCC for each simulation. Mock velocity maps are shown at the full simulation resolution with black dashed lines indicating the calculated rotational orientation of each velocity dipole. Prior to the pericenter passage, the velocity dipoles are well-aligned. As the merger progresses, the DM velocity dipole rotates more quickly than the ICM velocity dipole, which lags the DM velocity dipole in the inner regions before catching up and then propagating outwards over several $0.1$ Gyr post-pericenter passage, when the large-scale dipoles again approach alignment. This leads to a continuous sampling of intermediate rotational offset values, including angles that are semi-stable for some geometries (e.g., values near $\sim90$ degrees in the bottom example for $b = 500$ kpc).} \vspace{1em}\label{fig:movie_snaps} 
\end{figure*}

In the same range of epochs post-pericenter passage, the transition of the ICM and DM velocity dipoles from well-aligned to anti-aligned happens rapidly (i.e., almost entirely within 100 Myr of the pericenter passage) for low impact parameter simulations ($b \lesssim 100$ kpc). For simulations initialized with larger impact parameters, ($250 \lesssim b \lesssim 500$ kpc), this transition spreads over intermediate values of $\Delta \theta_{v, \text{sim}}$ much more prominently, and a simple 2D schematic is no longer a good approximation. In Figure \ref{fig:movie_snaps}, we illustrate the evolution of the ICM/DM velocity dipole decoupling as a function of epoch for simulations initialized with impact parameters of $b = 250$ and $500$ kpc (with $R = 1.5$, $v_i = 3000$ km s$^{-1}$, and gas profile: Int$+$NCC). Here, we utilize the idealized mock observables before any processing in order to highlight the decoupling at the full simulation resolution. Each snapshot is projected along a viewing angle of $|{\bf L} \cdot \; \hat x| = 0.88$, which is favored by the matching criteria. 

For the $b = 250$ kpc simulation, the gas and DM LOS velocity dipoles are aligned at epochs well prior to the pericenter passage (e.g., at $t - t_{\text{peri}} \approx -0.40$ Gyr), with $\Delta \theta_{v, \text{sim}} \approx 0$ (measured as described in Section \ref{sec:vgal_subsec}). As the $b = 250$ system approaches and undergoes pericenter passage, the DM velocity dipole rapidly rotates, since the DM is governed exclusively by gravitational interactions. The the gas dipole rotates more slowly at small radii as the gas undergoes strong collisional interactions, while it remains largely unchanged at large radii. This induces a global DM/ICM velocity dipole offset that reaches $\approx 145$ degrees at $t - t_{\text{peri}} \approx 0.20$ Gyr. As the DM dipole approaches anti-alignment with the original orientation of the dipole, the gas in the inner cluster cores takes several $0.1$ Gyr to `catch up' to the DM rotationally, and several additional $0.1$ Gyr are required for the rotational motion to propagate outwards to large radii, so that the global gas velocity dipole again becomes roughly aligned with the DM velocity dipole by $t - t_{\text{peri}} \approx 0.80$ Gyr. 

This general phenomenon is illustrated more dramatically in the $b = 500$ kpc simulation. In this case, the gas and DM velocity dipoles similarly begin aligned ($\Delta \theta_{v, \text{sim}} \approx 0$) at epochs well prior to the pericenter passage. As the $b = 500$ kpc system approaches and undergoes pericenter passage, the DM velocity dipole rapidly rotates towards $180$ degrees out of phase with the initial dipole orientation. The gas again rotates more slowly than the DM at the center due to collisional interactions, and it roughly retains its original dipole orientation on large scales. For several $0.1$ Gyr post-pericenter passage, the DM velocity dipole approaches and retains an anti-aligned orientation relative to the original DM velocity dipole, and the global gas dipole slowly progresses towards alignment with the DM velocity dipole, with the motion of the outer gas again being driven from the central inner gas outwards. As a result, there is a semi-stable rotational offset at $\Delta \theta_{v, \text{sim}} \approx 90$ degrees for simulations initialized with $b = 500$ kpc and viewed at $|{\bf L} \cdot \; \hat x| = 0.88$. This is emphasized in Figure \ref{fig:2dhist}, where there are pileups of rotational offsets as a function of $|{\bf L} \cdot \; \hat x|$, particularly for large impact parameters. 

These trends persist regardless of the sign of $|{\bf L} \cdot \; \hat x|$ and for simulations which are projected along viewing angles sampled an order of magnitude more finely in $|{\bf L} \cdot \; \hat x|$. Movies sampled more finely in epoch space illustrating this velocity decoupling, in addition to the evolution of the gas temperature and the gas/DM densities, are shown for the above cases in Figures \ref{fig:movie2} and \ref{fig:movie3}.

\section{Conclusions} \label{sec:conclusions}
In this work, we have demonstrated the constraining power of the ICM-SHOX pipeline to jointly infer multiple galaxy cluster merger parameters quantitatively via comparison of a novel combination of multi-probe observables to mock observables generated from idealized hydrodynamical simulations tailored to the \macs merger. We constrain the \macs cluster component masses and gas profiles, as well as global system properties: viewing angle, epoch, initial impact parameter, and initial relative velocity. The epoch is likely in the range $0$--$60$ Myr after the pericenter passage, and the viewing angle is likely inclined $\approx 27$--$40$ degrees from the merger axis along which the relative velocity between the cluster components is initialized when the clusters are separated by 3 Mpc at the initial infall. The most likely \macs impact parameter is $b \lesssim 250$ kpc, the mass ratio is $R \approx 1.5$--$3.0$, the initial relative velocity is $v_i \approx 1700$--$3000$ km s$^{-1}$, and the primary and secondary cluster components initially have gas distributions that are moderately and strongly disturbed, respectively. 

Our application of the ICM-SHOX pipeline to \macs has revealed a most curious observable: a misalignment of the velocity maps of the gas and the stellar components of the merging system. This misalignment is manifested by a separation in phase space between DM and ICM LOS velocity maps generated with cluster-member $z_{\text{spec}}$ and kSZ, respectively, which our simulations indicate is entirely dependent on the different collisional properties of the gas, DM, and stars. The decoupling is initiated as the clusters approach the pericenter passage, when the ICM begins to be affected by merger-driven hydrodynamical processes while the DM does not. Systematically, the behavior of the velocity space decoupling is, therefore, highly sensitive to the merger epoch, system geometry, and the orientation of the merger relative to our LOS. Such a discovery would not be possible without our multi-probe approach, especially incorporating the kSZ measurements, and its interpretation would be difficult without careful comparison to simulations. 

In this demonstration of the ICM-SHOX multi-probe data and pipeline, we identified a set of matching criteria that were manually calibrated on each \macs cluster observable (e.g., distance between GL mass peaks). However, due to the low S/N of the ICM v$_{\text{pec}}$ map, identifying a categorical matching criterion was not feasible, and we thus developed a more formal statistical comparison for the ICM v$_{\text{pec}}$ map. In future studies, we plan to apply this methodology to the full set of multi-probe data by empirically calibrating a simulated $\chi^2$ distribution for each mock observable. By utilizing these distributions, we can more objectively determine the matching thresholds without relying on the human eye to identify patterns in the observables. After implementing these improvements, we plan to apply this framework to the full ICM-SHOX sample.

\begin{acknowledgements}
Some of the data presented herein were obtained at the W. M. Keck Observatory, which is operated as a scientific partnership among the California Institute of Technology (Caltech), the University of California, and the National Aeronautics and Space Administration (NASA). The Observatory was made possible by the generous financial support of the W. M. Keck Foundation. The authors wish to recognize and acknowledge the very significant cultural role and reverence that the summit of Maunakea has always had within the indigenous Hawaiian community.  We are most fortunate to have the opportunity to conduct observations from this mountain. 

The X-ray observations in this analysis were obtained from the Chandra Data Archive. The hydrodynamical simulations were run on the Pleiades supercomputer at NASA/Ames Research Center. The simulation post-processing computations were conducted in the Resnick High Performance Computing Center, a facility supported by Resnick Sustainability Institute at Caltech. 

EMS acknowledges support from a National Science Foundation Graduate Research Fellowship (NSF GRFP) under Grant No. DGE‐1745301, the Wallace L. W. Sargent Graduate Fellowship at Caltech, and NSF/AST-2206082. JS acknowledges support from NSF/AST-2206082. JAZ is funded by the Chandra X-ray Center, which is operated by the Smithsonian Astrophysical Observatory for and on behalf of NASA under contract NAS8-03060. EB acknowledges support from NSF/AST-2206083. AZ acknowledges support by Grant No. 2020750 from the United States-Israel Binational Science Foundation (BSF) and Grant No. 2109066 from the United States NSF, and by the Ministry of Science \& Technology, Israel. TM acknowledges support from the AtLAST project, which has received funding from the European Union’s Horizon 2020 research and innovation program under grant agreement No 951815. AM thanks support from Consejo Nacional de Humanidades Ciencias y Technolog\'ias (CONAHCYT) project A1-S-45680. We thank the anonymous reviewer for their detailed comments, which helped improve this manuscript. 

\end{acknowledgements}

\facility{Keck/DEIMOS, Chandra, Bolocam, AzTEC (installed on the 10m Atacama Submillimeter Telescope Experiment), Planck, HST}

\software{GAMER-2 \citep{Schive2018}, Xspec 12.12.0 \citep{arnaud1996}, yt \citep{Turk2011}, pyXSIM \citep{zuhone2016}, SOXS \citep{soxs}, mpi4py \citep{dalcin2021}, CIAO \citep{ciao}, PypeIt \citep{pypeit}, spec2d \citep{spec2d_cooper, spec2d_newman}, SpecPro \citep{specpro}, Pumpkin \citep{rhea2020}, Astropy \citep{astropy2013, astropy2018}, NumPy \citep{vanderwalt2011, numpy}, Matplotlib \citep{matplotlib}, lmfit \citep{lmfit}, Colossus \citep{diemer2018}, contbin \citep{Sanders2006}, scikit-image; \citep{scikit-image}}

\clearpage
\bibliography{refs}

\begin{thebibliography}{}
\expandafter\ifx\csname natexlab\endcsname\relax\def\natexlab#1{#1}\fi
\providecommand{\url}[1]{\href{#1}{#1}}
\providecommand{\dodoi}[1]{doi:~\href{http://doi.org/#1}{\nolinkurl{#1}}}
\providecommand{\doeprint}[1]{\href{http://ascl.net/#1}{\nolinkurl{http://ascl.net/#1}}}
\providecommand{\doarXiv}[1]{\href{https://arxiv.org/abs/#1}{\nolinkurl{https://arxiv.org/abs/#1}}}

\bibitem[{{Adam} {et~al.}(2017){Adam}, {Bartalucci}, {Pratt}, {Ade},
  {Andr{\'e}}, {Arnaud}, {Beelen}, {Beno{\^\i}t}, {Bideaud}, {Billot},
  {Bourdin}, {Bourrion}, {Calvo}, {Catalano}, {Coiffard}, {Comis}, {D'Addabbo},
  {De Petris}, {D{\'e}mocl{\`e}s}, {D{\'e}sert}, {Doyle}, {Egami}, {Ferrari},
  {Goupy}, {Kramer}, {Lagache}, {Leclercq}, {Mac{\'\i}as-P{\'e}rez},
  {Maurogordato}, {Mauskopf}, {Mayet}, {Monfardini}, {Mroczkowski}, {Pajot},
  {Pascale}, {Perotto}, {Pisano}, {Pointecouteau}, {Ponthieu}, {Rev{\'e}ret},
  {Ritacco}, {Rodriguez}, {Romero}, {Ruppin}, {Schuster}, {Sievers},
  {Triqueneaux}, {Tucker}, {Zemcov}, \& {Zylka}}]{Adam2017}
{Adam}, R., {Bartalucci}, I., {Pratt}, G.~W., {et~al.} 2017, \aap, 598, A115,
  \dodoi{10.1051/0004-6361/201629182}

\bibitem[{{Anbajagane} {et~al.}(2022){Anbajagane}, {Aung}, {Evrard}, {Farahi},
  {Nagai}, {Barnes}, {Cui}, {Dolag}, {McCarthy}, {Rasia}, \&
  {Yepes}}]{Anbajagane2022}
{Anbajagane}, D., {Aung}, H., {Evrard}, A.~E., {et~al.} 2022, \mnras, 510,
  2980, \dodoi{10.1093/mnras/stab3587}

\bibitem[{{Anders} \& {Grevesse}(1989)}]{angrabunds}
{Anders}, E., \& {Grevesse}, N. 1989, \gca, 53, 197,
  \dodoi{10.1016/0016-7037(89)90286-X}

\bibitem[{{Araya-Melo} {et~al.}(2009){Araya-Melo}, {Reisenegger}, {Meza}, {van
  de Weygaert}, {D{\"u}nner}, \& {Quintana}}]{largestobjs}
{Araya-Melo}, P.~A., {Reisenegger}, A., {Meza}, A., {et~al.} 2009, \mnras, 399,
  97, \dodoi{10.1111/j.1365-2966.2009.15292.x}

\bibitem[{{Arnaud}(1996)}]{arnaud1996}
{Arnaud}, K.~A. 1996, in Astronomical Society of the Pacific Conference Series,
  Vol. 101, Astronomical Data Analysis Software and Systems V, ed. G.~H.
  {Jacoby} \& J.~{Barnes}, 17

\bibitem[{{Astropy Collaboration} {et~al.}(2013){Astropy Collaboration},
  {Robitaille}, {Tollerud}, {Greenfield}, {Droettboom}, {Bray}, {Aldcroft},
  {Davis}, {Ginsburg}, {Price-Whelan}, {Kerzendorf}, {Conley}, {Crighton},
  {Barbary}, {Muna}, {Ferguson}, {Grollier}, {Parikh}, {Nair}, {Unther},
  {Deil}, {Woillez}, {Conseil}, {Kramer}, {Turner}, {Singer}, {Fox}, {Weaver},
  {Zabalza}, {Edwards}, {Azalee Bostroem}, {Burke}, {Casey}, {Crawford},
  {Dencheva}, {Ely}, {Jenness}, {Labrie}, {Lim}, {Pierfederici}, {Pontzen},
  {Ptak}, {Refsdal}, {Servillat}, \& {Streicher}}]{astropy2013}
{Astropy Collaboration}, {Robitaille}, T.~P., {Tollerud}, E.~J., {et~al.} 2013,
  \aap, 558, A33, \dodoi{10.1051/0004-6361/201322068}

\bibitem[{{Astropy Collaboration} {et~al.}(2018){Astropy Collaboration},
  {Price-Whelan}, {Sip{\H{o}}cz}, {G{\"u}nther}, {Lim}, {Crawford}, {Conseil},
  {Shupe}, {Craig}, {Dencheva}, {Ginsburg}, {VanderPlas}, {Bradley},
  {P{\'e}rez-Su{\'a}rez}, {de Val-Borro}, {Aldcroft}, {Cruz}, {Robitaille},
  {Tollerud}, {Ardelean}, {Babej}, {Bach}, {Bachetti}, {Bakanov}, {Bamford},
  {Barentsen}, {Barmby}, {Baumbach}, {Berry}, {Biscani}, {Boquien}, {Bostroem},
  {Bouma}, {Brammer}, {Bray}, {Breytenbach}, {Buddelmeijer}, {Burke},
  {Calderone}, {Cano Rodr{\'\i}guez}, {Cara}, {Cardoso}, {Cheedella}, {Copin},
  {Corrales}, {Crichton}, {D'Avella}, {Deil}, {Depagne}, {Dietrich}, {Donath},
  {Droettboom}, {Earl}, {Erben}, {Fabbro}, {Ferreira}, {Finethy}, {Fox},
  {Garrison}, {Gibbons}, {Goldstein}, {Gommers}, {Greco}, {Greenfield},
  {Groener}, {Grollier}, {Hagen}, {Hirst}, {Homeier}, {Horton}, {Hosseinzadeh},
  {Hu}, {Hunkeler}, {Ivezi{\'c}}, {Jain}, {Jenness}, {Kanarek}, {Kendrew},
  {Kern}, {Kerzendorf}, {Khvalko}, {King}, {Kirkby}, {Kulkarni}, {Kumar},
  {Lee}, {Lenz}, {Littlefair}, {Ma}, {Macleod}, {Mastropietro}, {McCully},
  {Montagnac}, {Morris}, {Mueller}, {Mumford}, {Muna}, {Murphy}, {Nelson},
  {Nguyen}, {Ninan}, {N{\"o}the}, {Ogaz}, {Oh}, {Parejko}, {Parley}, {Pascual},
  {Patil}, {Patil}, {Plunkett}, {Prochaska}, {Rastogi}, {Reddy Janga},
  {Sabater}, {Sakurikar}, {Seifert}, {Sherbert}, {Sherwood-Taylor}, {Shih},
  {Sick}, {Silbiger}, {Singanamalla}, {Singer}, {Sladen}, {Sooley},
  {Sornarajah}, {Streicher}, {Teuben}, {Thomas}, {Tremblay}, {Turner},
  {Terr{\'o}n}, {van Kerkwijk}, {de la Vega}, {Watkins}, {Weaver}, {Whitmore},
  {Woillez}, {Zabalza}, \& {Astropy Contributors}}]{astropy2018}
{Astropy Collaboration}, {Price-Whelan}, A.~M., {Sip{\H{o}}cz}, B.~M., {et~al.}
  2018, \aj, 156, 123, \dodoi{10.3847/1538-3881/aabc4f}

\bibitem[{{Baltz} {et~al.}(2009){Baltz}, {Marshall}, \& {Oguri}}]{truncatedNFW}
{Baltz}, E.~A., {Marshall}, P., \& {Oguri}, M. 2009, \jcap, 2009, 015,
  \dodoi{10.1088/1475-7516/2009/01/015}

\bibitem[{{Barret} {et~al.}(2020){Barret}, {Decourchelle}, {Fabian},
  {Guainazzi}, {Nandra}, {Smith}, \& {den Herder}}]{athena}
{Barret}, D., {Decourchelle}, A., {Fabian}, A., {et~al.} 2020, Astronomische
  Nachrichten, 341, 224, \dodoi{10.1002/asna.202023782}

\bibitem[{{Bartalucci} {et~al.}(2014){Bartalucci}, {Mazzotta}, {Bourdin}, \&
  {Vikhlinin}}]{acis_bkg}
{Bartalucci}, I., {Mazzotta}, P., {Bourdin}, H., \& {Vikhlinin}, A. 2014, \aap,
  566, A25, \dodoi{10.1051/0004-6361/201423443}

\bibitem[{{Beers} {et~al.}(1990){Beers}, {Flynn}, \& {Gebhardt}}]{Beers1990}
{Beers}, T.~C., {Flynn}, K., \& {Gebhardt}, K. 1990, \aj, 100, 32,
  \dodoi{10.1086/115487}

\bibitem[{{Boschin} {et~al.}(2013){Boschin}, {Girardi}, \&
  {Barrena}}]{Boschin2013}
{Boschin}, W., {Girardi}, M., \& {Barrena}, R. 2013, \mnras, 434, 772,
  \dodoi{10.1093/mnras/stt1070}

\bibitem[{{Boschin} {et~al.}(2006){Boschin}, {Girardi}, {Spolaor}, \&
  {Barrena}}]{Boschin2006}
{Boschin}, W., {Girardi}, M., {Spolaor}, M., \& {Barrena}, R. 2006, \aap, 449,
  461, \dodoi{10.1051/0004-6361:20054408}

\bibitem[{{Breuer} {et~al.}(2020){Breuer}, {Werner}, {Mernier}, {Mroczkowski},
  {Simionescu}, {Clarke}, {ZuHone}, \& {Di Mascolo}}]{Breuer2020}
{Breuer}, J.~P., {Werner}, N., {Mernier}, F., {et~al.} 2020, \mnras, 495, 5014,
  \dodoi{10.1093/mnras/staa1492}

\bibitem[{{Cavagnolo} {et~al.}(2009){Cavagnolo}, {Donahue}, {Voit}, \&
  {Sun}}]{entropy_accept}
{Cavagnolo}, K.~W., {Donahue}, M., {Voit}, G.~M., \& {Sun}, M. 2009, \apjs,
  182, 12, \dodoi{10.1088/0067-0049/182/1/12}

\bibitem[{{Chadayammuri} {et~al.}(2022){Chadayammuri}, {ZuHone}, {Nulsen},
  {Nagai}, {Felix}, {Andrade-Santos}, {King}, \& {Russell}}]{Chadayammuri22}
{Chadayammuri}, U., {ZuHone}, J., {Nulsen}, P., {et~al.} 2022, \mnras, 509,
  1201, \dodoi{10.1093/mnras/stab2629}

\bibitem[{{Clowe} {et~al.}(2006){Clowe}, {Brada{\v{c}}}, {Gonzalez},
  {Markevitch}, {Randall}, {Jones}, \& {Zaritsky}}]{clowe2006}
{Clowe}, D., {Brada{\v{c}}}, M., {Gonzalez}, A.~H., {et~al.} 2006, \apjl, 648,
  L109, \dodoi{10.1086/508162}

\bibitem[{{Coe} {et~al.}(2019){Coe}, {Salmon}, {Brada{\v{c}}}, {Bradley},
  {Sharon}, {Zitrin}, {Acebron}, {Cerny}, {Cibirka}, {Strait},
  {Paterno-Mahler}, {Mahler}, {Avila}, {Ogaz}, {Huang}, {Pelliccia}, {Stark},
  {Mainali}, {Oesch}, {Trenti}, {Carrasco}, {Dawson}, {Rodney}, {Strolger},
  {Riess}, {Jones}, {Frye}, {Czakon}, {Umetsu}, {Vulcani}, {Graur}, {Jha},
  {Graham}, {Molino}, {Nonino}, {Hjorth}, {Selsing}, {Christensen},
  {Kikuchihara}, {Ouchi}, {Oguri}, {Welch}, {Lemaux}, {Andrade-Santos}, {Hoag},
  {Johnson}, {Peterson}, {Past}, {Fox}, {Agulli}, {Livermore}, {Ryan}, {Lam},
  {Sendra-Server}, {Toft}, {Lovisari}, \& {Su}}]{Coe2019}
{Coe}, D., {Salmon}, B., {Brada{\v{c}}}, M., {et~al.} 2019, \apj, 884, 85,
  \dodoi{10.3847/1538-4357/ab412b}

\bibitem[{{Coe, D.}(2016)}]{relics_ref}
{Coe, D.} 2016, Reionization Lensing Cluster Survey (``RELICS"),  STScI/MAST,
  \dodoi{10.17909/T9SP45}

\bibitem[{{Cooper} {et~al.}(2012){Cooper}, {Newman}, {Davis}, {Finkbeiner}, \&
  {Gerke}}]{spec2d_cooper}
{Cooper}, M.~C., {Newman}, J.~A., {Davis}, M., {Finkbeiner}, D.~P., \& {Gerke},
  B.~F. 2012, {spec2d: DEEP2 DEIMOS Spectral Pipeline}, Astrophysics Source
  Code Library, record ascl:1203.003.
\newblock \doeprint{1203.003}

\bibitem[{{Crawford} {et~al.}(2011){Crawford}, {Wirth}, {Bershady}, \&
  {Hon}}]{obsz_crawford}
{Crawford}, S.~M., {Wirth}, G.~D., {Bershady}, M.~A., \& {Hon}, K. 2011, \apj,
  741, 98, \dodoi{10.1088/0004-637X/741/2/98}

\bibitem[{{Dalcin} \& {Fang}(2021)}]{dalcin2021}
{Dalcin}, L., \& {Fang}, Y.-L.~L. 2021, Computing in Science and Engineering,
  23, 47, \dodoi{10.1109/MCSE.2021.3083216}

\bibitem[{{Dehghan} {et~al.}(2017){Dehghan}, {Johnston-Hollitt}, {Colless}, \&
  {Miller}}]{Dehghan2017}
{Dehghan}, S., {Johnston-Hollitt}, M., {Colless}, M., \& {Miller}, R. 2017,
  \mnras, 468, 2645, \dodoi{10.1093/mnras/stx582}

\bibitem[{{Di Mascolo} {et~al.}(2021){Di Mascolo}, {Mroczkowski}, {Perrott},
  {Rudnick}, {James Jee}, {HyeongHan}, {Churazov}, {Collier}, {Diego},
  {Hopkins}, {Kim}, {Koribalski}, {Marvil}, {van der Burg}, \&
  {West}}]{dimascolo2021}
{Di Mascolo}, L., {Mroczkowski}, T., {Perrott}, Y., {et~al.} 2021, \aap, 650,
  A153, \dodoi{10.1051/0004-6361/202040260}

\bibitem[{{Diehl} \& {Statler}(2006)}]{Diehl2006}
{Diehl}, S., \& {Statler}, T.~S. 2006, \mnras, 368, 497,
  \dodoi{10.1111/j.1365-2966.2006.10125.x}

\bibitem[{{Diemer}(2018)}]{diemer2018}
{Diemer}, B. 2018, \apjs, 239, 35, \dodoi{10.3847/1538-4365/aaee8c}

\bibitem[{{Diemer} \& {Joyce}(2019)}]{diemer2019}
{Diemer}, B., \& {Joyce}, M. 2019, \apj, 871, 168,
  \dodoi{10.3847/1538-4357/aafad6}

\bibitem[{{Dressler} \& {Gunn}(1992)}]{obsz_dressler}
{Dressler}, A., \& {Gunn}, J.~E. 1992, \apjs, 78, 1, \dodoi{10.1086/191620}

\bibitem[{{Ebeling} {et~al.}(2007){Ebeling}, {Barrett}, {Donovan}, {Ma},
  {Edge}, \& {van Speybroeck}}]{redshifts}
{Ebeling}, H., {Barrett}, E., {Donovan}, D., {et~al.} 2007, \apjl, 661, L33,
  \dodoi{10.1086/518603}

\bibitem[{{Eddington}(1916)}]{eddington1916}
{Eddington}, A.~S. 1916, \mnras, 76, 572, \dodoi{10.1093/mnras/76.7.572}

\bibitem[{{Ellingson} {et~al.}(1998){Ellingson}, {Yee}, {Abraham}, {Morris}, \&
  {Carlberg}}]{obsz_ellingson}
{Ellingson}, E., {Yee}, H.~K.~C., {Abraham}, R.~G., {Morris}, S.~L., \&
  {Carlberg}, R.~G. 1998, \apjs, 116, 247, \dodoi{10.1086/313106}

\bibitem[{{Faber} {et~al.}(2003){Faber}, {Phillips}, {Kibrick}, {Alcott},
  {Allen}, {Burrous}, {Cantrall}, {Clarke}, {Coil}, {Cowley}, {Davis}, {Deich},
  {Dietsch}, {Gilmore}, {Harper}, {Hilyard}, {Lewis}, {McVeigh}, {Newman},
  {Osborne}, {Schiavon}, {Stover}, {Tucker}, {Wallace}, {Wei}, {Wirth}, \&
  {Wright}}]{faber2003}
{Faber}, S.~M., {Phillips}, A.~C., {Kibrick}, R.~I., {et~al.} 2003, in Society
  of Photo-Optical Instrumentation Engineers (SPIE) Conference Series, Vol.
  4841, Instrument Design and Performance for Optical/Infrared Ground-based
  Telescopes, ed. M.~{Iye} \& A.~F.~M. {Moorwood}, 1657--1669,
  \dodoi{10.1117/12.460346}

\bibitem[{{Fakhouri} {et~al.}(2010){Fakhouri}, {Ma}, \&
  {Boylan-Kolchin}}]{Fakhouri2010}
{Fakhouri}, O., {Ma}, C.-P., \& {Boylan-Kolchin}, M. 2010, \mnras, 406, 2267,
  \dodoi{10.1111/j.1365-2966.2010.16859.x}

\bibitem[{{Freeman} {et~al.}(2002){Freeman}, {Kashyap}, {Rosner}, \&
  {Lamb}}]{Freeman2002}
{Freeman}, P.~E., {Kashyap}, V., {Rosner}, R., \& {Lamb}, D.~Q. 2002, \apjs,
  138, 185, \dodoi{10.1086/324017}

\bibitem[{{Fruscione} {et~al.}(2006){Fruscione}, {McDowell}, {Allen},
  {Brickhouse}, {Burke}, {Davis}, {Durham}, {Elvis}, {Galle}, {Harris},
  {Huenemoerder}, {Houck}, {Ishibashi}, {Karovska}, {Nicastro}, {Noble},
  {Nowak}, {Primini}, {Siemiginowska}, {Smith}, \& {Wise}}]{ciao}
{Fruscione}, A., {McDowell}, J.~C., {Allen}, G.~E., {et~al.} 2006, in Society
  of Photo-Optical Instrumentation Engineers (SPIE) Conference Series, Vol.
  6270, Society of Photo-Optical Instrumentation Engineers (SPIE) Conference
  Series, ed. D.~R. {Silva} \& R.~E. {Doxsey}, 62701V,
  \dodoi{10.1117/12.671760}

\bibitem[{{Furtak} {et~al.}(2022){Furtak}, {Plat}, {Zitrin}, {Topping},
  {Stark}, {Strait}, {Charlot}, {Coe}, {Andrade-Santos}, {Brada{\v{c}}},
  {Bradley}, {Lemaux}, \& {Sharon}}]{furtak2022}
{Furtak}, L.~J., {Plat}, A., {Zitrin}, A., {et~al.} 2022, \mnras, 516, 1373,
  \dodoi{10.1093/mnras/stac2169}

\bibitem[{{Harris} {et~al.}(2020){Harris}, {Millman}, {van der Walt},
  {Gommers}, {Virtanen}, {Cournapeau}, {Wieser}, {Taylor}, {Berg}, {Smith},
  {Kern}, {Picus}, {Hoyer}, {van Kerkwijk}, {Brett}, {Haldane}, {del R{\'\i}o},
  {Wiebe}, {Peterson}, {G{\'e}rard-Marchant}, {Sheppard}, {Reddy}, {Weckesser},
  {Abbasi}, {Gohlke}, \& {Oliphant}}]{numpy}
{Harris}, C.~R., {Millman}, K.~J., {van der Walt}, S.~J., {et~al.} 2020, \nat,
  585, 357, \dodoi{10.1038/s41586-020-2649-2}

\bibitem[{{HI4PI Collaboration} {et~al.}(2016){HI4PI Collaboration}, {Ben
  Bekhti}, {Fl{\"o}er}, {Keller}, {Kerp}, {Lenz}, {Winkel}, {Bailin},
  {Calabretta}, {Dedes}, {Ford}, {Gibson}, {Haud}, {Janowiecki}, {Kalberla},
  {Lockman}, {McClure-Griffiths}, {Murphy}, {Nakanishi}, {Pisano}, \&
  {Staveley-Smith}}]{absmap}
{HI4PI Collaboration}, {Ben Bekhti}, N., {Fl{\"o}er}, L., {et~al.} 2016, \aap,
  594, A116, \dodoi{10.1051/0004-6361/201629178}

\bibitem[{{Hitomi Collaboration} {et~al.}(2016){Hitomi Collaboration},
  {Aharonian}, {Akamatsu}, {Akimoto}, {Allen}, {Anabuki}, {Angelini}, {Arnaud},
  {Audard}, {Awaki}, {Axelsson}, {Bamba}, {Bautz}, {Blandford}, {Brenneman},
  {Brown}, {Bulbul}, {Cackett}, {Chernyakova}, {Chiao}, {Coppi}, {Costantini},
  {de Plaa}, {den Herder}, {Done}, {Dotani}, {Ebisawa}, {Eckart}, {Enoto},
  {Ezoe}, {Fabian}, {Ferrigno}, {Foster}, {Fujimoto}, {Fukazawa}, {Furuzawa},
  {Galeazzi}, {Gallo}, {Gandhi}, {Giustini}, {Goldwurm}, {Gu}, {Guainazzi},
  {Haba}, {Hagino}, {Hamaguchi}, {Harrus}, {Hatsukade}, {Hayashi}, {Hayashi},
  {Hayashida}, {Hiraga}, {Hornschemeier}, {Hoshino}, {Hughes}, {Iizuka},
  {Inoue}, {Inoue}, {Ishibashi}, {Ishida}, {Ishikawa}, {Ishisaki}, {Itoh},
  {Iyomoto}, {Kaastra}, {Kallman}, {Kamae}, {Kara}, {Kataoka}, {Katsuda},
  {Katsuta}, {Kawaharada}, {Kawai}, {Kelley}, {Khangulyan}, {Kilbourne},
  {King}, {Kitaguchi}, {Kitamoto}, {Kitayama}, {Kohmura}, {Kokubun}, {Koyama},
  {Koyama}, {Kretschmar}, {Krimm}, {Kubota}, {Kunieda}, {Laurent}, {Lebrun},
  {Lee}, {Leutenegger}, {Limousin}, {Loewenstein}, {Long}, {Lumb}, {Madejski},
  {Maeda}, {Maier}, {Makishima}, {Markevitch}, {Matsumoto}, {Matsushita},
  {McCammon}, {McNamara}, {Mehdipour}, {Miller}, {Miller}, {Mineshige},
  {Mitsuda}, {Mitsuishi}, {Miyazawa}, {Mizuno}, {Mori}, {Mori}, {Moseley},
  {Mukai}, {Murakami}, {Murakami}, {Mushotzky}, {Nagino}, {Nakagawa},
  {Nakajima}, {Nakamori}, {Nakano}, {Nakashima}, {Nakazawa}, {Nobukawa},
  {Noda}, {Nomachi}, {O'Dell}, {Odaka}, {Ohashi}, {Ohno}, {Okajima}, {Ota},
  {Ozaki}, {Paerels}, {Paltani}, {Parmar}, {Petre}, {Pinto}, {Pohl}, {Porter},
  {Pottschmidt}, {Ramsey}, {Reynolds}, {Russell}, {Safi-Harb}, {Saito},
  {Sakai}, {Sameshima}, {Sato}, {Sato}, {Sato}, {Sawada}, {Schartel},
  {Serlemitsos}, {Seta}, {Shidatsu}, {Simionescu}, {Smith}, {Soong}, {Stawarz},
  {Sugawara}, {Sugita}, {Szymkowiak}, {Tajima}, {Takahashi}, {Takahashi},
  {Takeda}, {Takei}, {Tamagawa}, {Tamura}, {Tamura}, {Tanaka}, {Tanaka},
  {Tanaka}, {Tashiro}, {Tawara}, {Terada}, {Terashima}, {Tombesi}, {Tomida},
  {Tsuboi}, {Tsujimoto}, {Tsunemi}, {Tsuru}, {Uchida}, {Uchiyama}, {Uchiyama},
  {Ueda}, {Ueda}, {Ueno}, {Uno}, {Urry}, {Ursino}, {de Vries}, {Watanabe},
  {Werner}, {Wik}, {Wilkins}, {Williams}, {Yamada}, {Yamaguchi}, {Yamaoka},
  {Yamasaki}, {Yamauchi}, {Yamauchi}, {Yaqoob}, {Yatsu}, {Yonetoku}, {Yoshida},
  {Yuasa}, {Zhuravleva}, \& {Zoghbi}}]{hitomi2016}
{Hitomi Collaboration}, {Aharonian}, F., {Akamatsu}, H., {et~al.} 2016, \nat,
  535, 117, \dodoi{10.1038/nature18627}

\bibitem[{{Hitomi Collaboration} {et~al.}(2018){Hitomi Collaboration},
  {Aharonian}, {Akamatsu}, {Akimoto}, {Allen}, {Angelini}, {Audard}, {Awaki},
  {Axelsson}, {Bamba}, {Bautz}, {Blandford}, {Brenneman}, {Brown}, {Bulbul},
  {Cackett}, {Canning}, {Chernyakova}, {Chiao}, {Coppi}, {Costantini}, {de
  Plaa}, {de Vries}, {den Herder}, {Done}, {Dotani}, {Ebisawa}, {Eckart},
  {Enoto}, {Ezoe}, {Fabian}, {Ferrigno}, {Foster}, {Fujimoto}, {Fukazawa},
  {Furuzawa}, {Galeazzi}, {Gallo}, {Gandhi}, {Giustini}, {Goldwurm}, {Gu},
  {Guainazzi}, {Haba}, {Hagino}, {Hamaguchi}, {Harrus}, {Hatsukade}, {Hayashi},
  {Hayashi}, {Hayashi}, {Hayashida}, {Hiraga}, {Hornschemeier}, {Hoshino},
  {Hughes}, {Ichinohe}, {Iizuka}, {Inoue}, {Inoue}, {Inoue}, {Ishida},
  {Ishikawa}, {Ishisaki}, {Iwai}, {Kaastra}, {Kallman}, {Kamae}, {Kataoka},
  {Katsuda}, {Kawai}, {Kelley}, {Kilbourne}, {Kitaguchi}, {Kitamoto},
  {Kitayama}, {Kohmura}, {Kokubun}, {Koyama}, {Koyama}, {Kretschmar}, {Krimm},
  {Kubota}, {Kunieda}, {Laurent}, {Lee}, {Leutenegger}, {Limousin},
  {Loewenstein}, {Long}, {Lumb}, {Madejski}, {Maeda}, {Maier}, {Makishima},
  {Markevitch}, {Matsumoto}, {Matsushita}, {McCammon}, {McNamara}, {Mehdipour},
  {Miller}, {Miller}, {Mineshige}, {Mitsuda}, {Mitsuishi}, {Miyazawa},
  {Mizuno}, {Mori}, {Mori}, {Mukai}, {Murakami}, {Mushotzky}, {Nakagawa},
  {Nakajima}, {Nakamori}, {Nakashima}, {Nakazawa}, {Nobukawa}, {Nobukawa},
  {Noda}, {Odaka}, {Ohashi}, {Ohno}, {Okajima}, {Ota}, {Ozaki}, {Paerels},
  {Paltani}, {Petre}, {Pinto}, {Porter}, {Pottschmidt}, {Reynolds},
  {Safi-Harb}, {Saito}, {Sakai}, {Sasaki}, {Sato}, {Sato}, {Sato}, {Sawada},
  {Schartel}, {Serlemtsos}, {Seta}, {Shidatsu}, {Simionescu}, {Smith}, {Soong},
  {Stawarz}, {Sugawara}, {Sugita}, {Szymkowiak}, {Tajima}, {Takahashi},
  {Takahashi}, {Takeda}, {Takei}, {Tamagawa}, {Tamura}, {Tanaka}, {Tanaka},
  {Tanaka}, {Tanaka}, {Tashiro}, {Tawara}, {Terada}, {Terashima}, {Tombesi},
  {Tomida}, {Tsuboi}, {Tsujimoto}, {Tsunemi}, {Tsuru}, {Uchida}, {Uchiyama},
  {Uchiyama}, {Ueda}, {Ueda}, {Uno}, {Urry}, {Ursino}, {Wang}, {Watanabe},
  {Werner}, {Wilkins}, {Williams}, {Yamada}, {Yamaguchi}, {Yamaoka},
  {Yamasaki}, {Yamauchi}, {Yamauchi}, {Yaqoob}, {Yatsu}, {Yonetoku},
  {Zhuravleva}, \& {Zoghbi}}]{hitomi2018}
---. 2018, \pasj, 70, 9, \dodoi{10.1093/pasj/psx138}

\bibitem[{{Hunter}(2007)}]{matplotlib}
{Hunter}, J.~D. 2007, Computing in Science and Engineering, 9, 90,
  \dodoi{10.1109/MCSE.2007.55}

\bibitem[{{Kaiser}(1986)}]{kaiser1986}
{Kaiser}, N. 1986, \mnras, 222, 323, \dodoi{10.1093/mnras/222.2.323}

\bibitem[{{Kazantzidis} {et~al.}(2004){Kazantzidis}, {Magorrian}, \&
  {Moore}}]{Kazantzidis2004}
{Kazantzidis}, S., {Magorrian}, J., \& {Moore}, B. 2004, \apj, 601, 37,
  \dodoi{10.1086/380192}

\bibitem[{{Kraft} {et~al.}(2022){Kraft}, {Markevitch}, {Kilbourne}, {Adams},
  {Akamatsu}, {Ayromlou}, {Bandler}, {Barbera}, {Bennett}, {Bhardwaj}, {Biffi},
  {Bodewits}, {Bogdan}, {Bonamente}, {Borgani}, {Branduardi-Raymont},
  {Bregman}, {Burchett}, {Cann}, {Carter}, {Chakraborty}, {Churazov}, {Crain},
  {Cumbee}, {Dave}, {DiPirro}, {Dolag}, {Bertrand Doriese}, {Drake}, {Dunn},
  {Eckart}, {Eckert}, {Ettori}, {Forman}, {Galeazzi}, {Gall}, {Gatuzz}, {Hell},
  {Hodges-Kluck}, {Jackman}, {Jahromi}, {Jennings}, {Jones}, {Kaaret},
  {Kavanagh}, {Kelley}, {Khabibullin}, {Kim}, {Koutroumpa}, {Kovacs}, {Kuntz},
  {Lau}, {Lee}, {Leutenegger}, {Lin}, {Lisse}, {Lo Cicero}, {Lovisari},
  {McCammon}, {McEntee}, {Mernier}, {Miller}, {Nagai}, {Negro}, {Nelson},
  {Ness}, {Nulsen}, {Ogorzalek}, {Oppenheimer}, {Oskinova}, {Patnaude},
  {Pfeifle}, {Pillepich}, {Plucinsky}, {Pooley}, {Porter}, {Randall}, {Rasia},
  {Raymond}, {Ruszkowski}, {Sakai}, {Sarkar}, {Sasaki}, {Sato},
  {Schellenberger}, {Schaye}, {Simionescu}, {Smith}, {Steiner}, {Stern}, {Su},
  {Sun}, {Tremblay}, {Truong}, {Tutt}, {Ursino}, {Veilleux}, {Vikhlinin},
  {Vladutescu-Zopp}, {Vogelsberger}, {Walker}, {Weaver}, {Weigt}, {Werk},
  {Werner}, {Wolk}, {Zhang}, {Zhang}, {Zhuravleva}, \&
  {ZuHone}}]{lem_white_paper}
{Kraft}, R., {Markevitch}, M., {Kilbourne}, C., {et~al.} 2022, arXiv e-prints,
  arXiv:2211.09827, \dodoi{10.48550/arXiv.2211.09827}

\bibitem[{{Kravtsov} \& {Borgani}(2012)}]{kbreview}
{Kravtsov}, A.~V., \& {Borgani}, S. 2012, \araa, 50, 353,
  \dodoi{10.1146/annurev-astro-081811-125502}

\bibitem[{{Lacey} \& {Cole}(1993)}]{lacey1993}
{Lacey}, C., \& {Cole}, S. 1993, \mnras, 262, 627,
  \dodoi{10.1093/mnras/262.3.627}

\bibitem[{{Lage} \& {Farrar}(2014)}]{Lage2014}
{Lage}, C., \& {Farrar}, G. 2014, \apj, 787, 144,
  \dodoi{10.1088/0004-637X/787/2/144}

\bibitem[{{Landau} \& {Lifshitz}(1959)}]{LandL1959}
{Landau}, L.~D., \& {Lifshitz}, E.~M. 1959, {Fluid mechanics}

\bibitem[{{Li} {et~al.}(2020){Li}, {Zhao}, {Jing}, {Han}, \& {Dong}}]{Li2020}
{Li}, Z.-Z., {Zhao}, D.-H., {Jing}, Y.~P., {Han}, J., \& {Dong}, F.-Y. 2020,
  \apj, 905, 177, \dodoi{10.3847/1538-4357/abc481}

\bibitem[{{Ma} {et~al.}(2009){Ma}, {Ebeling}, \& {Barrett}}]{Ma2009}
{Ma}, C.-J., {Ebeling}, H., \& {Barrett}, E. 2009, \apjl, 693, L56,
  \dodoi{10.1088/0004-637X/693/2/L56}

\bibitem[{{Mantz} {et~al.}(2010){Mantz}, {Allen}, {Ebeling}, {Rapetti}, \&
  {Drlica-Wagner}}]{Mantz2010}
{Mantz}, A., {Allen}, S.~W., {Ebeling}, H., {Rapetti}, D., \& {Drlica-Wagner},
  A. 2010, \mnras, 406, 1773, \dodoi{10.1111/j.1365-2966.2010.16993.x}

\bibitem[{{Markevitch} {et~al.}(2002){Markevitch}, {Gonzalez}, {David},
  {Vikhlinin}, {Murray}, {Forman}, {Jones}, \& {Tucker}}]{Markevitch2002}
{Markevitch}, M., {Gonzalez}, A.~H., {David}, L., {et~al.} 2002, \apjl, 567,
  L27, \dodoi{10.1086/339619}

\bibitem[{{Markevitch} \& {Vikhlinin}(2007)}]{Markevitch2007}
{Markevitch}, M., \& {Vikhlinin}, A. 2007, \physrep, 443, 1,
  \dodoi{10.1016/j.physrep.2007.01.001}

\bibitem[{{Masters} \& {Capak}(2011)}]{specpro}
{Masters}, D., \& {Capak}, P. 2011, \pasp, 123, 638, \dodoi{10.1086/660023}

\bibitem[{{Mastropietro} \& {Burkert}(2008)}]{Mastropietro2008}
{Mastropietro}, C., \& {Burkert}, A. 2008, \mnras, 389, 967,
  \dodoi{10.1111/j.1365-2966.2008.13626.x}

\bibitem[{{Merten} {et~al.}(2011){Merten}, {Coe}, {Dupke}, {Massey}, {Zitrin},
  {Cypriano}, {Okabe}, {Frye}, {Braglia}, {Jim{\'e}nez-Teja}, {Ben{\'\i}tez},
  {Broadhurst}, {Rhodes}, {Meneghetti}, {Moustakas}, {Sodr{\'e}}, {Krick}, \&
  {Bregman}}]{merten2011}
{Merten}, J., {Coe}, D., {Dupke}, R., {et~al.} 2011, \mnras, 417, 333,
  \dodoi{10.1111/j.1365-2966.2011.19266.x}

\bibitem[{{Milosavljevi{\'c}} {et~al.}(2007){Milosavljevi{\'c}}, {Koda},
  {Nagai}, {Nakar}, \& {Shapiro}}]{Milosavljevic2007}
{Milosavljevi{\'c}}, M., {Koda}, J., {Nagai}, D., {Nakar}, E., \& {Shapiro},
  P.~R. 2007, \apjl, 661, L131, \dodoi{10.1086/518960}

\bibitem[{{Molnar}(2016)}]{Molnar2016}
{Molnar}, S. 2016, Frontiers in Astronomy and Space Sciences, 2, 7,
  \dodoi{10.3389/fspas.2015.00007}

\bibitem[{{Molnar} {et~al.}(2012){Molnar}, {Hearn}, \& {Stadel}}]{Molnar2012}
{Molnar}, S.~M., {Hearn}, N.~C., \& {Stadel}, J.~G. 2012, \apj, 748, 45,
  \dodoi{10.1088/0004-637X/748/1/45}

\bibitem[{{Mroczkowski} {et~al.}(2019){Mroczkowski}, {Nagai}, {Basu}, {Chluba},
  {Sayers}, {Adam}, {Churazov}, {Crites}, {Di Mascolo}, {Eckert},
  {Macias-Perez}, {Mayet}, {Perotto}, {Pointecouteau}, {Romero}, {Ruppin},
  {Scannapieco}, \& {ZuHone}}]{Mroczkowski2019}
{Mroczkowski}, T., {Nagai}, D., {Basu}, K., {et~al.} 2019, \ssr, 215, 17,
  \dodoi{10.1007/s11214-019-0581-2}

\bibitem[{{Muldrew} {et~al.}(2015){Muldrew}, {Hatch}, \& {Cooke}}]{muldrew15}
{Muldrew}, S.~I., {Hatch}, N.~A., \& {Cooke}, E.~A. 2015, \mnras, 452, 2528,
  \dodoi{10.1093/mnras/stv1449}

\bibitem[{{Nagai} {et~al.}(2007){Nagai}, {Kravtsov}, \&
  {Vikhlinin}}]{nagai2007}
{Nagai}, D., {Kravtsov}, A.~V., \& {Vikhlinin}, A. 2007, \apj, 668, 1,
  \dodoi{10.1086/521328}

\bibitem[{{Newman} {et~al.}(2013){Newman}, {Cooper}, {Davis}, {Faber}, {Coil},
  {Guhathakurta}, {Koo}, {Phillips}, {Conroy}, {Dutton}, {Finkbeiner}, {Gerke},
  {Rosario}, {Weiner}, {Willmer}, {Yan}, {Harker}, {Kassin}, {Konidaris},
  {Lai}, {Madgwick}, {Noeske}, {Wirth}, {Connolly}, {Kaiser}, {Kirby},
  {Lemaux}, {Lin}, {Lotz}, {Luppino}, {Marinoni}, {Matthews}, {Metevier}, \&
  {Schiavon}}]{spec2d_newman}
{Newman}, J.~A., {Cooper}, M.~C., {Davis}, M., {et~al.} 2013, \apjs, 208, 5,
  \dodoi{10.1088/0067-0049/208/1/5}

\bibitem[{{Newville} {et~al.}(2014){Newville}, {Stensitzki}, {Allen}, \&
  {Ingargiola}}]{lmfit}
{Newville}, M., {Stensitzki}, T., {Allen}, D.~B., \& {Ingargiola}, A. 2014,
  {LMFIT: Non-Linear Least-Square Minimization and Curve-Fitting for Python},
  0.8.0, Zenodo,  Zenodo, \dodoi{10.5281/zenodo.11813}

\bibitem[{{Owers} {et~al.}(2011){Owers}, {Randall}, {Nulsen}, {Couch}, {David},
  \& {Kempner}}]{Owers2011}
{Owers}, M.~S., {Randall}, S.~W., {Nulsen}, P. E.~J., {et~al.} 2011, \apj, 728,
  27, \dodoi{10.1088/0004-637X/728/1/27}

\bibitem[{{Piffaretti} {et~al.}(2003){Piffaretti}, {Jetzer}, \&
  {Schindler}}]{Piffaretti2003}
{Piffaretti}, R., {Jetzer}, P., \& {Schindler}, S. 2003, \aap, 398, 41,
  \dodoi{10.1051/0004-6361:20021648}

\bibitem[{{Planck Collaboration} {et~al.}(2020){Planck Collaboration},
  {Aghanim}, {Akrami}, {Ashdown}, {Aumont}, {Baccigalupi}, {Ballardini},
  {Banday}, {Barreiro}, {Bartolo}, {Basak}, {Battye}, {Benabed}, {Bernard},
  {Bersanelli}, {Bielewicz}, {Bock}, {Bond}, {Borrill}, {Bouchet}, {Boulanger},
  {Bucher}, {Burigana}, {Butler}, {Calabrese}, {Cardoso}, {Carron},
  {Challinor}, {Chiang}, {Chluba}, {Colombo}, {Combet}, {Contreras}, {Crill},
  {Cuttaia}, {de Bernardis}, {de Zotti}, {Delabrouille}, {Delouis}, {Di
  Valentino}, {Diego}, {Dor{\'e}}, {Douspis}, {Ducout}, {Dupac}, {Dusini},
  {Efstathiou}, {Elsner}, {En{\ss}lin}, {Eriksen}, {Fantaye}, {Farhang},
  {Fergusson}, {Fernandez-Cobos}, {Finelli}, {Forastieri}, {Frailis},
  {Fraisse}, {Franceschi}, {Frolov}, {Galeotta}, {Galli}, {Ganga},
  {G{\'e}nova-Santos}, {Gerbino}, {Ghosh}, {Gonz{\'a}lez-Nuevo}, {G{\'o}rski},
  {Gratton}, {Gruppuso}, {Gudmundsson}, {Hamann}, {Handley}, {Hansen},
  {Herranz}, {Hildebrandt}, {Hivon}, {Huang}, {Jaffe}, {Jones}, {Karakci},
  {Keih{\"a}nen}, {Keskitalo}, {Kiiveri}, {Kim}, {Kisner}, {Knox},
  {Krachmalnicoff}, {Kunz}, {Kurki-Suonio}, {Lagache}, {Lamarre}, {Lasenby},
  {Lattanzi}, {Lawrence}, {Le Jeune}, {Lemos}, {Lesgourgues}, {Levrier},
  {Lewis}, {Liguori}, {Lilje}, {Lilley}, {Lindholm}, {L{\'o}pez-Caniego},
  {Lubin}, {Ma}, {Mac{\'\i}as-P{\'e}rez}, {Maggio}, {Maino}, {Mandolesi},
  {Mangilli}, {Marcos-Caballero}, {Maris}, {Martin}, {Martinelli},
  {Mart{\'\i}nez-Gonz{\'a}lez}, {Matarrese}, {Mauri}, {McEwen}, {Meinhold},
  {Melchiorri}, {Mennella}, {Migliaccio}, {Millea}, {Mitra},
  {Miville-Desch{\^e}nes}, {Molinari}, {Montier}, {Morgante}, {Moss}, {Natoli},
  {N{\o}rgaard-Nielsen}, {Pagano}, {Paoletti}, {Partridge}, {Patanchon},
  {Peiris}, {Perrotta}, {Pettorino}, {Piacentini}, {Polastri}, {Polenta},
  {Puget}, {Rachen}, {Reinecke}, {Remazeilles}, {Renzi}, {Rocha}, {Rosset},
  {Roudier}, {Rubi{\~n}o-Mart{\'\i}n}, {Ruiz-Granados}, {Salvati}, {Sandri},
  {Savelainen}, {Scott}, {Shellard}, {Sirignano}, {Sirri}, {Spencer},
  {Sunyaev}, {Suur-Uski}, {Tauber}, {Tavagnacco}, {Tenti}, {Toffolatti},
  {Tomasi}, {Trombetti}, {Valenziano}, {Valiviita}, {Van Tent}, {Vibert},
  {Vielva}, {Villa}, {Vittorio}, {Wandelt}, {Wehus}, {White}, {White},
  {Zacchei}, \& {Zonca}}]{planck18}
{Planck Collaboration}, {Aghanim}, N., {Akrami}, Y., {et~al.} 2020, \aap, 641,
  A6, \dodoi{10.1051/0004-6361/201833910}

\bibitem[{{Poole} {et~al.}(2006){Poole}, {Fardal}, {Babul}, {McCarthy},
  {Quinn}, \& {Wadsley}}]{Poole2006}
{Poole}, G.~B., {Fardal}, M.~A., {Babul}, A., {et~al.} 2006, \mnras, 373, 881,
  \dodoi{10.1111/j.1365-2966.2006.10916.x}

\bibitem[{{Prochaska} {et~al.}(2020){Prochaska}, {Hennawi}, {Westfall},
  {Cooke}, {Wang}, {Hsyu}, {Davies}, {Farina}, \& {Pelliccia}}]{pypeit}
{Prochaska}, J., {Hennawi}, J., {Westfall}, K., {et~al.} 2020, The Journal of
  Open Source Software, 5, 2308, \dodoi{10.21105/joss.02308}

\bibitem[{{Rhea} {et~al.}(2020){Rhea}, {Hlavacek-Larrondo},
  {Perreault-Levasseur}, {Gendron-Marsolais}, \& {Kraft}}]{rhea2020}
{Rhea}, C., {Hlavacek-Larrondo}, J., {Perreault-Levasseur}, L.,
  {Gendron-Marsolais}, M.-L., \& {Kraft}, R. 2020, \aj, 160, 202,
  \dodoi{10.3847/1538-3881/abb468}

\bibitem[{{Ricker} \& {Sarazin}(2001)}]{Ricker2001}
{Ricker}, P.~M., \& {Sarazin}, C.~L. 2001, \apj, 561, 621,
  \dodoi{10.1086/323365}

\bibitem[{Russell {et~al.}(2012)Russell, McNamara, Sanders, Fabian, Nulsen,
  Canning, Baum, Donahue, Edge, King, \& O’Dea}]{Russell12}
Russell, H.~R., McNamara, B.~R., Sanders, J.~S., {et~al.} 2012, Monthly Notices
  of the Royal Astronomical Society, 423, 236,
  \dodoi{10.1111/j.1365-2966.2012.20808.x}

\bibitem[{{Russell} {et~al.}(2022){Russell}, {Nulsen}, {Caprioli},
  {Chadayammuri}, {Fabian}, {Kunz}, {McNamara}, {Sanders},
  {Richard-Laferri{\`e}re}, {Beleznay}, {Canning}, {Hlavacek-Larrondo}, \&
  {King}}]{russell22}
{Russell}, H.~R., {Nulsen}, P.~E.~J., {Caprioli}, D., {et~al.} 2022, \mnras,
  514, 1477, \dodoi{10.1093/mnras/stac1055}

\bibitem[{{Sanders}(2006)}]{Sanders2006}
{Sanders}, J.~S. 2006, \mnras, 371, 829,
  \dodoi{10.1111/j.1365-2966.2006.10716.x}

\bibitem[{{Sanders} {et~al.}(2016{\natexlab{a}}){Sanders}, {Fabian}, {Russell},
  {Walker}, \& {Blundell}}]{ggm2}
{Sanders}, J.~S., {Fabian}, A.~C., {Russell}, H.~R., {Walker}, S.~A., \&
  {Blundell}, K.~M. 2016{\natexlab{a}}, \mnras, 460, 1898,
  \dodoi{10.1093/mnras/stw1119}

\bibitem[{{Sanders} {et~al.}(2016{\natexlab{b}}){Sanders}, {Fabian}, {Taylor},
  {Russell}, {Blundell}, {Canning}, {Hlavacek-Larrondo}, {Walker}, \&
  {Grimes}}]{sanders2016}
{Sanders}, J.~S., {Fabian}, A.~C., {Taylor}, G.~B., {et~al.}
  2016{\natexlab{b}}, \mnras, 457, 82, \dodoi{10.1093/mnras/stv2972}

\bibitem[{{Sayers} {et~al.}(2013){Sayers}, {Mroczkowski}, {Zemcov}, {Korngut},
  {Bock}, {Bulbul}, {Czakon}, {Egami}, {Golwala}, {Koch}, {Lin}, {Mantz},
  {Molnar}, {Moustakas}, {Pierpaoli}, {Rawle}, {Reese}, {Rex}, {Shitanishi},
  {Siegel}, \& {Umetsu}}]{Sayers2013}
{Sayers}, J., {Mroczkowski}, T., {Zemcov}, M., {et~al.} 2013, \apj, 778, 52,
  \dodoi{10.1088/0004-637X/778/1/52}

\bibitem[{{Sayers} {et~al.}(2019){Sayers}, {Monta{\~n}a}, {Mroczkowski},
  {Wilson}, {Zemcov}, {Zitrin}, {Cibirka}, {Golwala}, {Hughes}, {Nagai},
  {Reese}, {S{\'a}nchez}, \& {Zuhone}}]{sayers2019}
{Sayers}, J., {Monta{\~n}a}, A., {Mroczkowski}, T., {et~al.} 2019, \apj, 880,
  45, \dodoi{10.3847/1538-4357/ab29ef}

\bibitem[{{Schive} {et~al.}(2018){Schive}, {ZuHone}, {Goldbaum}, {Turk},
  {Gaspari}, \& {Cheng}}]{Schive2018}
{Schive}, H.-Y., {ZuHone}, J.~A., {Goldbaum}, N.~J., {et~al.} 2018, \mnras,
  481, 4815, \dodoi{10.1093/mnras/sty2586}

\bibitem[{{Smith} {et~al.}(2001){Smith}, {Brickhouse}, {Liedahl}, \&
  {Raymond}}]{Smith2001}
{Smith}, R.~K., {Brickhouse}, N.~S., {Liedahl}, D.~A., \& {Raymond}, J.~C.
  2001, \apjl, 556, L91, \dodoi{10.1086/322992}

\bibitem[{{Solovyeva} {et~al.}(2007){Solovyeva}, {Anokhin}, {Sauvageot},
  {Teyssier}, \& {Neumann}}]{Solovyeva2007}
{Solovyeva}, L., {Anokhin}, S., {Sauvageot}, J.~L., {Teyssier}, R., \&
  {Neumann}, D. 2007, \aap, 476, 63, \dodoi{10.1051/0004-6361:20077966}

\bibitem[{{Springel} \& {Farrar}(2007)}]{Springel2007}
{Springel}, V., \& {Farrar}, G.~R. 2007, \mnras, 380, 911,
  \dodoi{10.1111/j.1365-2966.2007.12159.x}

\bibitem[{{Springel} {et~al.}(2005){Springel}, {White}, {Jenkins}, {Frenk},
  {Yoshida}, {Gao}, {Navarro}, {Thacker}, {Croton}, {Helly}, {Peacock}, {Cole},
  {Thomas}, {Couchman}, {Evrard}, {Colberg}, \& {Pearce}}]{millenium}
{Springel}, V., {White}, S. D.~M., {Jenkins}, A., {et~al.} 2005, \nat, 435,
  629, \dodoi{10.1038/nature03597}

\bibitem[{{Sunyaev} \& {Zeldovich}(1980)}]{Sunyaev1980}
{Sunyaev}, R.~A., \& {Zeldovich}, I.~B. 1980, \mnras, 190, 413

\bibitem[{{Thompson} {et~al.}(2015){Thompson}, {Dav{\'e}}, \&
  {Nagamine}}]{thompson2015}
{Thompson}, R., {Dav{\'e}}, R., \& {Nagamine}, K. 2015, \mnras, 452, 3030,
  \dodoi{10.1093/mnras/stv1433}

\bibitem[{{Thompson} \& {Nagamine}(2012)}]{Thompson2012}
{Thompson}, R., \& {Nagamine}, K. 2012, \mnras, 419, 3560,
  \dodoi{10.1111/j.1365-2966.2011.20000.x}

\bibitem[{{Turk} {et~al.}(2011){Turk}, {Smith}, {Oishi}, {Skory}, {Skillman},
  {Abel}, \& {Norman}}]{Turk2011}
{Turk}, M.~J., {Smith}, B.~D., {Oishi}, J.~S., {et~al.} 2011, \apjs, 192, 9,
  \dodoi{10.1088/0067-0049/192/1/9}

\bibitem[{{van der Walt} {et~al.}(2011){van der Walt}, {Colbert}, \&
  {Varoquaux}}]{vanderwalt2011}
{van der Walt}, S., {Colbert}, S.~C., \& {Varoquaux}, G. 2011, Computing in
  Science and Engineering, 13, 22, \dodoi{10.1109/MCSE.2011.37}

\bibitem[{van~der Walt {et~al.}(2014)van~der Walt, {S}ch\"onberger,
  {Nunez-Iglesias}, {B}oulogne, {W}arner, {Y}ager, {G}ouillart, {Y}u, \& the
  scikit-image contributors}]{scikit-image}
van~der Walt, S., {S}ch\"onberger, J.~L., {Nunez-Iglesias}, J., {et~al.} 2014,
  PeerJ, 2, e453, \dodoi{10.7717/peerj.453}

\bibitem[{{van Weeren} {et~al.}(2010){van Weeren}, {R{\"o}ttgering},
  {Br{\"u}ggen}, \& {Hoeft}}]{vanWeeren2010}
{van Weeren}, R.~J., {R{\"o}ttgering}, H. J.~A., {Br{\"u}ggen}, M., \& {Hoeft},
  M. 2010, Science, 330, 347, \dodoi{10.1126/science.1194293}

\bibitem[{{Vikhlinin} {et~al.}(2006){Vikhlinin}, {Kravtsov}, {Forman}, {Jones},
  {Markevitch}, {Murray}, \& {Van Speybroeck}}]{gasprof}
{Vikhlinin}, A., {Kravtsov}, A., {Forman}, W., {et~al.} 2006, \apj, 640, 691,
  \dodoi{10.1086/500288}

\bibitem[{{Voit}(2005)}]{voitrev}
{Voit}, G.~M. 2005, Reviews of Modern Physics, 77, 207,
  \dodoi{10.1103/RevModPhys.77.207}

\bibitem[{{Wilms} {et~al.}(2000){Wilms}, {Allen}, \& {McCray}}]{tbabs}
{Wilms}, J., {Allen}, A., \& {McCray}, R. 2000, \apj, 542, 914,
  \dodoi{10.1086/317016}

\bibitem[{{XRISM Science Team}(2020)}]{xrism}
{XRISM Science Team}. 2020, arXiv e-prints, arXiv:2003.04962,
  \dodoi{10.48550/arXiv.2003.04962}

\bibitem[{{Zitrin} {et~al.}(2011){Zitrin}, {Broadhurst}, {Barkana}, {Rephaeli},
  \& {Ben{\'\i}tez}}]{zitrin2011}
{Zitrin}, A., {Broadhurst}, T., {Barkana}, R., {Rephaeli}, Y., \&
  {Ben{\'\i}tez}, N. 2011, \mnras, 410, 1939,
  \dodoi{10.1111/j.1365-2966.2010.17574.x}

\bibitem[{{Zitrin} {et~al.}(2015){Zitrin}, {Fabris}, {Merten}, {Melchior},
  {Meneghetti}, {Koekemoer}, {Coe}, {Maturi}, {Bartelmann}, {Postman},
  {Umetsu}, {Seidel}, {Sendra}, {Broadhurst}, {Balestra}, {Biviano}, {Grillo},
  {Mercurio}, {Nonino}, {Rosati}, {Bradley}, {Carrasco}, {Donahue}, {Ford},
  {Frye}, \& {Moustakas}}]{zitrin2015}
{Zitrin}, A., {Fabris}, A., {Merten}, J., {et~al.} 2015, \apj, 801, 44,
  \dodoi{10.1088/0004-637X/801/1/44}

\bibitem[{{ZuHone}(2011)}]{ZuHone2011}
{ZuHone}, J.~A. 2011, \apj, 728, 54, \dodoi{10.1088/0004-637X/728/1/54}

\bibitem[{{ZuHone} \& {Hallman}(2016)}]{zuhone2016}
{ZuHone}, J.~A., \& {Hallman}, E.~J. 2016, {pyXSIM: Synthetic X-ray
  observations generator}.
\newblock \doeprint{1608.002}

\bibitem[{{ZuHone} {et~al.}(2018){ZuHone}, {Kowalik}, {{\"O}hman}, {Lau}, \&
  {Nagai}}]{ZuHone2018}
{ZuHone}, J.~A., {Kowalik}, K., {{\"O}hman}, E., {Lau}, E., \& {Nagai}, D.
  2018, \apjs, 234, 4, \dodoi{10.3847/1538-4365/aa99db}

\bibitem[{{ZuHone} {et~al.}(2023){ZuHone}, {Vikhlinin}, {Tremblay}, {Randall},
  {Andrade-Santos}, \& {Bourdin}}]{soxs}
{ZuHone}, J.~A., {Vikhlinin}, A., {Tremblay}, G.~R., {et~al.} 2023, {SOXS:
  Simulated Observations of X-ray Sources}, Astrophysics Source Code Library,
  record ascl:2301.024.
\newblock \doeprint{2301.024}

\end{thebibliography}

\appendix 
\restartappendixnumbering
\section{Merger progression movie} \label{sec:appA}
\begin{figure}[h]
\begin{interactive}{animation}{run6_EB_R1.5_v1695_b250_L_0.92_0.30_-0.25_rot0_trim}
        \includegraphics[width=0.8\textwidth]{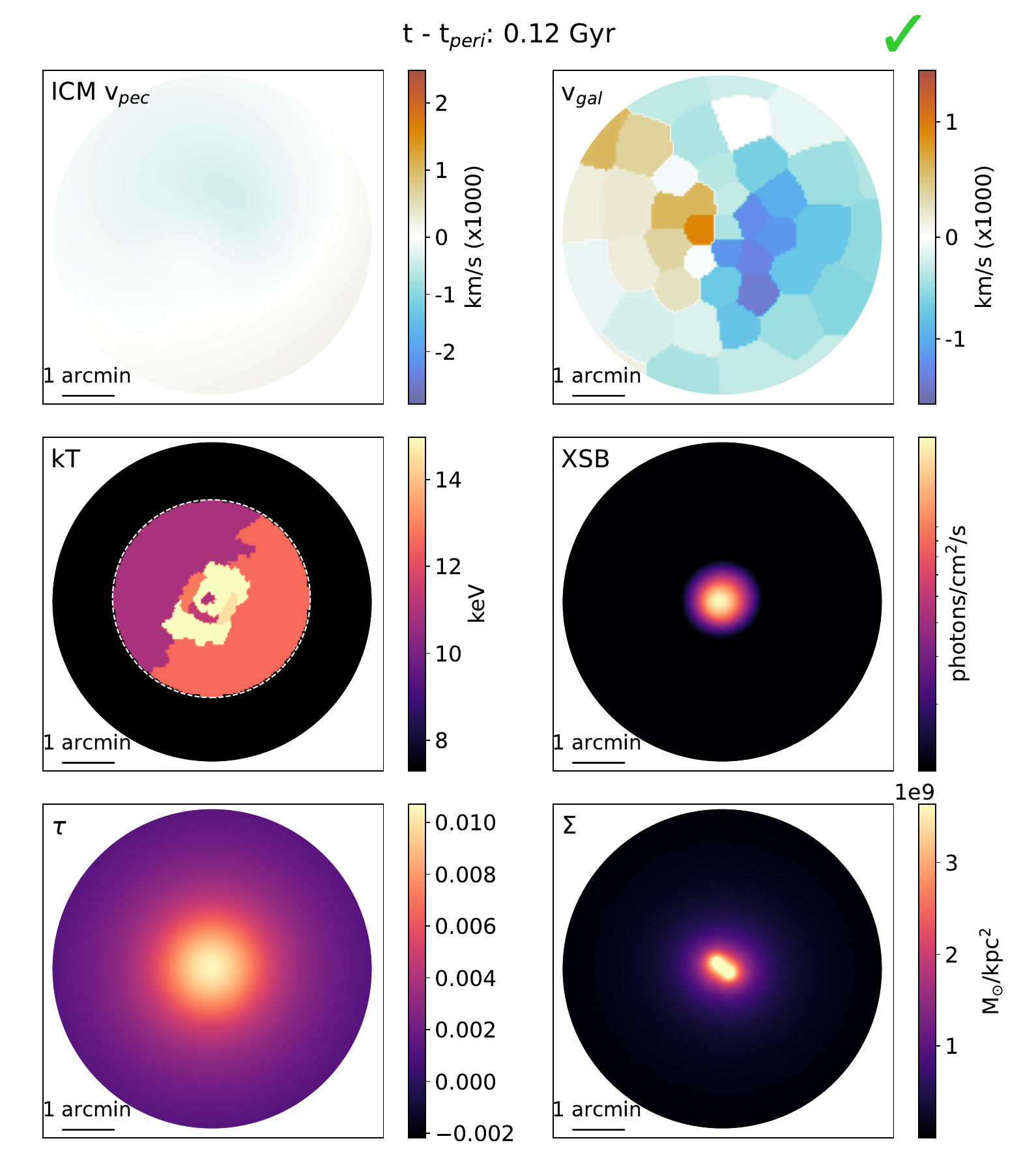}
\end{interactive}
\caption{Simulated cluster merger progression as a function of epoch for a simulation initialized with $R = 1.5$, $b = 250$ kpc, $v_i = 1700$ km s$^{-1}$, gas profile: Int$+$NCC, and $|{\bf L} \cdot \; \hat x| = 0.92$. The red `X' or green check-mark in the top right corner indicates whether or not a given snapshot passes the matching criteria. Those which pass the matching criteria correspond to epochs near the pericenter passage, when the mock observables best resemble the \macs observational data.} \label{fig:movie}
\end{figure}

\section{Velocity decoupling movies} \label{sec:appB}
\begin{figure}[h]
\begin{interactive}{animation}{mocks_short_250}
        \includegraphics[width=0.96\textwidth]{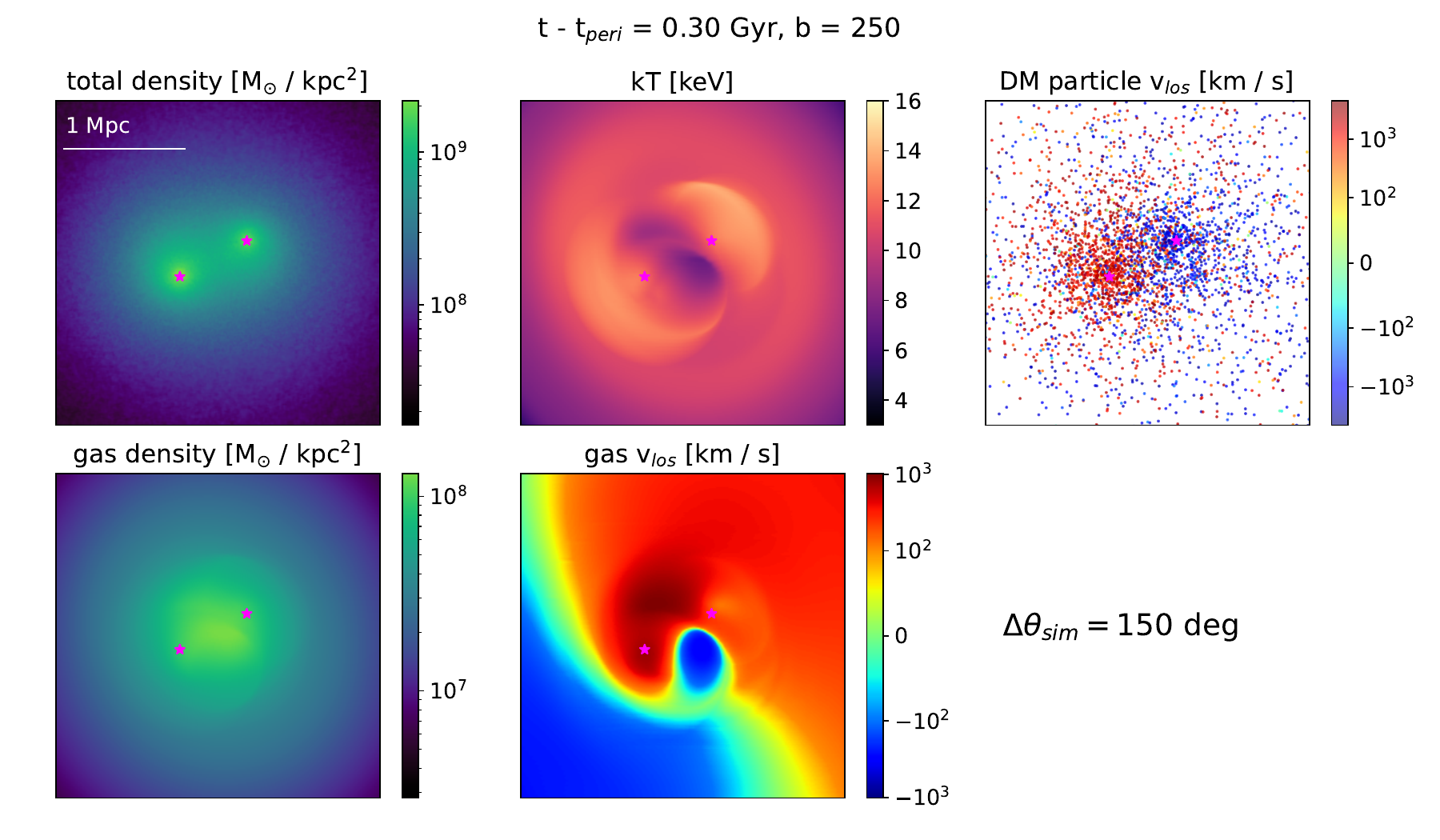}
\end{interactive}
\caption{Simulated evolution of the ICM/DM velocity dipole decoupling, as well as the total/gas densities and gas temperature, as a function of epoch for simulations initialized with $b = 250$, $R = 1.5$, $v_i = 3000$ km s$^{-1}$, and gas profile: Int$+$NCC viewed at $|{\bf L} \cdot \; \hat x| = 0.88$. Each mock map is shown at the full simulation resolution for clarity. Prior to the pericenter passage, the velocity dipoles are aligned ($\Delta \theta_{v, \text{sim}} \approx 0$). As the merger progresses, the DM velocity dipole in the inner regions rotates more quickly than the ICM velocity dipole, which lags the DM velocity dipole. The gas velocity dipole catches up to the DM velocity dipole within several $0.1$ Gyr post-pericenter passage, and several additional $0.1$ Gyr are required for the rotational gas motion to propagate outwards to large radii, so that the global gas velocity dipole again becomes roughly aligned with the DM velocity dipole by $t - t_{\text{peri}} \approx 0.80$ Gyr.} \label{fig:movie3}
\end{figure}

\begin{figure}[h]
\begin{interactive}{animation}{mocks_short_500}
        \includegraphics[width=0.96\textwidth]{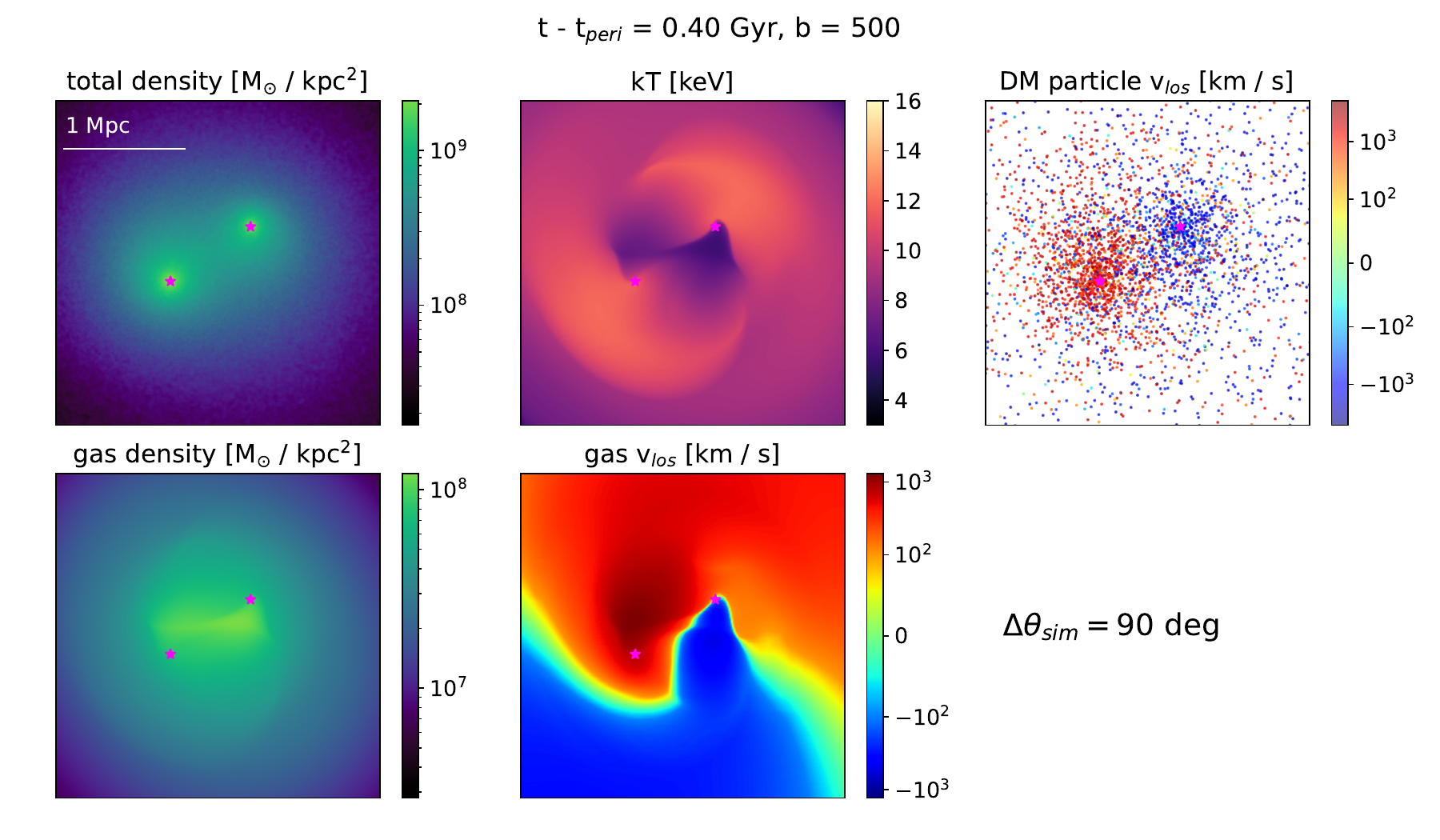}
\end{interactive}
\caption{Simulated evolution of the ICM/DM velocity dipole decoupling, as well as the total/gas densities and gas temperature, as a function of epoch for simulations initialized with $b = 500$, $R = 1.5$, $v_i = 3000$ km s$^{-1}$, and gas profile: Int$+$NCC viewed at $|{\bf L} \cdot \; \hat x| = 0.88$. Each mock map is shown at the full simulation resolution for clarity. Prior to the pericenter passage, the velocity dipoles are aligned ($\Delta \theta_{v, \text{sim}} \approx 0$). As the merger progresses, the DM velocity dipole in the inner regions rotates more quickly than the ICM velocity dipole, which lags the DM velocity dipole. The gas velocity dipole catches up to the DM velocity dipole within several $0.1$ Gyr post-pericenter passage, and as the rotational gas motions propagate outwards to large radii, there is a semi-stable rotational offset at $\Delta \theta_{v, \text{sim}} \approx 90$ degrees for several $0.1$ Gyr post-pericenter passage before reaching an offset of $\approx 20$ degrees by $t - t_{\text{peri}} \approx 0.80$ Gyr.} \label{fig:movie2}
\end{figure}

\clearpage

\section{Spectroscopic redshifts} \label{sec:appB} \vspace{-1em}
\startlongtable
\begin{deluxetable*}{llccc} \label{tab:lit_specz}
\tablehead{\colhead{RA} & \colhead{Dec} & \colhead{$z$} & \colhead{Reference} & \colhead{Duplicate}}
\tablecaption{Literature cluster-member spectroscopic redshifts for \macsper Objects which are observed as duplicates, and therefore excluded from our analysis, are indicated. Only objects within a $6.7' \times 6.7'$ region centered on the cluster center are used in the $v_{\text{gal}}$ map generation.}
\startdata
00:18:01.63 & +16:27:01.02 & 0.5515 & \citep{obsz_crawford} &  \\
00:18:03.83 & +16:24:16.13 & 0.5410 & \citep{obsz_crawford} &  \\
00:18:05.98 & +16:27:26.28 & 0.5536 & \citep{obsz_crawford} &  \\
00:18:06.66 & +16:26:56.73 & 0.5658 & \citep{obsz_crawford} &  \\
00:18:06.84 & +16:24:27.81 & 0.5500 & \citep{obsz_crawford} &  \\
00:18:08.46 & +16:26:12.34 & 0.5448 & \citep{obsz_crawford} &  \\
00:18:08.58 & +16:24:30.19 & 0.5654 & \citep{obsz_crawford} &  \\
00:18:10.00 & +16:24:59.25 & 0.5408 & \citep{obsz_crawford} &  \\
00:18:13.32 & +16:24:40.80 & 0.5450 & \citep{obsz_crawford} &  \\
00:18:13.61 & +16:24:55.09 & 0.5454 & \citep{obsz_crawford} &  \\
00:18:14.69 & +16:24:26.29 & 0.5491 & \citep{obsz_crawford} & * \\
00:18:14.93 & +16:25:46.28 & 0.5355 & \citep{obsz_crawford} &  \\
00:18:16.44 & +16:24:33.44 & 0.5447 & \citep{obsz_ellingson} &  \\
00:18:16.48 & +16:25:47.25 & 0.5467 & \citep{obsz_crawford} &  \\
00:18:16.75 & +16:24:50.76 & 0.5468 & \citep{obsz_ellingson} &  \\
00:18:17.00 & +16:24:13.35 & 0.5462 & \citep{obsz_crawford} &  \\
00:18:17.33 & +16:24:56.05 & 0.5317 & \citep{obsz_ellingson} &  \\
00:18:17.56 & +16:27:01.93 & 0.5449 & \citep{obsz_crawford} &  \\
00:18:17.57 & +16:27:50.65 & 0.5472 & \citep{obsz_ellingson} &  \\
00:18:17.85 & +16:23:22.16 & 0.5515 & \citep{obsz_ellingson} &  \\
00:18:18.22 & +16:24:55.98 & 0.5476 & \citep{obsz_ellingson} &  \\
00:18:18.36 & +16:22:23.05 & 0.5410 & \citep{obsz_ellingson} &  \\
00:18:18.49 & +16:24:01.24 & 0.5331 & \citep{obsz_crawford} &  \\
00:18:18.50 & +16:29:13.56 & 0.5448 & \citep{obsz_ellingson} &  \\
00:18:18.87 & +16:27:06.97 & 0.5656 & \citep{obsz_crawford} &  \\
00:18:19.21 & +16:24:15.66 & 0.5463 & \citep{obsz_crawford} &  \\
00:18:19.24 & +16:27:04.75 & 0.5647 & \citep{obsz_ellingson} &  \\
00:18:20.09 & +16:23:18.90 & 0.5728 & \citep{obsz_crawford} &  \\
00:18:21.40 & +16:22:01.88 & 0.5500 & \citep{obsz_ellingson} &  \\
00:18:21.79 & +16:22:02.78 & 0.5383 & \citep{obsz_ellingson} &  \\
00:18:22.44 & +16:23:30.59 & 0.5464 & \citep{obsz_ellingson} &  \\
00:18:22.44 & +16:25:31.65 & 0.5442 & \citep{obsz_crawford} &  \\
00:18:22.71 & +16:25:50.99 & 0.5450 & \citep{obsz_ellingson} &  \\
00:18:23.02 & +16:26:04.56 & 0.5500 & \citep{obsz_crawford} &  \\
00:18:23.24 & +16:23:50.40 & 0.5544 & \citep{obsz_crawford} &  \\
00:18:23.58 & +16:26:27.67 & 0.5522 & \citep{obsz_ellingson} &  \\
00:18:23.64 & +16:26:27.65 & 0.5534 & \citep{obsz_crawford} & * \\
00:18:23.68 & +16:25:02.64 & 0.5516 & \citep{obsz_crawford} &  \\
00:18:23.97 & +16:27:46.80 & 0.5501 & \citep{obsz_ellingson} & * \\
00:18:24.04 & +16:22:00.48 & 0.5465 & \citep{obsz_ellingson} & * \\
00:18:24.09 & +16:26:16.82 & 0.5452 & \citep{obsz_crawford} &  \\
00:18:24.22 & +16:25:13.20 & 0.5526 & \citep{obsz_crawford} &  \\
00:18:24.40 & +16:24:00.59 & 0.5511 & \citep{obsz_crawford} &  \\
00:18:25.08 & +16:24:12.10 & 0.5339 & \citep{obsz_ellingson} &  \\
00:18:25.13 & +16:22:07.90 & 0.5488 & \citep{obsz_ellingson} & * \\
00:18:26.22 & +16:24:57.79 & 0.5379 & \citep{obsz_crawford} &  \\
00:18:26.23 & +16:25:07.44 & 0.5539 & \citep{obsz_crawford} &  \\
00:18:26.57 & +16:22:18.19 & 0.5514 & \citep{obsz_ellingson} & * \\
00:18:26.65 & +16:24:50.62 & 0.5470 & \citep{obsz_crawford} &  \\
00:18:26.81 & +16:25:45.00 & 0.5635 & \citep{obsz_crawford} &  \\
00:18:26.97 & +16:22:59.81 & 0.5361 & \citep{obsz_ellingson} &  \\
00:18:27.09 & +16:25:13.89 & 0.5389 & \citep{obsz_crawford} &  \\
00:18:27.79 & +16:23:09.38 & 0.5319 & \citep{obsz_ellingson} &  \\
00:18:27.84 & +16:26:05.24 & 0.5425 & \citep{obsz_crawford} &  \\
00:18:28.08 & +16:24:31.34 & 0.5352 & \citep{obsz_crawford} &  \\
00:18:28.43 & +16:26:15.76 & 0.5442 & \citep{obsz_crawford} &  \\
00:18:29.07 & +16:23:08.81 & 0.5551 & \citep{obsz_ellingson} & * \\
00:18:29.15 & +16:26:53.46 & 0.5378 & \citep{obsz_crawford} &  \\
00:18:29.25 & +16:23:12.41 & 0.5500 & \citep{obsz_ellingson} &  \\
00:18:29.44 & +16:25:08.33 & 0.5487 & \citep{obsz_dressler} &  \\
00:18:29.45 & +16:27:03.20 & 0.5538 & \citep{obsz_ellingson} &  \\
00:18:29.49 & +16:26:56.80 & 0.5373 & \citep{obsz_dressler} &  \\
00:18:29.91 & +16:26:09.20 & 0.5581 & \citep{obsz_ellingson} &  \\
00:18:29.96 & +16:26:08.43 & 0.5592 & \citep{obsz_crawford} &  \\
00:18:30.05 & +16:26:08.20 & 0.5448 & \citep{obsz_ellingson} & * \\
00:18:30.14 & +16:25:51.38 & 0.5463 & \citep{obsz_crawford} &  \\
00:18:30.36 & +16:22:04.12 & 0.5567 & \citep{obsz_ellingson} &  \\
00:18:30.42 & +16:26:44.05 & 0.5480 & \citep{obsz_dressler} &  \\
00:18:30.67 & +16:25:37.86 & 0.5630 & \citep{obsz_crawford} &  \\
00:18:30.98 & +16:25:40.84 & 0.5561 & \citep{obsz_dressler} &  \\
00:18:31.12 & +16:26:35.70 & 0.5429 & \citep{obsz_ellingson} &  \\
00:18:31.14 & +16:22:05.02 & 0.5509 & \citep{obsz_ellingson} &  \\
00:18:31.14 & +16:26:20.72 & 0.5499 & \citep{obsz_dressler} &  \\
00:18:31.26 & +16:25:43.54 & 0.5523 & \citep{obsz_dressler} &  \\
00:18:31.31 & +16:22:46.88 & 0.5456 & \citep{obsz_ellingson} &  \\
00:18:31.56 & +16:26:55.57 & 0.5304 & \citep{obsz_dressler} &  \\
00:18:31.78 & +16:25:41.59 & 0.5415 & \citep{obsz_ellingson} &  \\
00:18:31.90 & +16:24:41.11 & 0.5464 & \citep{obsz_crawford} &  \\
00:18:32.21 & +16:24:43.60 & 0.5485 & \citep{obsz_dressler} &  \\
00:18:32.23 & +16:28:52.21 & 0.5464 & \citep{obsz_ellingson} &  \\
00:18:32.24 & +16:25:06.72 & 0.5490 & \citep{obsz_crawford} &  \\
00:18:32.45 & +16:24:35.89 & 0.5499 & \citep{obsz_ellingson} &  \\
00:18:32.46 & +16:24:34.61 & 0.5467 & \citep{obsz_crawford} & * \\
00:18:32.50 & +16:25:18.95 & 0.5498 & \citep{obsz_dressler} &  \\
00:18:32.53 & +16:25:09.37 & 0.5447 & \citep{obsz_dressler} &  \\
00:18:32.66 & +16:25:06.28 & 0.5507 & \citep{obsz_dressler} &  \\
00:18:32.80 & +16:25:50.41 & 0.5504 & \citep{obsz_ellingson} &  \\
00:18:32.82 & +16:26:03.42 & 0.5468 & \citep{obsz_crawford} &  \\
00:18:32.83 & +16:27:41.62 & 0.5511 & \citep{obsz_ellingson} &  \\
00:18:32.86 & +16:26:03.05 & 0.5382 & \citep{obsz_dressler} &  \\
00:18:32.93 & +16:27:21.28 & 0.5300 & \citep{obsz_dressler} &  \\
00:18:33.00 & +16:27:47.30 & 0.5429 & \citep{obsz_dressler} &  \\
00:18:33.01 & +16:24:30.58 & 0.5447 & \citep{obsz_crawford} & * \\
00:18:33.52 & +16:26:17.02 & 0.5383 & \citep{obsz_ellingson} &  \\
00:18:33.61 & +16:26:15.25 & 0.5433 & \citep{obsz_dressler} &  \\
00:18:33.83 & +16:26:20.29 & 0.5471 & \citep{obsz_ellingson} &  \\
00:18:33.93 & +16:26:18.56 & 0.5324 & \citep{obsz_dressler} &  \\
00:18:34.23 & +16:26:21.88 & 0.5405 & \citep{obsz_dressler} &  \\
00:18:34.40 & +16:26:41.92 & 0.5442 & \citep{obsz_crawford} &  \\
00:18:34.71 & +16:25:46.35 & 0.5426 & \citep{obsz_crawford} &  \\
00:18:34.81 & +16:27:19.91 & 0.5392 & \citep{obsz_crawford} &  \\
00:18:35.15 & +16:25:36.98 & 0.5519 & \citep{obsz_ellingson} &  \\
00:18:35.15 & +16:27:22.07 & 0.5391 & \citep{obsz_dressler} &  \\
00:18:35.20 & +16:25:46.49 & 0.5523 & \citep{obsz_ellingson} &  \\
00:18:35.31 & +16:22:32.48 & 0.5573 & \citep{obsz_ellingson} &  \\
00:18:35.40 & +16:26:34.76 & 0.5528 & \citep{obsz_dressler} &  \\
00:18:35.53 & +16:25:08.29 & 0.5532 & \citep{obsz_ellingson} &  \\
00:18:35.92 & +16:25:59.20 & 0.5545 & \citep{obsz_ellingson} &  \\
00:18:35.99 & +16:26:51.25 & 0.5601 & \citep{obsz_dressler} &  \\
00:18:36.00 & +16:25:57.78 & 0.5540 & \citep{obsz_crawford} & * \\
00:18:36.05 & +16:22:38.39 & 0.5455 & \citep{obsz_ellingson} &  \\
00:18:36.05 & +16:29:06.00 & 0.5509 & \citep{obsz_ellingson} &  \\
00:18:36.15 & +16:26:35.50 & 0.5469 & \citep{obsz_crawford} &  \\
00:18:36.23 & +16:25:31.55 & 0.5549 & \citep{obsz_dressler} &  \\
00:18:36.24 & +16:25:06.42 & 0.5304 & \citep{obsz_dressler} &  \\
00:18:36.31 & +16:27:07.88 & 0.5531 & \citep{obsz_ellingson} &  \\
00:18:36.54 & +16:25:15.11 & 0.5532 & \citep{obsz_crawford} &  \\
00:18:36.55 & +16:22:19.09 & 0.5371 & \citep{obsz_ellingson} &  \\
00:18:36.62 & +16:27:36.40 & 0.5471 & \citep{obsz_crawford} &  \\
00:18:36.69 & +16:26:45.63 & 0.5571 & \citep{obsz_crawford} &  \\
00:18:36.85 & +16:25:17.11 & 0.5555 & \citep{obsz_dressler} &  \\
00:18:36.99 & +16:27:38.38 & 0.5477 & \citep{obsz_dressler} &  \\
00:18:37.23 & +16:24:41.51 & 0.5502 & \citep{obsz_ellingson} & * \\
00:18:38.40 & +16:26:19.10 & 0.5348 & \citep{obsz_crawford} &  \\
00:18:38.45 & +16:25:15.31 & 0.5363 & \citep{obsz_ellingson} &  \\
00:18:38.48 & +16:27:02.70 & 0.5642 & \citep{obsz_ellingson} &  \\
00:18:38.51 & +16:25:13.73 & 0.5371 & \citep{obsz_crawford} & * \\
00:18:38.92 & +16:27:02.95 & 0.5595 & \citep{obsz_dressler} &  \\
00:18:39.00 & +16:27:23.37 & 0.5339 & \citep{obsz_crawford} &  \\
00:18:39.13 & +16:21:50.00 & 0.5653 & \citep{obsz_ellingson} &  \\
00:18:39.47 & +16:24:14.23 & 0.5559 & \citep{obsz_crawford} &  \\
00:18:39.67 & +16:28:53.40 & 0.5485 & \citep{obsz_ellingson} &  \\
00:18:40.15 & +16:25:08.08 & 0.5376 & \citep{obsz_ellingson} &  \\
00:18:40.20 & +16:25:06.61 & 0.5381 & \citep{obsz_crawford} & * \\
00:18:40.51 & +16:27:21.87 & 0.5460 & \citep{obsz_crawford} &  \\
00:18:40.59 & +16:27:19.48 & 0.5500 & \citep{obsz_ellingson} &  \\
00:18:41.03 & +16:26:20.32 & 0.5467 & \citep{obsz_crawford} &  \\
00:18:41.07 & +16:27:19.30 & 0.5432 & \citep{obsz_dressler} &  \\
00:18:41.12 & +16:24:59.05 & 0.5400 & \citep{obsz_crawford} &  \\
00:18:41.21 & +16:28:42.28 & 0.5370 & \citep{obsz_ellingson} &  \\
00:18:41.94 & +16:26:25.48 & 0.5583 & \citep{obsz_ellingson} &  \\
00:18:41.98 & +16:25:24.84 & 0.5573 & \citep{obsz_crawford} &  \\
00:18:42.07 & +16:25:26.68 & 0.5475 & \citep{obsz_crawford} &  \\
00:18:42.65 & +16:27:01.41 & 0.5420 & \citep{obsz_crawford} &  \\
00:18:42.87 & +16:27:25.20 & 0.5544 & \citep{obsz_ellingson} &  \\
00:18:43.46 & +16:26:08.27 & 0.5540 & \citep{obsz_ellingson} &  \\
00:18:43.56 & +16:26:06.03 & 0.5541 & \citep{obsz_crawford} & * \\
00:18:43.85 & +16:23:01.97 & 0.5484 & \citep{obsz_ellingson} &  \\
00:18:45.13 & +16:27:08.68 & 0.5389 & \citep{obsz_ellingson} &  \\
00:18:45.26 & +16:24:55.41 & 0.5518 & \citep{obsz_crawford} &  \\
00:18:45.29 & +16:27:06.53 & 0.5385 & \citep{obsz_crawford} &  \\
00:18:45.83 & +16:26:43.31 & 0.5393 & \citep{obsz_crawford} &  \\
00:18:45.90 & +16:27:48.48 & 0.5394 & \citep{obsz_crawford} &  \\
00:18:47.60 & +16:26:14.82 & 0.5528 & \citep{obsz_crawford} &  \\
00:18:48.15 & +16:25:56.10 & 0.5325 & \citep{obsz_crawford} &  \\
00:18:48.48 & +16:25:27.98 & 0.5486 & \citep{obsz_ellingson} &  \\
00:18:48.79 & +16:26:08.25 & 0.5443 & \citep{obsz_crawford} &  \\
00:18:49.14 & +16:29:25.55 & 0.5457 & \citep{obsz_ellingson} &  \\
00:18:49.38 & +16:27:14.87 & 0.5400 & \citep{obsz_ellingson} &  \\
00:18:49.50 & +16:27:12.51 & 0.5402 & \citep{obsz_crawford} & * \\
00:18:50.70 & +16:21:56.84 & 0.5677 & \citep{obsz_ellingson} &  \\
00:18:56.00 & +16:27:28.14 & 0.5414 & \citep{obsz_crawford} &  \\
00:18:56.61 & +16:27:21.11 & 0.5427 & \citep{obsz_crawford} & * \\
00:18:56.78 & +16:27:21.36 & 0.5431 & \citep{obsz_crawford} &  \\
00:19:01.17 & +16:27:19.94 & 0.5433 & \citep{obsz_crawford} &  \\
00:19:01.69 & +16:25:28.10 & 0.5418 & \citep{obsz_crawford} &  \\
00:19:03.29 & +16:24:56.11 & 0.5428 & \citep{obsz_crawford} &  \\
00:19:04.54 & +16:24:58.74 & 0.5463 & \citep{obsz_crawford} &     
\enddata 
\end{deluxetable*} \vspace{-4em}

\startlongtable
\begin{deluxetable*}{llccc} \label{tab:cl_members}
\tablehead{\colhead{RA} & \colhead{Dec} & \colhead{$z$} & \colhead{$\;\;\;\;$Observation date$\;\;\;\;$} & \colhead{Duplicate}}
\tablecaption{Cluster-member spectroscopic redshifts for \macs observed with \textit{DEIMOS} in our \textit{Keck} program. Objects which are observed as duplicates, and therefore excluded from our analysis, are indicated. Only objects within a $6.7' \times 6.7'$ region centered on the cluster center are used in the $v_{\text{gal}}$ map generation.}
\startdata
00:17:59.90 & +16:33:06.00 & 0.5645 & 16 August 2021 &  \\
00:18:02.56 & +16:17:34.60 & 0.5416 & 13 August 2020 &  \\
00:18:02.69 & +16:17:22.70 & 0.5539 & 13 August 2020 &  \\
00:18:08.32 & +16:18:05.30 & 0.5507 & 13 August 2020 &  \\
00:18:08.77 & +16:30:26.90 & 0.5455 & 16 August 2021 &  \\
00:18:09.54 & +16:14:07.50 & 0.5504 & 13 August 2020 &  \\
00:18:09.64 & +16:17:43.40 & 0.5512 & 13 August 2020 &  \\
00:18:09.84 & +16:20:14.10 & 0.5426 & 13 August 2020 &  \\
00:18:10.94 & +16:17:14.60 & 0.5475 & 13 August 2020 &  \\
00:18:11.99 & +16:16:51.00 & 0.5489 & 13 August 2020 &  \\
00:18:12.37 & +16:17:45.20 & 0.5473 & 13 August 2020 &  \\
00:18:12.38 & +16:19:18.50 & 0.5474 & 13 August 2020 &  \\
00:18:13.03 & +16:18:25.40 & 0.5551 & 13 August 2020 &  \\
00:18:13.69 & +16:32:01.30 & 0.5490 & 16 August 2021 &  \\
00:18:14.48 & +16:23:12.40 & 0.5473 & 13 August 2020 &  \\
00:18:14.64 & +16:21:44.30 & 0.5464 & 13 August 2020 &  \\
00:18:14.70 & +16:24:25.80 & 0.5488 & 13 August 2020 &  \\
00:18:15.78 & +16:19:39.00 & 0.5403 & 13 August 2020 &  \\
00:18:16.68 & +16:18:04.10 & 0.5511 & 13 August 2020 &  \\
00:18:17.24 & +16:26:53.20 & 0.5487 & 16 August 2021 &  \\
00:18:18.75 & +16:22:29.70 & 0.5390 & 13 August 2020 &  \\
00:18:19.93 & +16:19:34.90 & 0.5520 & 13 August 2020 &  \\
00:18:20.29 & +16:23:46.00 & 0.5370 & 13 August 2020 &  \\
00:18:20.58 & +16:20:07.80 & 0.5706 & 13 August 2020 &  \\
00:18:20.82 & +16:19:19.10 & 0.5451 & 13 August 2020 &  \\
00:18:20.92 & +16:27:02.90 & 0.5481 & 16 August 2021 &  \\
00:18:21.33 & +16:20:48.40 & 0.5468 & 13 August 2020 &  \\
00:18:21.54 & +16:20:12.60 & 0.5510 & 13 August 2020 &  \\
00:18:21.55 & +16:20:56.70 & 0.5513 & 13 August 2020 &  \\
00:18:21.70 & +16:24:52.40 & 0.5253 & 13 August 2020 &  \\
00:18:21.71 & +16:24:52.40 & 0.5253 & 13 August 2020 & * \\
00:18:21.84 & +16:27:55.40 & 0.5275 & 16 August 2021 &  \\
00:18:22.00 & +16:26:54.40 & 0.5477 & 16 August 2021 &  \\
00:18:22.08 & +16:29:27.30 & 0.5510 & 16 August 2021 &  \\
00:18:22.30 & +16:23:36.90 & 0.5452 & 13 August 2020 &  \\
00:18:22.90 & +16:19:16.00 & 0.5479 & 13 August 2020 &  \\
00:18:22.97 & +16:26:53.40 & 0.5275 & 16 August 2021 &  \\
00:18:23.63 & +16:26:27.50 & 0.5533 & 13 August 2020 &  \\
00:18:23.65 & +16:26:56.00 & 0.5537 & 13 August 2020 &  \\
00:18:24.01 & +16:22:00.30 & 0.5475 & 13 August 2020 & * \\
00:18:24.01 & +16:22:00.30 & 0.5481 & 13 August 2020 &  \\
00:18:24.05 & +16:27:46.70 & 0.5502 & 16 August 2021 &  \\
00:18:24.42 & +16:24:00.30 & 0.5480 & 13 August 2020 &  \\
00:18:24.77 & +16:26:43.10 & 0.5421 & 16 August 2021 &  \\
00:18:24.82 & +16:27:30.60 & 0.5363 & 13 August 2020 &  \\
00:18:25.06 & +16:26:34.10 & 0.5472 & 13 August 2020 &  \\
00:18:25.09 & +16:24:11.60 & 0.5349 & 13 August 2020 &  \\
00:18:25.10 & +16:22:07.70 & 0.5491 & 13 August 2020 &  \\
00:18:25.10 & +16:22:07.70 & 0.5479 & 13 August 2020 &  \\
00:18:25.55 & +16:23:47.30 & 0.5429 & 13 August 2020 &  \\
00:18:25.55 & +16:27:48.40 & 0.5420 & 16 August 2021 &  \\
00:18:25.71 & +16:25:21.30 & 0.5430 & 13 August 2020 &  \\
00:18:26.22 & +16:24:57.70 & 0.5500 & 13 August 2020 &  \\
00:18:26.23 & +16:23:24.80 & 0.5600 & 13 August 2020 &  \\
00:18:26.53 & +16:22:17.60 & 0.5523 & 13 August 2020 &  \\
00:18:26.67 & +16:26:43.70 & 0.5300 & 13 August 2020 &  \\
00:18:26.74 & +16:27:26.00 & 0.5393 & 13 August 2020 &  \\
00:18:27.04 & +16:28:28.60 & 0.5430 & 16 August 2021 &  \\
00:18:27.23 & +16:28:21.90 & 0.5523 & 16 August 2021 &  \\
00:18:27.44 & +16:23:02.90 & 0.5351 & 13 August 2020 &  \\
00:18:27.46 & +16:24:50.90 & 0.5511 & 13 August 2020 &  \\
00:18:28.08 & +16:24:31.10 & 0.5340 & 13 August 2020 &  \\
00:18:28.13 & +16:23:40.90 & 0.5419 & 13 August 2020 &  \\
00:18:28.29 & +16:24:17.30 & 0.5431 & 13 August 2020 &  \\
00:18:29.06 & +16:23:07.80 & 0.5557 & 13 August 2020 &  \\
00:18:29.10 & +16:24:11.90 & 0.5550 & 13 August 2020 &  \\
00:18:29.12 & +16:25:05.00 & 0.5541 & 16 August 2021 &  \\
00:18:29.77 & +16:25:47.70 & 0.5441 & 16 August 2021 &  \\
00:18:29.84 & +16:26:21.70 & 0.5451 & 13 August 2020 &  \\
00:18:30.00 & +16:25:17.00 & 0.5364 & 13 August 2020 &  \\
00:18:30.09 & +16:26:07.20 & 0.5455 & 13 August 2020 &  \\
00:18:30.45 & +16:26:35.30 & 0.5476 & 13 August 2020 &  \\
00:18:30.61 & +16:24:51.20 & 0.5466 & 16 August 2021 &  \\
00:18:31.18 & +16:28:33.00 & 0.5454 & 13 August 2020 &  \\
00:18:31.37 & +16:24:31.80 & 0.5600 & 13 August 2020 &  \\
00:18:31.83 & +16:25:52.80 & 0.5425 & 16 August 2021 &  \\
00:18:32.06 & +16:28:26.80 & 0.5439 & 16 August 2021 &  \\
00:18:32.46 & +16:24:34.40 & 0.5476 & 13 August 2020 &  \\
00:18:32.49 & +16:30:54.30 & 0.5522 & 13 August 2020 &  \\
00:18:32.53 & +16:26:00.20 & 0.5440 & 13 August 2020 &  \\
00:18:32.69 & +16:26:55.70 & 0.5410 & 13 August 2020 &  \\
00:18:32.88 & +16:24:30.30 & 0.5450 & 16 August 2021 &  \\
00:18:33.07 & +16:24:04.80 & 0.5426 & 13 August 2020 &  \\
00:18:33.33 & +16:27:01.20 & 0.5500 & 13 August 2020 &  \\
00:18:33.99 & +16:26:36.50 & 0.5398 & 13 August 2020 &  \\
00:18:34.14 & +16:30:03.50 & 0.5507 & 13 August 2020 &  \\
00:18:34.93 & +16:28:37.00 & 0.5403 & 13 August 2020 &  \\
00:18:35.01 & +16:24:17.20 & 0.5520 & 13 August 2020 &  \\
00:18:36.17 & +16:29:54.00 & 0.5542 & 16 August 2021 &  \\
00:18:37.07 & +16:32:30.50 & 0.5484 & 13 August 2020 &  \\
00:18:37.27 & +16:24:40.00 & 0.5511 & 13 August 2020 &  \\
00:18:37.43 & +16:31:08.20 & 0.5480 & 13 August 2020 &  \\
00:18:37.79 & +16:30:16.60 & 0.5543 & 13 August 2020 &  \\
00:18:38.77 & +16:28:57.70 & 0.5400 & 13 August 2020 &  \\
00:18:39.22 & +16:23:50.80 & 0.5542 & 16 August 2021 &  \\
00:18:39.23 & +16:30:11.50 & 0.5530 & 16 August 2021 &  \\
00:18:39.66 & +16:23:45.80 & 0.5393 & 16 August 2021 &  \\
00:18:39.78 & +16:28:56.90 & 0.5373 & 13 August 2020 &  \\
00:18:40.58 & +16:26:38.50 & 0.5315 & 13 August 2020 &  \\
00:18:42.51 & +16:29:55.87 & 0.5521 & 16 August 2021 &  \\
00:18:42.64 & +16:29:56.70 & 0.5541 & 16 August 2021 &  \\
00:18:47.06 & +16:31:59.70 & 0.5490 & 16 August 2021 &  \\
00:18:48.12 & +16:29:23.50 & 0.5505 & 16 August 2021 &  \\
00:18:48.73 & +16:32:10.00 & 0.5502 & 13 August 2020 &  \\
00:18:50.05 & +16:31:51.00 & 0.5456 & 16 August 2021 &  \\
00:18:50.36 & +16:31:42.50 & 0.5430 & 13 August 2020 &  \\
00:18:50.76 & +16:32:13.20 & 0.5416 & 16 August 2021 &  \\
00:18:51.58 & +16:32:41.80 & 0.5420 & 13 August 2020 &  \\
00:18:55.41 & +16:33:07.00 & 0.5414 & 13 August 2020 &  \\
00:18:55.81 & +16:31:50.10 & 0.5409 & 13 August 2020 &  \\
00:18:58.12 & +16:32:07.70 & 0.5512 & 16 August 2021 &  \\
00:19:00.94 & +16:32:28.60 & 0.5468 & 16 August 2021 &  \\
00:19:01.18 & +16:31:11.00 & 0.5472 & 16 August 2021 &  \\
00:19:01.24 & +16:32:59.80 & 0.5405 & 16 August 2021 &  \\
00:19:02.20 & +16:34:16.00 & 0.5480 & 16 August 2021 &  \\
00:19:05.21 & +16:30:32.50 & 0.5367 & 16 August 2021 &  \\
00:19:06.62 & +16:34:23.60 & 0.5507 & 16 August 2021 & 
\enddata 
\end{deluxetable*} \vspace{-4em}

\startlongtable
\begin{deluxetable*}{llcc} \label{tab:noncl_members}
\tablehead{\colhead{RA} & \colhead{Dec} & \colhead{$z$} & \colhead{$\;\;\;\;$Observation date$\;\;\;\;$}}
\tablecaption{Non-cluster-member spectroscopic redshifts observed with \textit{DEIMOS} in our \textit{Keck} program. These redshifts were not used in the analysis, but we include this table for completeness.}
\startdata
00:18:00.17 & +16:32:50.60 & 0.3295 & 16 August 2021 \\
00:18:01.19 & +16:34:55.90 & 0.4750 & 16 August 2021 \\
00:18:03.04 & +16:30:54.50 & 0.2375 & 16 August 2021 \\
00:18:03.37 & +16:34:11.30 & 0.7404 & 16 August 2021 \\
00:18:03.44 & +16:32:06.90 & 0.4870 & 16 August 2021 \\
00:18:04.92 & +16:33:06.90 & 0.4050 & 16 August 2021 \\
00:18:05.84 & +16:33:42.10 & 0.1935 & 16 August 2021 \\
00:18:07.66 & +16:15:46.10 & 0.4604 & 13 August 2020 \\
00:18:08.42 & +16:31:55.30 & 0.9875 & 16 August 2021 \\
00:18:11.39 & +16:31:41.00 & 0.2810 & 16 August 2021 \\
00:18:11.89 & +16:32:19.70 & 0.3876 & 16 August 2021 \\
00:18:13.25 & +16:18:15.20 & 0.7253 & 13 August 2020 \\
00:18:14.15 & +16:29:28.20 & 0.6550 & 16 August 2021 \\
00:18:14.47 & +16:27:13.10 & 0.4700 & 16 August 2021 \\
00:18:15.90 & +16:18:32.50 & 0.7253 & 13 August 2020 \\
00:18:17.38 & +16:32:40.30 & 0.7400 & 16 August 2021 \\
00:18:19.43 & +16:32:36.00 & 0.2670 & 16 August 2021 \\
00:18:20.37 & +16:31:40.70 & 0.6250 & 16 August 2021 \\
00:18:21.12 & +16:29:26.80 & 0.3280 & 16 August 2021 \\
00:18:21.46 & +16:20:30.40 & 0.7732 & 13 August 2020 \\
00:18:21.55 & +16:20:56.70 & 0.7733 & 13 August 2020 \\
00:18:22.00 & +16:29:02.30 & 0.3263 & 16 August 2021 \\
00:18:22.26 & +16:29:20.40 & 0.3300 & 16 August 2021 \\
00:18:23.20 & +16:31:40.10 & 0.4066 & 16 August 2021 \\
00:18:27.53 & +16:28:47.90 & 0.3926 & 16 August 2021 \\
00:18:28.15 & +16:28:08.20 & 0.6559 & 16 August 2021 \\
00:18:28.15 & +16:28:08.20 & 0.6558 & 13 August 2020 \\
00:18:31.12 & +16:26:40.80 & 0.6555 & 13 August 2020 \\
00:18:31.37 & +16:28:08.70 & 0.6220 & 13 August 2020 \\
00:18:31.38 & +16:28:08.70 & 0.6220 & 16 August 2021 \\
00:18:33.88 & +16:28:11.30 & 0.1650 & 16 August 2021 \\
00:18:34.42 & +16:28:53.90 & 0.6210 & 16 August 2021 \\
00:18:34.43 & +16:28:53.90 & 0.6213 & 13 August 2020 \\
00:18:34.54 & +16:28:09.30 & 0.8086 & 13 August 2020 \\
00:18:34.81 & +16:28:21.70 & 0.6221 & 13 August 2020 \\
00:18:37.54 & +16:27:50.60 & 0.6200 & 13 August 2020 \\
00:18:38.37 & +16:28:04.20 & 0.6170 & 13 August 2020 \\
00:18:39.33 & +16:30:31.70 & 0.6410 & 16 August 2021 \\
00:18:39.79 & +16:26:41.80 & 0.6560 & 13 August 2020 \\
00:18:40.51 & +16:28:14.20 & 0.6197 & 13 August 2020 \\
00:18:40.53 & +16:23:29.70 & 0.8987 & 16 August 2021 \\
00:18:42.82 & +16:28:53.20 & 0.7284 & 13 August 2020 \\
00:18:44.17 & +16:32:16.50 & 0.6239 & 13 August 2020 \\
00:18:44.91 & +16:29:33.40 & 0.6219 & 16 August 2021 \\
00:18:44.91 & +16:29:33.40 & 0.6226 & 13 August 2020 \\
00:18:45.02 & +16:31:45.00 & 0.7165 & 13 August 2020 \\
00:18:45.54 & +16:30:20.20 & 0.8002 & 16 August 2021 \\
00:18:45.77 & +16:31:42.00 & 0.7157 & 16 August 2021 \\
00:18:45.87 & +16:30:57.30 & 0.3300 & 16 August 2021 \\
00:18:46.63 & +16:25:47.50 & 0.4388 & 16 August 2021 \\
00:18:48.87 & +16:32:52.30 & 0.6575 & 16 August 2021 \\
00:18:50.08 & +16:25:01.70 & 0.1757 & 16 August 2021 \\
00:18:50.97 & +16:33:10.40 & 0.7163 & 16 August 2021 \\
00:18:53.69 & +16:32:03.50 & 0.3870 & 16 August 2021 \\
00:18:53.86 & +16:32:41.70 & 0.6547 & 16 August 2021 \\
00:18:54.32 & +16:30:38.10 & 0.5820 & 16 August 2021 \\
00:18:55.70 & +16:33:03.40 & 0.9952 & 16 August 2021 \\
00:19:00.83 & +16:33:56.30 & 0.6262 & 16 August 2021 \\
00:19:01.57 & +16:34:15.10 & 0.3050 & 16 August 2021 \\
00:19:02.77 & +16:35:06.70 & 0.6249 & 16 August 2021 \\
00:19:06.51 & +16:31:14.70 & 0.2620 & 16 August 2021 \\
00:19:09.99 & +16:30:53.80 & 0.3300 & 16 August 2021 \\
00:19:10.93 & +16:33:39.40 & 0.4780 & 16 August 2021 \\
00:19:12.25 & +16:31:51.30 & 0.3610 & 16 August 2021 
\enddata 
\end{deluxetable*}
\end{document}